\documentclass[aps,prb,reprint,floatfix,superscriptaddress,letterpaper]{revtex4-2}

\usepackage{graphicx}
\usepackage{bm}
\usepackage{amsmath}
\usepackage{amssymb}
\usepackage{amsfonts}
\usepackage{xfrac}
\usepackage{xcolor}

\begin{document}
\nocite{Ultramicroscopy}

\title{Fast, efficient, and accurate dielectric screening using a local, real-space approach}

\author{John Vinson}
\email{john.vinson@nist.gov}
\affiliation{Material Measurement Laboratory, National Institute of Standards and Technology, Gaithersburg, MD 20899}

\author{Eric L. Shirley}
\affiliation{Physical Measurement Laboratory, National Institute of Standards and Technology, Gaithersburg, MD 20899}

\date{\today}

\begin{abstract}
Various many-body perturbation theory techniques for calculating electron behavior rely on {\it W}, the screened Coulomb interaction. Computing {\it W}
requires complete knowledge of the dielectric response of the electronic system, and the fidelity of the calculated dielectric response limits the reliability of predicted electronic and structural properties. 
As a simplification, calculations often begin with the random-phase approximation (RPA). 
However, even RPA calculations are costly and scale poorly, typically as $N^4$ ($N$ representing the system size). 
A local approach has been shown to be efficient while maintaining accuracy for screening core-level excitations [Ultramicroscopy {\bf 106}, 986 (2006)]. 
We extend this method to valence-level excitations. 
We present improvements to the accuracy and execution of this scheme, including reconstruction of the all-electron character of the pseudopotential-based wave functions, improved $N^2\log N$ scaling, and a parallelized implementation. 
We discuss applications to Bethe-Salpeter equation (BSE) calculations of core and valence spectroscopies. 
\end{abstract}

\maketitle

\section{Introduction}

For condensed matter systems, one of the fundamental properties is the dielectric response. 
This is closely related to the polarizability, the movement of the constituent, electrically charged
electrons and ions that 
is responsible for any difference between the applied and total potentials. 
Here we limit our investigation to the electronic behavior of the system, treating the ions as stationary.  
The dielectric response determines conductivity as well as frequency-dependent absorption and transmission of photons. 
The polarizability is used in many-body perturbation theory to more accurately determine electronic properties. 
In calculations of a variety of properties of condensed matter systems, band alignment, optical absorption, adsorption energies, etc., the determination of the electronic response plays a vital role. 

Calculations of electron polarizability in condensed systems are 
commonplace. 
For a periodic system (infinite crystal), the polarizability within the random-phase approximation (RPA) is typically calculated in reciprocal space based on the spectral representation of the Green's function or a sum over states \cite{PhysRev.129.62}. 
More recently, real-space approaches have also been considered \cite{PhysRevLett.74.1827,PhysRevB.52.R2225,PhysRevB.94.085125}. 
Alternative methods based on many-body perturbation theory have also been suggested \cite{PhysRevB.81.115105, PhysRevB.85.081101}. 
However, all of these approaches scale with system size $N$ roughly as $N^4$ \cite{DELBEN2019187}.
Imaginary-time techniques have been shown to reduce the scaling to $N^3$ \cite{RIEGER1999211,PhysRevB.90.054115,PhysRevB.101.035139}. 
Here we present improvements to the localized, real-space approach originally introduced in Ref.~\cite{Ultramicroscopy}. 
We revisit the local, real-space approach for two broad reasons: improvements of the accuracy of the calculations and improvements to the execution of the code, yielding a scaling of $N^2 \log N$. 
Highly accurate calculations are necessary to understand subtle changes in spectral features due to changes in crystal or electronic structure.  
Improved algorithmic scaling and an efficient implementation are necessary to make screening calculations practical for large systems.

In the first part of this paper, we extend the calculation of the polarizability to include all-electron projectors and examine the effect of the pseudopotential approximation on the screened Coulomb potential. 
We find that pseudopotential-based wave functions yield an incorrect polarizability for small distances, {\it e.g.}, inside the pseudopotential radius. 
The removal of the core-level electrons removes nodes from the valence atomic orbitals, 
changing the density distribution of the valence orbitals. 
The discrepancy is small, but has a noticeable effect on calculated near-edge x-ray absorption spectra, where the strength of excitonic binding is highly dependent on the screening provided by the valence electrons. 
We have extended this screening approach for use in valence excitations. 
In the case of valence excitations (UV/optical spectra), we find that augmentation is not necessary.

Second, we present a substantial improvement to the system-size scaling of RPA calculations. 
Our method scales as $N^2 \log N$, while still performing well for small systems sizes. 
It also does not require the dense {\it k}-point grids of traditional reciprocal-space methods. 
Even for small unit cells, a $2^3$ {\it k}-point grid can be sufficient.
This not only reduces the computational cost of the screening calculation itself, but also the cost of generating electron wave functions on a dense {\it k}-point grid. 
The calculation of the polarizability has been implemented as a hybrid {\sc openmp} + {\sc mpi} code, allowing screening calculations of large systems to be carried out quickly. 
The local, real-space approach also provides an ideal testbed for investigating higher-order screening methods.

In Section II, we review polarizability and screening and the main approximations of the real-space approach. 
In Section III, we detail how the use of pseudopotentials affects the orbitals near an atom which in turn modifies the electronic screening. 
We introduce the optimal projector functions which are used to augment the pseudopotential-based orbitals, restoring their all-electron character, and we examine the effect this has on calculations of the screened Coulomb potential. 
In Section IV, we illustrate the effect of augmentation on Bethe-Salpeter equation calculations of x-ray absorption through modifications of the core-hole screening. 
In Section V, we generalize the local, real-space approach for use in calculating valence (optical/UV) excitation spectra, and we show that augmentation is unnecessary for valence calculations. 
In Section VI, we detail the performance of our implementation, demonstrating the superior system-size scaling of our method. 
Lastly, in Section VII we summarize future enhancements and applications for our method, including an extension to beyond-RPA screening.

\section{Review}

\subsection{Polarizability and Screening}

We begin by reviewing the definitions of the polarizability and the dielectric response that screens an applied, external potential. 
We use Hartree atomic units throughout, such that the electron charge, electron mass, 
Planck's constant and Coulomb's constant are given by $e = m_e = \hbar = 4\pi\epsilon_0 = 1$.
We make use of the one-electron Green's function 
\begin{align}
\label{eq-g1}
g^{-1}(1,2) &= g_0^{-1}(1,2) -V(1,2) - \Sigma(1,2)
\end{align}
where each numerical index denotes position, time, and spin, $g_0$ is the non-interacting Green's function, $V$ is the total potential, which is local and so contains a factor $\delta(1-2)$, and $\Sigma$ is the self-energy encompassing many-body interactions.

The irreducible polarizability $\chi_0$ of the electron system is the change in electron density $n$ in response to a change in the total potential $V$,
\begin{equation}
\chi_0(1,2) = \delta n(1) / \delta V(2) \; .
\end{equation}
The density can be written in terms of the one-electron Green's function 
\begin{equation}
n(1) = -i g(1,1^+) ,
\end{equation}
where $1^+$ refers to a time infinitesimally later than $1$. 
Taking functional derivatives with respect to the potential in Eq.~\ref{eq-g1}, the polarizability can be written
\begin{align}
\chi_0(1,2) &= -i \delta g(1,1^+) / \delta V(2) \nonumber \\
&= \int \!\! d3d4 \; i g(1,3) \frac{ \delta g^{-1}(3, 4)}{\delta V(2)} g(4,1^+)
.
\end{align}
By approximating the Green's function using $g^{-1} = g_0^{-1} - V$, 
we arrive at the random phase approximation for the polarizability, 
\begin{align}
\chi_0^\textrm{RPA}(1,2) = -i g_0(1,2) g_0(2,1^+) .
\label{eq-rpa}
\end{align}
This can be transformed from the two-time representation to the response as a function of a single external energy $\omega$ and written in real-space:  
$\chi_0(\mathbf{r}, \mathbf{r}', \omega)$. 

Above, in Eq.~\ref{eq-g1}, the potential term includes both the external and Hartree terms, and the Hartree term itself will change with changes in the electron density. 
One should therefore use the reducible polarizability $\chi$ which is the response to only changes in the external potential, 
\begin{align}
\label{eq-reducible}
\chi(\mathbf{r},\mathbf{r}',\omega) &=  \chi_0(\mathbf{r},\mathbf{r'},\omega) \nonumber \\ &+ \chi_0(\mathbf{r},\mathbf{x},\omega) v(\mathbf{x},\mathbf{x'}) \chi(\mathbf{x'},\mathbf{r'},\omega) 
\end{align}
where repeated spatial indices $\mathbf{x}$ are integrated over. 
Here $v$ is the Coulomb operator. 
Most importantly, from the reducible polarizability one obtains the screened Coulomb operator $W(\mathbf{r},\mathbf{r}',\omega)$ 
\begin{align}
\label{eq-W} 
W(\mathbf{r},\mathbf{r}',\omega) &=  \epsilon^{-1}(\mathbf{r},\mathbf{x},\omega) v_\textit{ext}(\mathbf{x},\mathbf{r}') \\ 
&= v_\textit{ext}(\mathbf{r},\mathbf{r}') + v(\mathbf{r}, \mathbf{x}) \chi(\mathbf{x},\mathbf{x}',\omega) v_\textit{ext}(\mathbf{x}',\mathbf{r}'), 
 \nonumber
\end{align}
where $\epsilon$ is the dielectric tensor. 
This is central to many-body perturbation techniques, for instance, for treating single-particle self-energies via the {\it GW} method  
or electron-hole excitation calculations via the BSE \cite{RevModPhys.74.601}.

\subsection{Real-space decomposition of the screening}
\label{sec-localreview}

One can divide a potential into pieces and separately consider the screening of each piece. 
The two-coordinate external Coulomb operator in Eq.~\ref{eq-W} can, without loss of generality, be written with a strength $q$ and parametric dependance on the second spatial coordinate 
\begin{equation}
v_\textit{ext}^{[\mathbf{r}']}(\mathbf{x}) = q \vert \mathbf{x}\vert^{-1} = q \vert \mathbf{r} - \mathbf{r}' \vert^{-1}
\end{equation}
where $\mathbf{x=r-r'}$. Following Ref.~\cite{Ultramicroscopy}, this potential is partitioned by adding and subtracting a shell of charge:
\begin{align}
\label{eq-partition}
v_\textit{ext}(\mathbf{x}) &= v_1(\mathbf{x}) + v_2(\mathbf{x}) ,  \nonumber \\
v_1(\mathbf{x}) &= v_\textit{ext}(\mathbf{x}) - v_2(\mathbf{x}) ,  \\
v_2(\mathbf{x}) &= q R_S^{-1} \Theta[ R_S - x ] + q x^{-1} \Theta[ x - R_S ]  . \nonumber
\end{align}
Here $\Theta$ is the Heaviside theta function and $R_S$ is the shell radius. 
The screened potential is therefore 
\begin{align}
\label{W1}
W^{[\mathbf{r}']}(\mathbf{r},\omega) & = \epsilon^{-1}(\mathbf{r,x=r-r'},\omega) \, v^{[\mathbf{r}']}_\textit{ext}(\mathbf{x}) \nonumber \\
& = \epsilon^{-1}v_1 + \epsilon^{-1}v_2  \\
&= W^{(1)} \; + W^{(2)} .  \nonumber 
\end{align}
To this point no approximation has been made in the screening of $v_\textrm{ext}$. 
The full, two-coordinate $W(\mathbf{r,r'})$ is recovered by evaluating the decomposed, real-space screening at each $\mathbf{r'}$, as is shown for valence BSE calculations in section~V.

\subsection{Local, Real-space Approach}
\label{sec-localreview}

The utility of the local, real-space approach rests on treating the screening of $v_1$ and $v_2$ differently. 
Because $v_1$ is nonzero only within $R_S$, the dielectric response that screens $v_1$ is also localized. 
By only calculating the polarizability within a finite volume, we are able to reduce the computational cost of the calculation. 
In contrast, the screening of $v_2$ must still cover all space. 
We therefore treat the screening of $v_2$ only approximately, using a model for the dielectric response. 

Our method is therefore an approximation to the RPA screening, controlled by the shell radius $R_S$ and the model screening used to screen $v_2$. 
As $R_S$ increases, $v_2$ goes to zero, and treating $v_2$ approximately is a 
\emph{controlled} approximation.  
The effect of dividing the calculation at finite $R_S$ is addressed further in section~\ref{layered} and appendix~\ref{sec-truncation}.  
A modified Levine-Louie dielectric model was adopted in Ref.~\cite{Ultramicroscopy} and 
is used 
here \cite{PhysRevB.25.6310,Shirley2005}. 
This model screening is parametrized by the static, long-range dielectric constant $\epsilon_\infty$ and the average and local valence electron densities. 

This choice makes $\epsilon_\infty$ an input parameter, but  
small errors in  $1/\epsilon_\infty$ will only lead to small errors in the resulting x-ray absorption spectra, and we investigate this further in section \ref{layered} and appendix~\ref{sec-approx-eps}.
As suggested in Ref.~\cite{Ultramicroscopy}, the preferred source is an experimental measurement of the optical constants or index of refraction below the band gap. 
If such data are not available, several computational approaches can be used. 
The optical spectra can be calculated with the {\sc ocean} code \cite{ocean0,ocean1}, either within 
the RPA or within the Bethe-Salpeter equation (BSE) approximation, which formally requires iteration to equate the input and output $\epsilon_\infty$.

The short-ranged potential $v_1$ is screened by calculating the RPA polarizability. 
Typically we use a shell radius $R_S$ between 3~a.u.\ and 5~a.u. 
The polarizability is calculated within a spherical region of space given by a radius of 8~a.u.\ to 10~a.u. 
Per Eq.~\ref{eq-rpa}, this necessitates calculating the one-electron Green's functions for this region of space. 
Improvements in the fidelity, efficiency, and parallelization of calculating $g$, $\chi_0$, and $\chi$ in order to screen $v_1$ are the focus of this work.

\section{ Augmentation of electron orbitals }

\subsection{Pseudopotentials}

The screening method of Ref.~\cite{Ultramicroscopy}, adopted in the {\sc ocean} spectroscopy code, utilizes electron wave functions generated from a pseudopotential-based density-functional theory code. 
The pseudopotential approximation allows for a dramatic reduction in the computational cost of planewave codes.
The core-level electrons are removed, and the $-Z/r$ potential of each ion is replaced by a pseudopotential, such that the valence electrons are the most-bound states. 
This reduces the computational by treating fewer electrons and smoothing the remaining electron wave functions, reducing the required number of planewaves. 
More information on pseudopotential theory and methods can be found in \cite{RMartin}.

Consider a pseudopotential $V^\textit{ps}_\textrm{ion}$ for an element which treat angular-momentum-dependent effects separably 
following the Kleinman-Bylander form \cite{PhysRevLett.48.1425}. 
For each value of the principle angular momentum $l$ up to $l_\textrm{max}$ it consists of some local potential and optionally some non-local projectors. 
In contrast, the all-electron ionic potential is simply an attractive Coulomb potential set by the ion's atomic number $Z$. 
Both the all-electron ($ae$) and pseudized ($ps$) radial Schr{\" o}dinger equations can be solved numerically. 
\begin{align}
\label{eq-atom}
&H^{ae/ps} \phi_j^{ae/ps} = \varepsilon_j \phi_j^{ae/ps} \\
&H^{ae/ps} = 
 -\frac{\nabla^2}{2}  +\frac{l(l+1)}{2r^2}+ V_H[n] + V_{xc}[n] + V^{ae/ps}_{\textrm{ion}} \, \nonumber 
 \end{align}
The Hartree $V_H$ and DFT exchange-correlation $V_{xc}$ potentials both depend on the density $n$, and so the problem must be solved self-consistently. 
Outside of some cutoff radius $r_c$ the pseudopotential matches the all-electron ionic potential, and the pseudo and all-electron electron orbitals are equal for an isolated atom. 
Additional requirements are enforced at $r_c$.  
The all-electron and pseudo orbitals should have the same radial derivative and scattering length, although this is exact only for specific energies. 
Because the scattering properties are the same for the all-electron and pseudo wave functions, 
the behavior of the pseudo electrons in the interstitial regions is identical to that of the electrons in the all-electron system. 

While pseudopotentials are capable of reproducing all-electron results for many properties, 
such as band structures (Bloch-state energies) or structural properties, they fall short in others, such as core-level excitations. 
This is unsurprising because there are no core-level electrons in the pseudo-potential system. 
We point to localization and location of a perturbation or transition operator in determining if the pseudopotential wave functions are sufficient. 
In the case of core-level transitions, the operator is both highly-localized and within the pseudized region. 
Conversely, structural properties are not localized. They rely on force constants, 
which are a measure of the change in the distribution electron density as felt by one atom in response to the motion of another. 
In the case of a real-space screening approach, we are interested in the density response to a localized perturbation, 
and for screening a core hole this localized perturbation is within the pseudized region.

\begin{figure}
\begin{centering}
\includegraphics[height=3.1in,trim=0 35 0 60,angle=270]{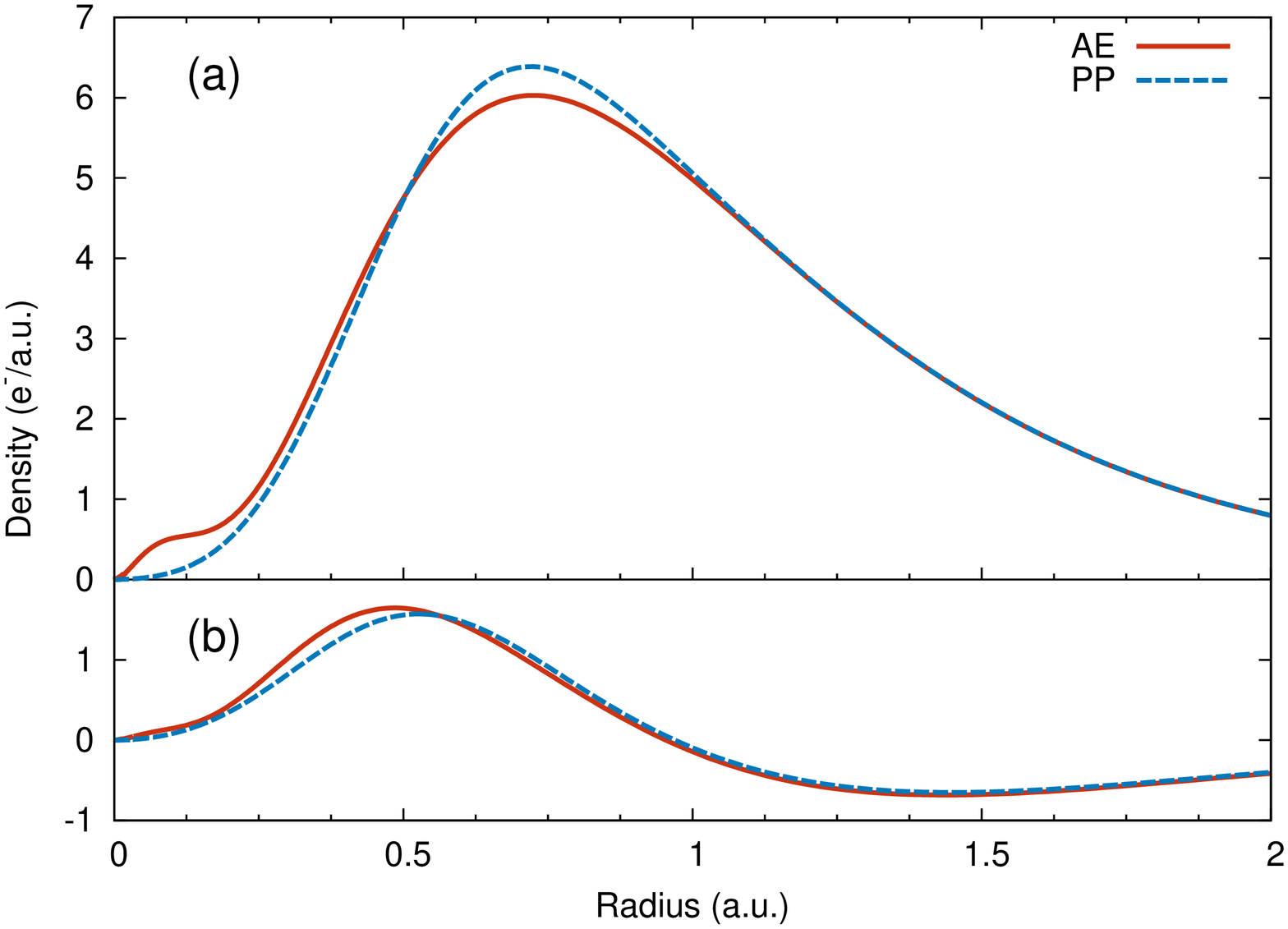} \\
\includegraphics[height=3.1in,trim=0 57 15 50,angle=270]{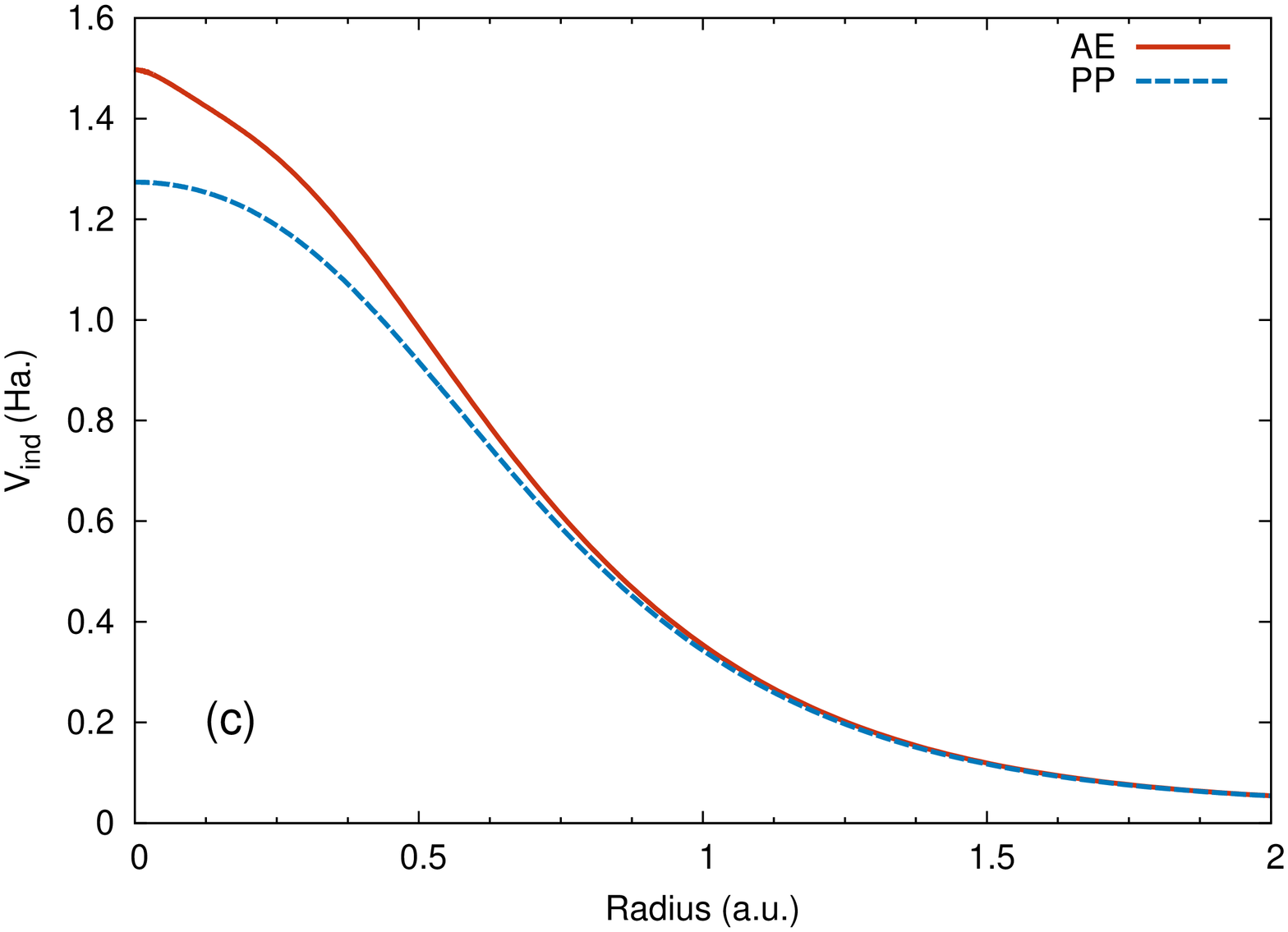}
\caption{ (a) Valence electron density for an isolated fluorine atom for both the all-electron (AE, solid red) and pseudopotential (PP, dashed blue) systems. In the pseudopotential system the 2{\it s} orbitals have no nodes. This is why the all-electron systems has more density at very small radii, inside the 2{\it s} node at 0.2~a.u. (b) The change in valence density of the fluorine atom in response to a 1{\it s} core hole. (c) The induced potential that arises from the change in the valence density. Even though the difference in densities is quite small, the subsequent difference in the induced potential is substantial at short ranges.   }
\label{plot-f}
\end{centering}
\end{figure}

To demonstrate the differences between an all-electron and pseudopotential screened core-hole, 
in Fig.~\ref{plot-f}(a) we show the difference in the valence (2{\it s} and 2{\it p}) electron densities between the all-electron and the pseudopotential in a fluorine atom. 
While the all-electron 2{\it s} orbital has a node (and is orthogonal to the 1{\it s} orbital), the pseudopotential 2{\it s} does not. 
The first anti-node  
of the all-electron 2{\it s} is responsible for the difference in densities below around 0.2~a.u. 
Around 0.7~a.u.\ the pseudopotential density exceeds the all-electron density, a consequence of norm conservation and the deficiency at small radii.  
In Fig.~\ref{plot-f}(b) we plot the change in valence electron density upon introduction of a 1{\it s} core hole in the system. 
For both systems the valence electron density moves to smaller radii (towards the positively charged core-hole), but for the all-electron system there is a larger change at short distances. 

The small difference in density response is magnified when converted into an induced potential as shown in Fig.~\ref{plot-f}(c). 
Clearly, at short distances the all-electron valence orbitals are more efficient at screening core-level excitations than the pseudopotential orbitals. 
In section \ref{xas-examples} we will explore how this difference in induced potential 
affects calculated x-ray absorption spectra that can be compared to measured ones.  

In the context of screening a core hole, the potential $v_{\textit {ext}}$ is already weakened by relaxation of the core orbitals that are not included explicitly when valence screening effects are computed.
This largely prevents problems that could arise from a frozen-core approximation regarding changes in the core level occupancies.  
Smaller, additional core-relaxation effects arise when there are changes in the chemical environment.  
These can include core polarization (mostly dipolar) 
and relaxational self-energies 
of valence electrons or holes added onto an atomic site.  
One method to treat many of these effects is presented by 
Shirley and Martin~\cite{PhysRevB.47.15404}, who also refer to previous work.  

\subsection{Optimal Projector Functions}

To include the correct all-electron behavior, we augment wave functions of the pseudopotential system.
This augmentation style was introduced for core-level transitions \cite{SHIRLEY20051187,ocean0} and borrows heavily from the projector-augmented wave (PAW) method \cite{PhysRevB.50.17953}. 
In the PAW method, Bloch states are augmented by projecting smoothened wave functions onto a basis set of projector functions.  For DFT ground-state calculations, it may be advantageous to optimize the projection scheme for describing the highest occupied and lowest unoccupied states, whereas our emphasis is obtaining realistic electron wave functions over a wide range of occupied and unoccupied bands.  The coefficient of each projector function weighs the correction of the wave function in the form of replacing the projector function with an all-electron counterpart.  

In {\sc ocean}, all-electron and pseudo versions of partial waves are evaluated and condensed into a set of optical projector functions (OPFs).  
Typically, partial waves are sampled at several dozen regularly spaced energies over a multi-hartree energy range. 
The partial waves are constructed within an augmentation radius the encompasses the the pseudotpotential cut-off radius $r_a \ge r_c$. 
The OPFs are the eigenvectors that have the largest eigenvalues of the overlap matrix between all of the partial waves, considering each angular momentum separately. 
As a result, the OPFs are not single-energy partial waves, in contrast to typical PAW construction. 
The eigenvalues fall of rapidly enough that over 99.9~\% of the wave function's degrees of freedom are sampled with only a few OPFs. 
This is described in further detail in appendix~\ref{app-OPFs}.

Augmentation exploits the same properties that we earlier asserted our pseudopotential would have. 
Namely, outside of the pseudopotential cut-off radius $r_c$ the pseudo wave function is identical to the all-electron wave function. 
Over a reasonable energy range, a bound or scattering state can be transformed from the pseudopotential system to the all-electron system by replacing the pseudo wave function inside of the cut-off radius with the all-electron solution.
The OPF method is only used for augmentation to reconstruct wave functions as a means of post-processing DFT results, and the implementation is discussed in section~\ref{implementation-augment}.

\subsection{ Approximate augmentation }
\label{aa}

Previously, the {\sc ocean} code relied on an approximation instead of carrying out augmentation of the wave functions. 
We document the old approach here, and in the next section we will compare it to the current method. 
As was shown in Fig.~\ref{plot-f}(c), even the isolated atom demonstrates the importance of all-electron wave functions when calculating the screening near the nucleus. 
The difference between the two induced potentials in Fig.~\ref{plot-f}(c) can be calculated purely within the isolated atomic case
\begin{equation}
\Delta v_\textrm{ind}(r) = v^{ae}_\textrm{ind}(r) - v^{ps}_\textrm{ind}(r) \; .
\end{equation}
Previously this correction was at times applied to the induced potential as calculated within the RPA using the {\it un-augmented} wave functions of the system.  
In this way the approximate effect of augmentation was included in the screening. 
However, because this method relies on calculations of an isolated atom, the valence orbitals and their occupations are only a rough approximation to the system of interest.

\subsection{Screening with augmented orbitals}

\begin{figure}
\begin{centering}
\includegraphics[height=3.1in,trim=0 35 15 70,angle=270]{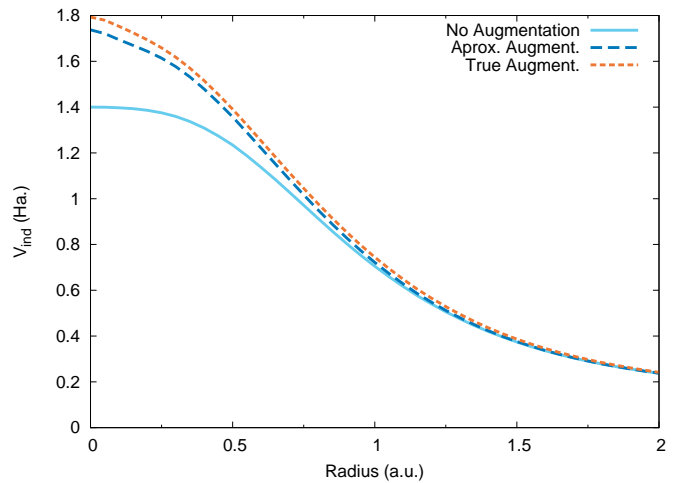}
\caption{ The spherically averaged induced potential in response to a F 1{\it s} hole in LiF. The same calculation is carried out three times using different treatments for correcting the electron orbitals near the F atom (see text). The difference between the no augmentation and true augmentation potentials has the same origin as in Fig.~\ref{plot-f}(c) for the case of an isolated F atom. }
\label{pot-f1s}
\end{centering}
\end{figure}

Before showing the effects of augmentation on calculated spectra, we examine the effects on the screening, using LiF as a model system.
In Fig.~\ref{pot-f1s} we show the spherically averaged induced potential in response to a F 1{\it s} core hole. 
As in the case of the isolated atom (Fig.~\ref{plot-f}), the induced potential is stronger near the origin when the orbitals have all-electron character, and the effect of augmentation is only significant near the atom where the all-electron and pseudo-potential orbitals are different. 
The approximate augmentation, while exact for an isolated atom, does not fully reproduce the screening using augmented orbitals within a solid. 
It is also only valid for atom-centered response such as for core-level spectroscopy.

\begin{figure}
\begin{centering}
\includegraphics[height=3.1in,trim=0 35 15 70,angle=270]{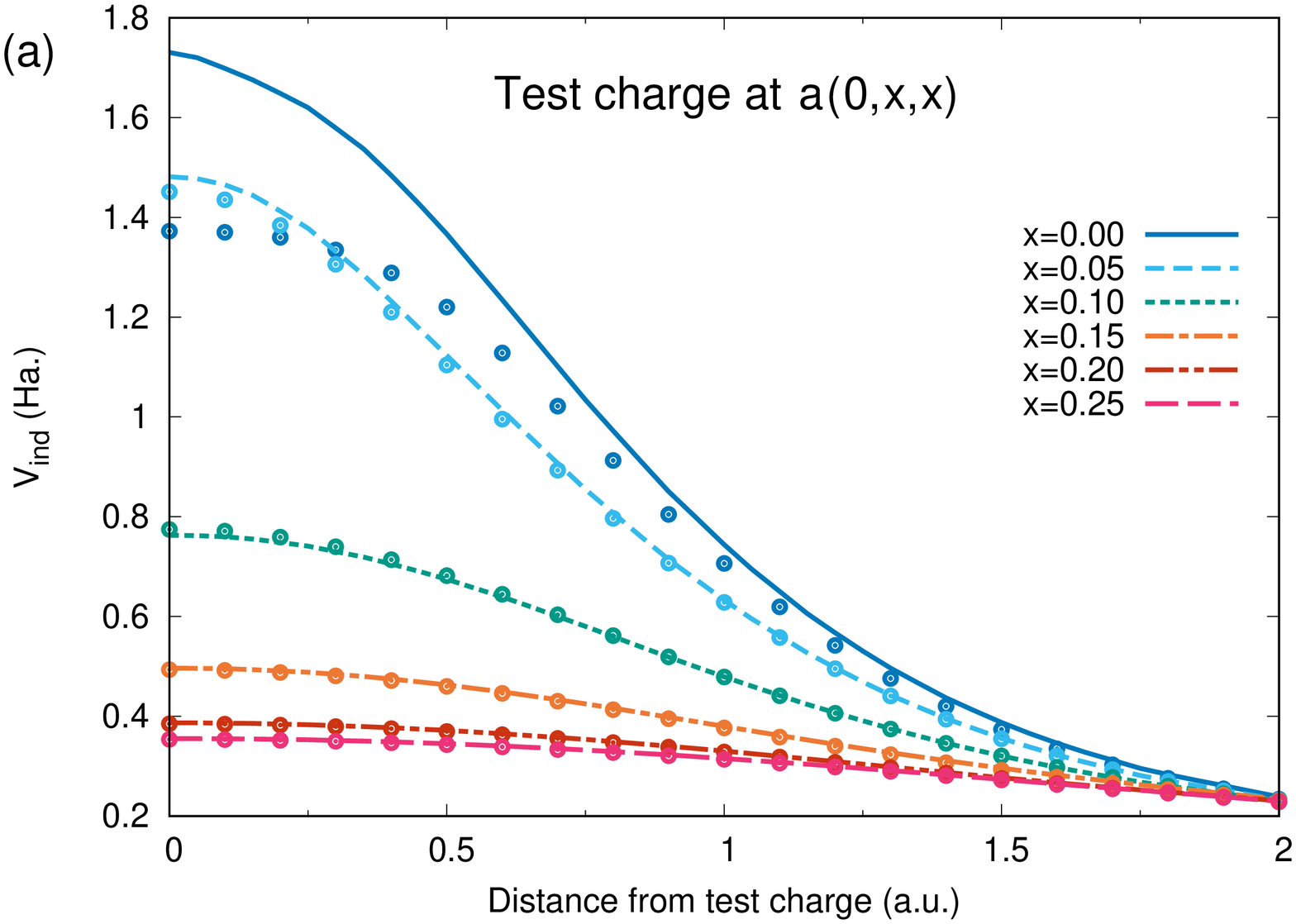}
\includegraphics[height=3.1in,trim=0 35 15 70,angle=270]{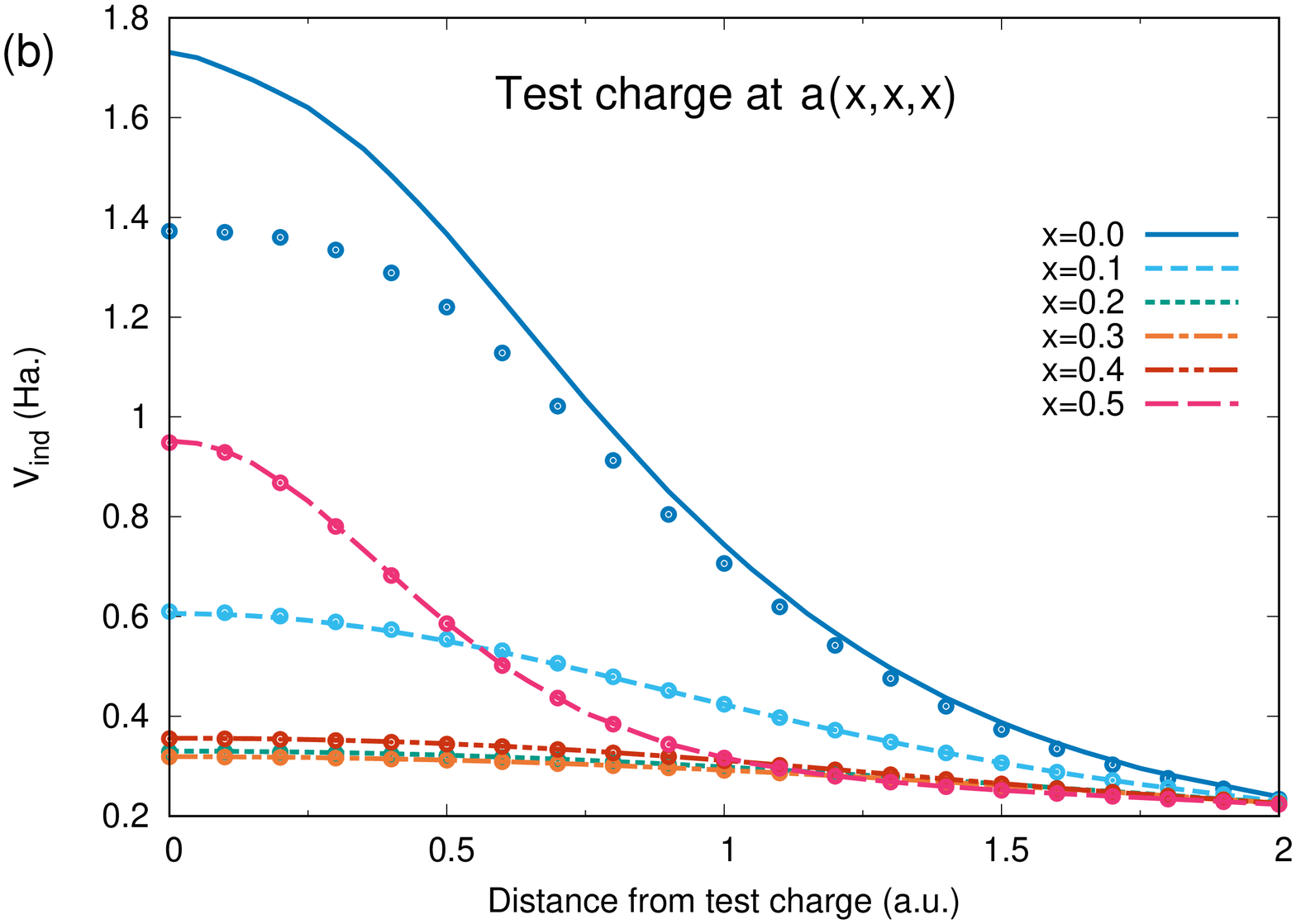}
\caption{ 
(a) The spherically averaged induced potentials in response to test charges in LiF as a function of the distance from the test charge. The potential is shown calculated both with augmentation (lines) and without (circles). 
The test charges are centered at points along the line segment from the F atom at the origin to halfway to the nearest-neighbor F atom (0,\sfrac{$a$}{2},\sfrac{$a$}{2}), where $a=4.0$~{\AA} is the lattice constant.
Only near the F atom are the effects of augmentation observable. 
(b) The same, but along 
($a$,$a$,$a$). 
The Li atom is located at $x=0.5$ [{\it i.e.}, (\sfrac{$a$}{2},\sfrac{$a$}{2},\sfrac{$a$}{2})].  
Because the 1{\it s} orbitals are included as valence there is no effect of augmentation around it.}
\label{val-pot}
\end{centering}
\end{figure}

For valence-level spectroscopy or other calculations, such as self-energy calculations using the {\it GW} method, the screening must be calculated throughout the unit cell.
In Fig.~\ref{val-pot} we show the induced potential in response to test charges centered along the $(a/2,a/2,a/2)$ (from F to Li) and $(0,a/4,a/4)$ (from F halfway to the nearest F) directions within the unit cell, where $a=4.0$~\AA. 
Unsurprisingly, the effect of augmentation is only noticeable near the F atom. 
Note that while the point $(a/2, a/2, a/2)$ is centered on a Li atom, there is no effect from augmentation because the Li 1{\it s} orbitals are included as valence. 

\section{Core-level spectroscopy}

\label{xas-examples}

We now will use calculations of x-ray absorption spectra (XAS) to investigate the effect of augmenting the orbitals and the robustness of the approximations in real-space method. 
Starting with LiF, we show the importance of the screened electron--core-hole interaction and demonstrate that the effect of augmentation is comparable to the changes observed due to thermal disorder. 
Using the series of lithium halides, we show that the approximate augmentation method is sufficient for heavier ions and that the importance of augmentation is reduced by including semi-core orbitals in the pseudopotential. 
Finally, using hexagonal boron nitride, we show that the real-space method is broadly applicable, including for layered materials with high levels of site anisotropy.

Within the BSE approach, absorption spectra are modeled by considering an interacting electron-hole pair \cite{ocean0,ocean1,RevModPhys.74.601}. 
The strong Coulomb attraction between the electron and the core hole is screened by the dielectric response of the material. 
The calculated XAS depends strongly on the strength of this attraction, and, therefore, the details of the screening. 
Within {\sc ocean}, the BSE calculations are carried out using a basis of electron orbitals calculated within density-functional theory (DFT).
Here we use the {\sc Quantum} ESPRESSO code \cite{espresso2,espresso1,*espresso0} 
and the local-density approximation for the density functional \cite{PhysRevB.45.13244}. 
Pseudopotentials are taken from PseudoDojo \cite{pspdojo1,*pspdojo0} and generated with {\sc oncvpsp} \cite{PhysRevB.88.085117,*oncvp}.
Various convergence parameters are summarized in appendix~\ref{conv-param}.

\begin{figure}
\begin{centering}
\includegraphics[height=3.1in,trim=0 35 15 70,angle=270]{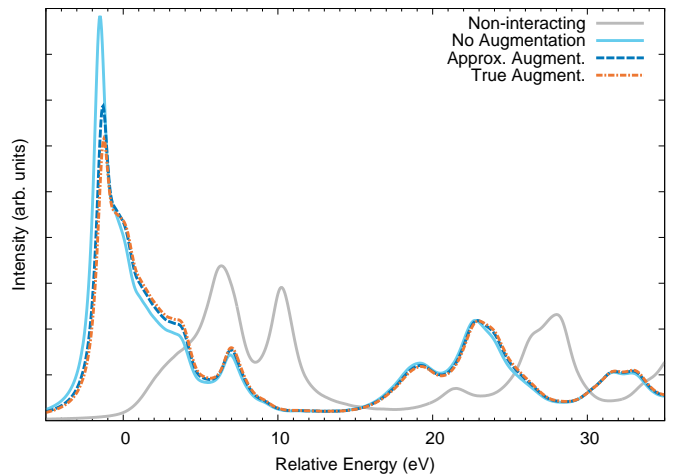}
\caption{ The fluorine K-edge absorption in LiF using different approximations for the screened Coulomb interaction. 
Augmentation of the electron wave functions leads to a stronger screening of the core-hole potential. This stronger screening reduces the strength of the exciton and blue-shifts the spectra.   }
\label{LiF_xas}
\end{centering}
\end{figure}

\subsection{Effect of augmentation on XAS calculations}

\begin{figure}
\begin{centering}
\includegraphics[height=3.1in,trim=0 35 15 70,angle=270]{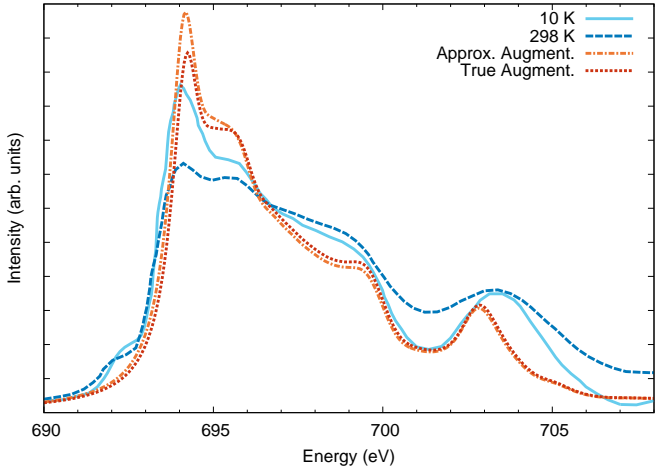}
\caption{ The fluorine K-edge near-edge absorption in LiF comparing calculations using approximate (orange, dot-dash) and true (red, dot) augmentation for the screening for measurements at 10~K (light blue, solid) and 298~K (dark blue, dashed).
The difference in strength of the main exciton at 694~eV between the two calculations is comparable to the difference seen in the two measurements due to increased vibrational disorder at room temperature. Measured data were taken from reference~\cite{SCHWARTZ201730}. 
 }
\label{LiF_near}
\end{centering}
\end{figure}

We first consider 
the fluorine K edge in LiF, which has been studied with {\sc ocean} previously \cite{ocean0,SCHWARTZ201730}. 
Fig.~\ref{LiF_xas} shows the effect of changes to the screened Coulomb attraction between the electron and core hole. 
In light grey, the non-interacting spectrum (neglecting electron-hole interactions) is dramatically different from any other approximation, and it shows the importance of the excitonic binding on the absorption spectrum. 
The three different approximations to the screening all give qualitatively the same results, capturing a strong exciton around 1.5~eV below the onset of the non-interacting spectrum (0~eV in the plot). 
The differences are mostly confined to the near-edge region, within 10~eV of the onset. 
Without any augmentation the exciton is 1.52~eV below the conduction band minimum. This reduces to 1.34~eV with approximate augmentation and to 1.29~eV with true augmentation. 
Correspondingly, the exciton strength is reduced with augmentation, improving agreement with experiment (Fig.~\ref{LiF_near}).

Small changes in intensity and spacing of peaks in the x-ray absorption spectra can be signs of changes in the local structure around the absorbing atom \cite{Pascal}.
To illustrate the importance of accurately determining the relative intensities of peaks, we reproduce measured XAS from reference~\cite{SCHWARTZ201730} which shows the change in the F K-edge absorption with temperature. 
LiF has a dipole-forbidden pre-edge that is observable in dipole-limited absorption spectra due to the vibration of the ions which breaks the symmetry around the F atoms \cite{SCHWARTZ201730,Pascal}. 
When simulating spectroscopy of condensed systems, disorder is typically treated within the Born-Oppenheimer approximation where the ions are stationary during the excitation process.  
The final calculated spectrum is an average over spectra from different atomic positions, {\it e.g.}, generated using molecular dynamics simulations \cite{Pascal} or statistical sampling of the vibrational modes of the system \cite{PhysRevB.90.205207}, or a statistical average of core-hole displacements \cite{PhysRevB.81.115125}.
Typically in calculations of extended system, the vibrational energy levels of the ionic system are treated as negligible and Frank-Condon effects are neglected \cite{PhysRevB.81.115125}. 

In Fig.~\ref{LiF_near} we compare the calculated changes in the F K near-edge spectra of LiF due to changes in the screening calculation to measured changes due to temperature. 
At increased temperature, the pre-edge feature near 692~eV  moves to lower energy, all of the features are broadened, and there is a noticeable weakening of the main exciton between 694~eV and 696~eV. 
The main exciton also differs in strength between the calculations using different augmentation methods, which is even more exaggerated if augmentation is neglected (Fig~\ref{LiF_xas}). 
While the error in the calculation from neglecting or approximating the augmentation has only a small qualitative effect on the spectrum, the differences are substantial when compared to small structural changes. 
High-fidelity screening calculations are necessary for correctly identifying local structure or assessing the accuracy of approximations used for incorporating disorder or vibrations.

\subsection{Influence of valence principle quantum number and semi-core orbitals}

\begin{figure}
\begin{centering}
\includegraphics[height=3.1in,trim=0 35 15 70,angle=270]{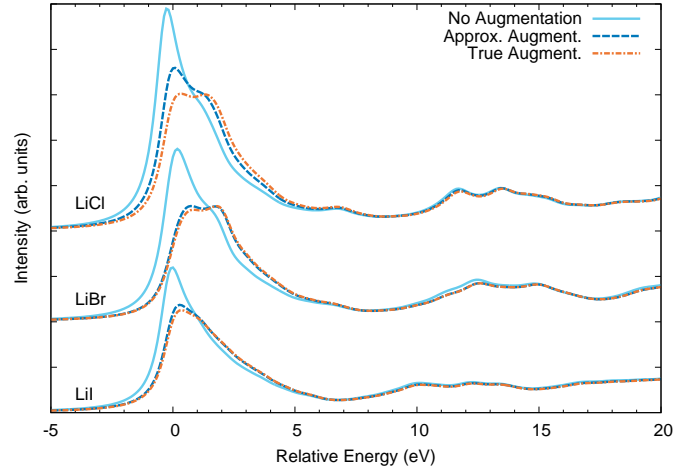}
\caption{ The halide K-edge absorption for LiCl, LiBr, and LiI using different approximations for the screened Coulomb interaction. Across the three materials proper augmentation leads to increased short-range screening and weaker excitonic peaks.   }
\label{LiHalides_xas}
\end{centering}
\end{figure}

Having shown the large effect of changes to the screening on the XAS of LiF, we next examine the effect of augmentation on heavier ions by exploring the range of lithium halides. 
All crystallize in the rocksalt Fm$\bar{3}$m structure, and for all four materials a uniform core-hole lifetime broadening of 0.5~eV was used. 
In Fig.~\ref{LiHalides_xas} we show the effects of different augmentation approaches on the halide K edges of LiCl, LiBr, and LiI. 
The trend between approximations for the same compound shown for LiF in Fig.~\ref{LiF_xas} holds for the heavier halides as well. 
Calculating the screening of the core-hole potential with wave functions from a pseudopotential calculation dramatically under-screens the core hole, resulting in an exaggerated excitonic peak. 
There is a trend towards smaller discrepancy with increasing atomic number in the halide series.
For the LiCl the exciton is approximately 0.56~eV (0.26~eV) more bound without any augmentation (with approximate augmentation), while for LiBr the over-binding is 0.65~eV (0.11~eV) and 0.05~eV (0.01~eV) for LiI. 
We conclude that some all-electron augmentation is necessary for the proper calculation of the screening even for heavier atoms, but for Br and I, it appears that the approximate augmentation method may be sufficient. 

The differences are primarily confined to the near-edge region, which could be expected from Fig.~\ref{plot-f}(c). 
As for LiF, the differences between the augmented and un-augmented induced potential are confined to a small region around the core hole. 
Only near the edge onset is the excited electron localized enough to be strongly affected by this very localized difference in core-hole potential.  
However, an increase in spectral weight near threshold necessitates a reduction in spectral weight higher in energy -- high energy states in the non-interacting system are pulled down by the core hole. 
The differences shown here would be mitigated somewhat by more realistic core-hole lifetime broadening for the Br and I K edges, though by observing partial fluorescence emission from the $2{\mathrm{p}}_{\sfrac{3}{2}}$, the 0.5~eV broadening used here remains realistic for the Br. 

\begin{figure}
\begin{centering}
\includegraphics[height=3.1in,trim=0 35 15 70,angle=270]{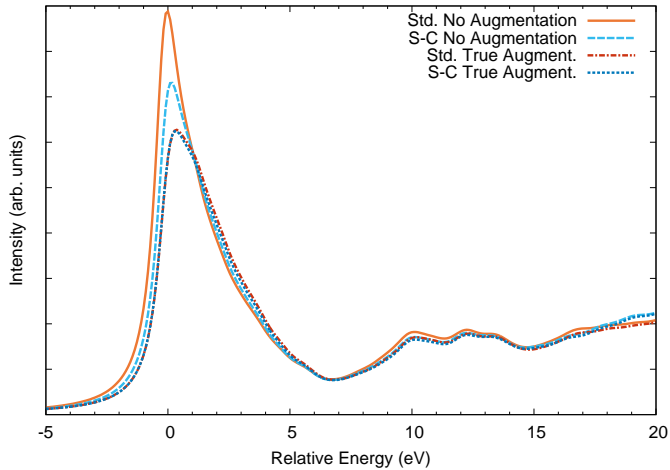}
\caption{ The iodine K-edge absorption for LiI using either the standard (Std.) iodine pseudopential (4{\it d}5{\it s}5{\it p}) or a semi-core (S-C) iodine pseudopotential (4{\it s}4{\it p}4{\it d}5{\it s}5{\it p}) and either no augmentation in the screening or true augmentation. 
The presence of semi-core orbitals substantially reduces the error if no augmentation is used. 
With augmentation the two pseudopotentials give equivalent spectra.    }
\label{I_xas}
\end{centering}
\end{figure}

Next, we examine the effects of augmentation by comparing calculations of LiI using two different iodine pseudopotentials. 
The standard iodine pseudopotentials uses a Kr core with the 4{\it d}, 5{\it s}, and 5{\it p} electrons in the valence bands while the semi-core pseudopotential also includes the 4{\it s} and 4{\it p} orbitals as valence. 
In Fig.~\ref{I_xas} we show that the calculated I K edge of LiI does not depend on the pseudopotential when the orbitals are properly augmented for the screening calculation.
However, without augmentation the screening calculated using the standard pseudopotential is notably weaker than that of the semi-core pseudopotential, leading to a stronger exciton.
The weaker effect of augmentation on the calculation using the semi-core pseudopotential is due to the node in the 5{\it s} and 5{\it p} orbitals (absent for the standard pseudopotential).
As discussed earlier (see Fig.~\ref{plot-f}), the absence of nodes in the pseudopotential orbitals shifts the electron density away from the atom. 
Even though without augmentation the $n=5$ orbitals only get a single node in the semi-core system the effect is already dramatic and the discrepancy due to neglecting augmentation is strongly reduced.

We have demonstrated that the pseudopotential-based orbitals are insufficiently accurate for calculating the screened electron-hole interaction for core-level spectra. 
This failure is independent of the quality of the pseudopotential used. 
Instead it is a straightforward consequence of the difference in nodal structure between valence orbitals in the all-electron and pseudopotential systems. 
As shown in the atomic case (Fig.~\ref{plot-f}), the absence of nodes in valence orbitals of the pseudopotential system affects the density response and hence the screening. 
This is mitigated somewhat by the use of semi-core pseudopotentials that include the next highest principle quantum number in the valence, {\it e.g.}, Fig.~\ref{I_xas}. 
The moderate success of including semi-core orbitals without augmentation may indicate that a kind of false augmentation could be used, where nodes are added without requiring that they mimic the true all-electron orbitals. 
These false projectors could be constructed such that they add nodes without increasing the required plane wave energy cutoff for use in reciprocal-space methods.  

\subsection{Layered materials, convergence, and errors}
\label{layered}

We next investigate the K edges of hexagonal boron nitride ({\it h}-BN). The XAS of {\it h}-BN has been the subject of recent investigations into the role that vibrational disorder and defects play in the spectra \cite{PhysRevB.96.205116,doi:10.1021/acs.jpcc.9b00179,PhysRevB.96.144106}. 
Because {\it h}-BN is layered, it has an anisotropic dielectric response, unlike the lithium halides. We will use it to showcase the real-space screening method on systems with reduced symmetry. 
The dielectric screening within the BN planes (perpendicular to the {\it c}-axis) is stronger than that parallel to the {\it c}-axis: $\epsilon_\infty^{\perp}=4.95$ versus  $\epsilon_\infty^{\parallel}=4.10$ \cite{PhysRev.146.543}. 
The anisotropy is also visible in the XAS. 
Both the B and N edges show strong excitonic features when the x-ray polarization vector is aligned with the {\it c}-axis and a delayed onset when the x-ray polarization is in-plane.

We compare our calculations of {\it h}-BN to two different computational approaches. 
First, the {\sc exciting} code (like {\sc ocean}) uses the BSE \cite{Gulans_2014,PhysRevB.95.155121}. 
The screening in {\sc exciting} is also carried out in the RPA, but calculated in reciprocal space with no modeled response. 
Additionally, it is an all-electron code making augmentation unnecessary. 
Second, within OptaDOS x-ray spectra are calculated using the $\Delta$SCF method \cite{MORRIS20141477}. In this approach the final states are calculated directly as the unoccupied states of a DFT calculation with a core hole. The initial state is a the core-level orbital, and the transition matrix elements can be calculated directly. The density response to the core hole is calculated self-consistently within DFT. 
The OptaDOS calculations reproduced here used the {\sc castep} DFT code \cite{CASTEP}. 

\begin{figure}
\begin{centering}
\includegraphics[height=3.1in,trim=0 50 5 60,angle=270]{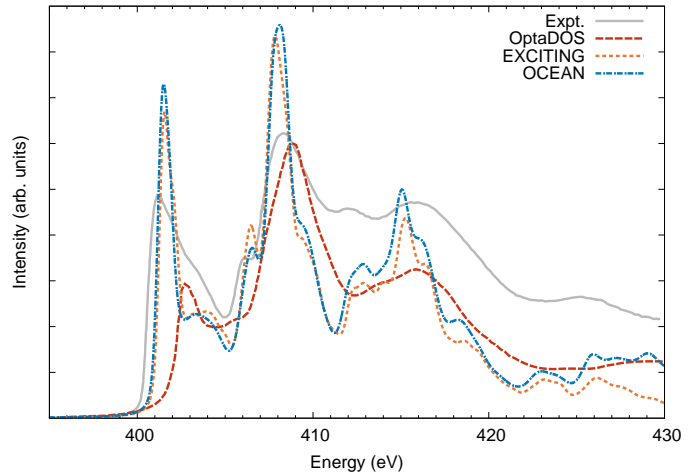}
\caption{ The measured nitrogen K-edge x-ray absorption of {\it h}-BN  compared to three different computational methods. The two BSE-based codes, {\sc exiciting} and {\sc ocean}, show good agreement with each other and experiment. The experiment is reproduced from Ref.~\cite{PREOBRAJENSKI200559} and the OptaDOS from Ref.~\cite{mcdougall_nicholls_partridge_mcculloch_2014}. }
\label{BN-xas}
\end{centering}
\end{figure}

In Fig.~\ref{BN-xas} we compare the N K-edge of {\it h}-BN calculated with {\sc ocean} to experiment \cite{PREOBRAJENSKI200559} and calculations using OptaDOS and {\sc exciting}. 
The experiment was taken with polarization at an angle of $50^\circ$ to the {\it c}-axis, giving nearly the same 2:1 in-plane to out-of-plane ratio of a disordered sample. All three calculations are averaged over polarization directions.  
The $\Delta$SCF OptaDOS spectrum is a clear outlier with a substantially reduced exciton strength and binding at the onset, a known characteristic of $\Delta$SCF methods \cite{PhysRevLett.118.096402}.
 The agreement between {\sc ocean} and {\sc exciting} is very good. The {\sc exciting} calculation includes fewer conduction bands, leading to an artificial die-off in spectral intensity at higher energies.

\begin{figure}
\begin{centering}
\includegraphics[height=3.1in,trim=0 50 5 60,angle=270]{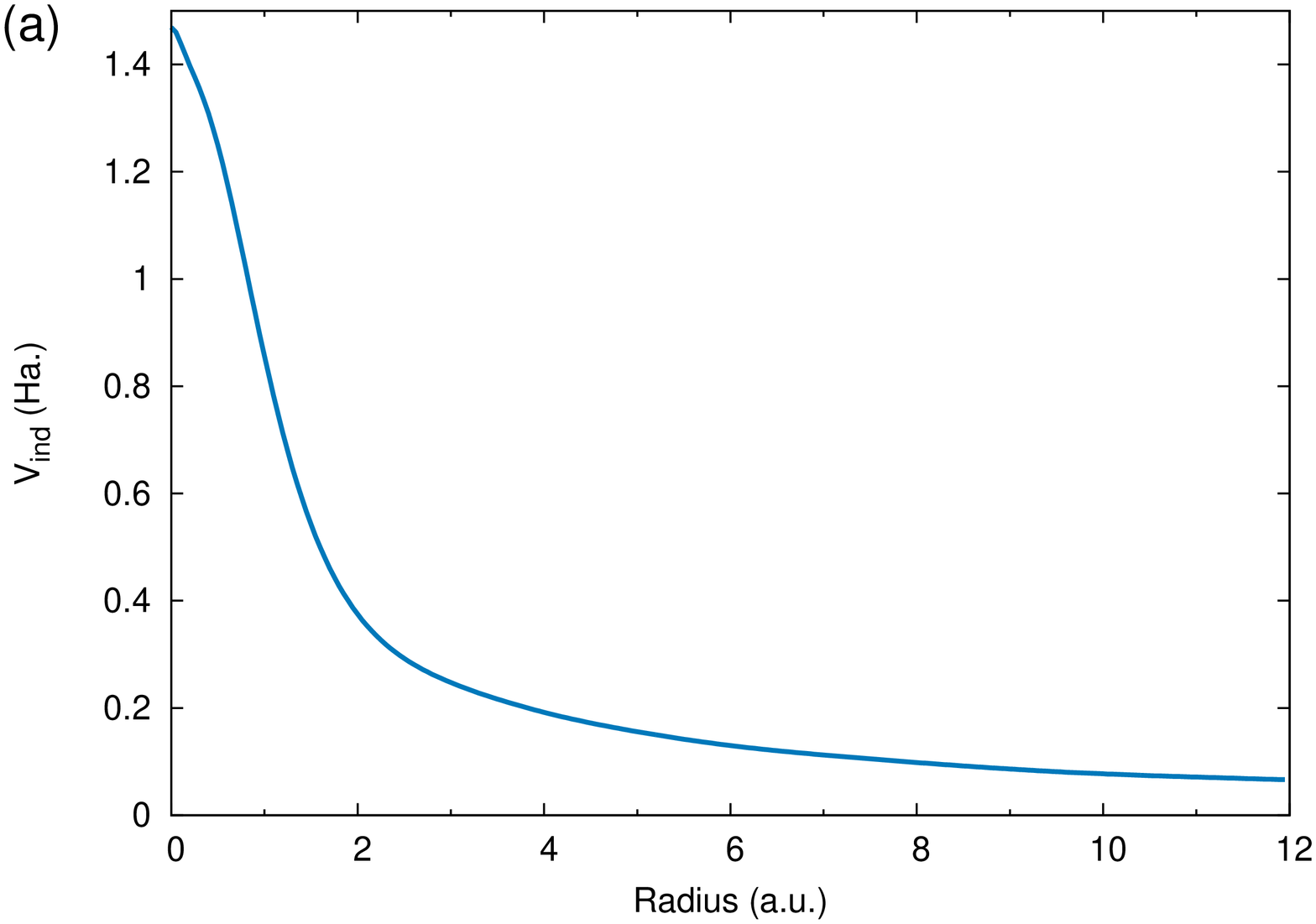}
\includegraphics[height=3.1in,trim=0 30 15 60,angle=270]{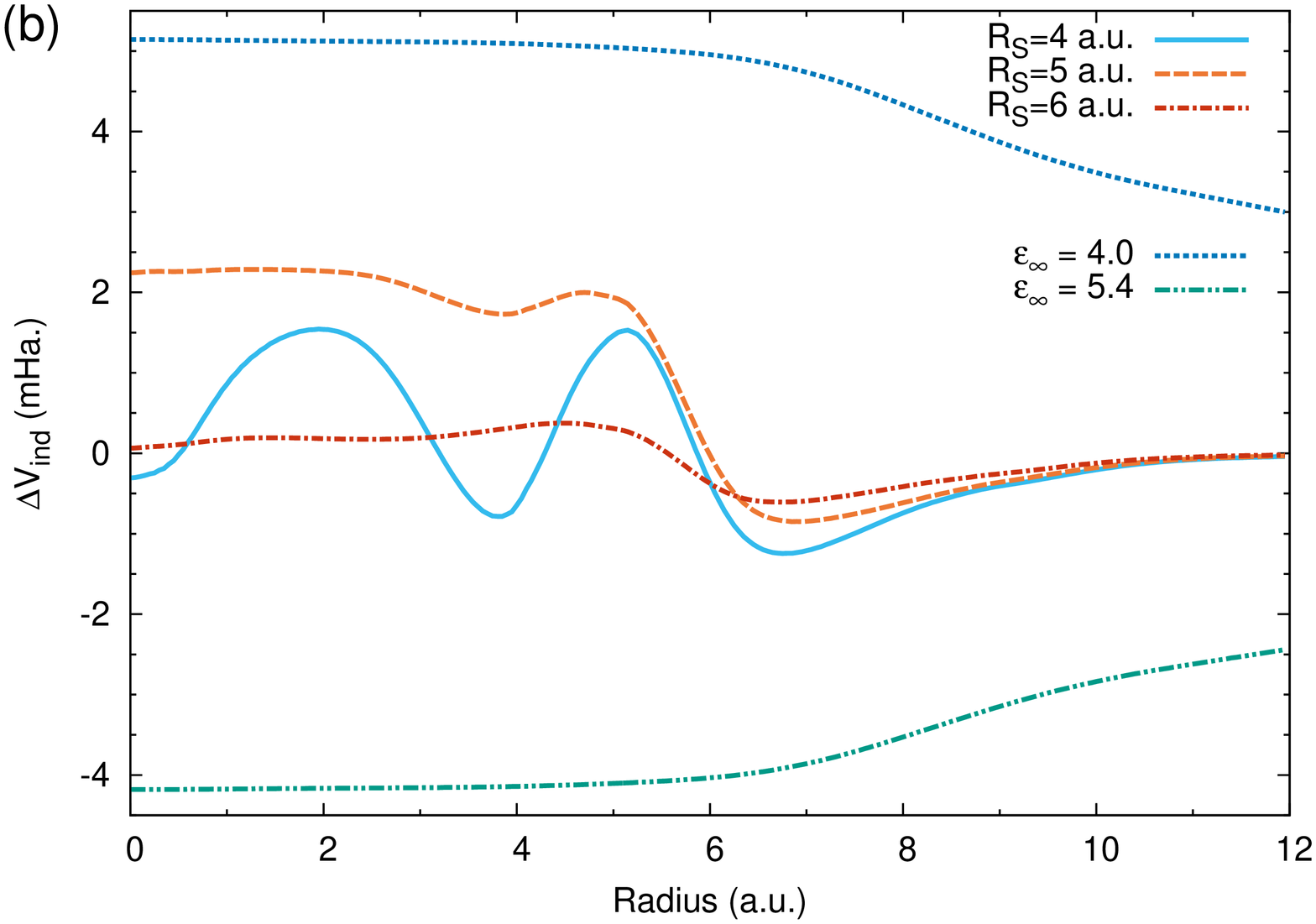}
\includegraphics[height=3.1in,trim=0 10 5 60,angle=270]{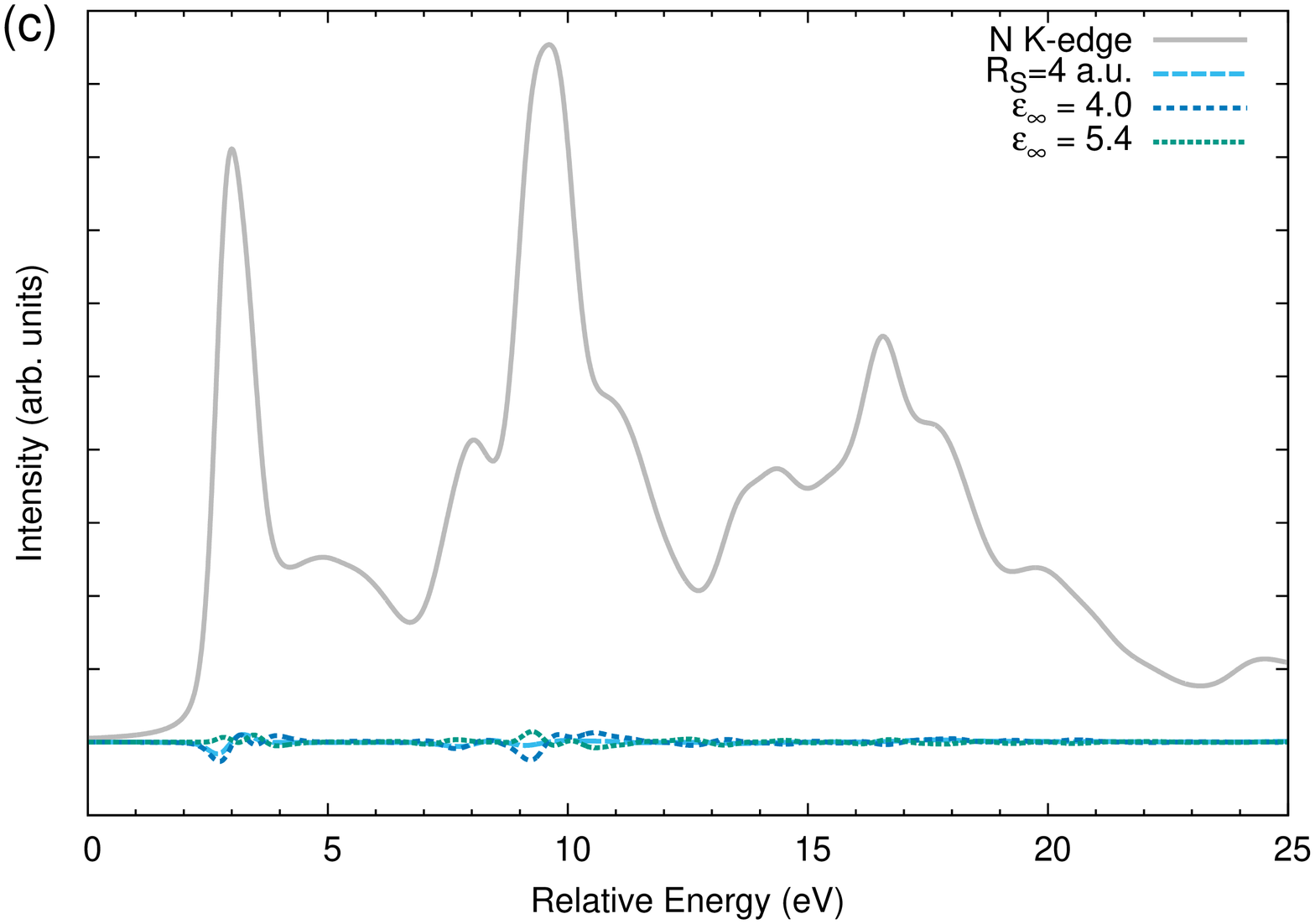}
\caption{ (a) The induced potential due to a 1{\it s} core hole on a nitrogen atom in {\it h}-BN as a function of radius, using a spherical charge radius of $R_S=7$~a.u.\ and a dielectric constant of $\epsilon_\infty=4.667$. (b) The differences in the calculated induced potential due to changes in either the shell radius or input dielectric constant. (c) The calculated N K-edge XAS of {\it h}-BN (grey, solid). The differences between the spectrum calculated from the converged screening and the spectra using different changes to the screening calculation. Before taking the differences, the XAS calculations are aligned to correct for red or blue shifts (see text).}
\label{hBN-pot}
\end{centering}
\end{figure}

Per section~\ref{sec-localreview}, the real-space screening method is only an approximation to the RPA response. 
The approximation is controlled by $R_S$, the radius of the shell of charge that is screened with a model dielectric response, and $\epsilon_\infty$, the value of the static dielectric constant which is input into the Levine-Louie dielectric model  (see appendix~\ref{appendixLL}). 
In Fig.~\ref{hBN-pot}(a) we show the total induced potential in {\it h}-BN due to a core hole on the nitrogen atom. 
We also show the difference in the potential calculated using a shell radius of 7~a.u.\ versus smaller radii, and we find that the errors are less than 3~mHa. 
Next we vary the value of the input dielectric constant by approximately $\pm15\%$, plotting the difference in the potentials in Fig.~\ref{hBN-pot}(b). 
As expected, increasing the dielectric constant input into the model decreases the induced potential. 
However, we find the differences are $\lesssim 5$~mHa.
The screening calculation is relatively insensitive to changes in the sphere radius or errors in the input dielectric constant. 
Further examination is presented in appendix~\ref{app-errors}.

The primary effect of small errors in the screening is a shift of the spectra, reflecting a change in the excitonic binding energy. 
For the nitrogen K edge, using $\epsilon_\infty=5.4$ blue-shifts the x-ray absorption by 0.10~eV, while using $\epsilon_\infty=4.0$ red-shifts the x-ray absorption by 0.12~eV. 
In Fig.~\ref{hBN-pot}(c) we show how the nitrogen K-edge spectra change with these changes in the calculation of the screening. 
The changes are minor, and so only the differences between the spectra are plotted, after taking into account the aforementioned red or blue shifts.

\section{Valence excitation spectroscopy}

The {\sc ocean} code is also capable of calculating valence optical/UV spectroscopy within the BSE, following earlier work \cite{PhysRevB.59.5441,PhysRevB.78.205108}. 
In moving from core to valence excitations, the hole is no longer confined to a local basis around the atom (the core-level orbital), but instead spans the unit cell. 
This introduces a trade-off. For a valence calculation, the screening must be calculated throughout the unit cell, but the more delocalized nature of a valence exciton means that the importance of the screening calculation at any given point in space is diminished. 
Previous valence calculations have forgone RPA calculations and used a model dielectric function while still achieving good agreement with experiment \cite{PhysRevB.59.5441,PhysRevB.78.205108}.
Conversely, using only a model dielectric to screen core-level excitations had been found to be inaccurate \cite{Shirley_2005}. 
The strong co-locality of the core hole and photo-electron in near-edge x-ray absorption makes the spectra very sensitive to the accuracy of the screened Coulomb potential at small distances where the accuracy of 
an electron-gas-based model 
model breaks down.

\subsection{Screening for valence-level BSE}

Previously in {\sc ocean}, the screened potential for valence calculations was approximated using the Hybertsen-Levine-Louie dielectric model \cite{PhysRevB.25.6310,PhysRevB.37.2733}, which depends parametrically on the local density $\rho(\mathbf{x})$ and static dielectric constant $\epsilon_\infty$.
The wave functions for the electrons and holes are sampled on regular real-space grids $\mathbf{x}$, and therefore we need a description of the screened Coulomb potential for each set of grid points $W(\mathbf{x},\mathbf{x}')$. 
Following Ref.~\cite{PhysRevB.59.5441}, the screened Coulomb potential is given by,
\begin{align}
W_\textrm{HLL}(\mathbf{x},\mathbf{x}') = \sfrac{1}{2} &\left[ W_\textrm{hom} (\vert \mathbf{x}-\mathbf{x}' \vert ; \rho(\mathbf{x}), \epsilon_\infty ) \right. \nonumber \\
 & \left. W_\textrm{hom} (\vert \mathbf{x}-\mathbf{x}' \vert ; \rho(\mathbf{x}'), \epsilon_\infty )  \right]
 \end{align}
which simply averages the results using the density at points $\mathbf{x}$ and $\mathbf{x}'$. 
To avoid the divergence at $\mathbf{x} \rightarrow \mathbf{x}'$, a spherical average over the discretization volume is used when $\mathbf{x}=\mathbf{x}'$.  

To improve this, we substitute the more accurate local-RPA result for the short-range part of $W$. 
Using the previously introduced method, we can calculate the screened Coulomb from Eq.~\ref{W1} for each grid point $\mathbf{x}$, ie, $W^{[\mathbf{x}]}$. 
In the case of core-level excitations the external potential in Eq.~\ref{eq-W} was given by the core-hole potential from an atomic calculation. 
For valence calculations we use the potential from a spherical charge centered at $\mathbf{x}$.
The volume of the spherical charge is set by the discretization volume of the real-space grid, $V_x = \Omega / N_x$, the unit cell volume $\Omega$ divided by the number of grid points $N_x$. 
The screening calculation can be carried out with or without augmentation.

As above, we enforce the symmetry in interchanging $\mathbf{x}$ and $\mathbf{x}'$. 
\begin{equation}
\label{rpa-val}
    W(\mathbf{x},\mathbf{x}') = 
\begin{cases}
    \frac{1}{2} \left[ W^{[\mathbf{x}]}(\mathbf{x}') + W^{[\mathbf{x'}]}(\mathbf{x}) \right],& \!\!\text{if } \vert \mathbf{x}- \mathbf{x}' \vert \leq r_m\\
    W_\textrm{HLL}(\mathbf{x},\mathbf{x}'),              & \text{otherwise}
\end{cases}
\end{equation}
As for the core-level, the RPA screening is calculated only within a finite space and parametrized by the shell radius $R_S$ and dielectric constant $\epsilon_\infty$. 
At large distances the HLL approximation is still used, governed by the parameter $r_m$. 
Much like the shell radius, this constitutes a {\it controlled} approximation.

Like the HLL model, care must be taken when evaluating the real-space $W$ in the limit of $\mathbf{x} \rightarrow \mathbf{x}'$. 
We numerically integrate the $l=0$ component of the calculated screened potential over the discretization volume 
\begin{equation}
W^{[\mathbf{x}]}(\mathbf{x}) = \frac{3}{ R_x^3 } \int_0^{R_x} W^{[\mathbf{x}]}(r) r^2 dr \; ,
\end{equation}
where $R_x = [ 3 V_x / 4 \pi ]^{1/3}$.

The system-size scaling behavior is the same for the valence screening as it was for the core-level case --- the number of grid points $\mathbf{x}$ scales linearly with volume the same as the expected number of atomic sites. 
(The scaling is discussed in section~\ref{sec-implementation} and illustrated in \ref{sec-time-scale}.)
There are two major differences between calculating the screening for the valence and core cases with negative and positive impacts on run time.
First, the number of real-space grid points is much larger than the number of atoms. 
For example, in a unit cell of LiF an $10^3$ x-point mesh is necessary, resulting in 1000 screening sites instead of the 1 needed for the fluorine x-ray absorption calculation.  
The dramatic increase in the number of sites is offset by the use of coarser real-space grids in the calculation of $\chi$. 
The perturbing potential for the core case is the core-hole potential, and, like the core-hole density, it is strongly localized. 
In the valence case, the perturbing potential is taken to be a uniform ball of charge whose volume is set by the discretization volume $V_x$ defined above. 
Therefore, the valence screening calculations converge with a coarser radial mesh than is needed for the core-level screening. 

\subsection{Effect of model screening and augmentation on optical calculations}

 \begin{figure}
\begin{centering}
\includegraphics[height=3.1in,trim=0 35 15 70,angle=270]{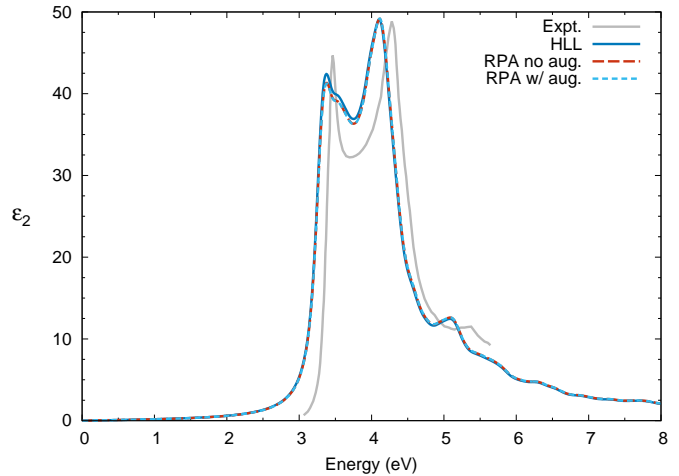}
\caption{ The calculated imaginary part of the dielectric function for silicon as a function of energy compared with experiment taken from Ref.~\cite{PhysRevB.36.4821}.
The spectra are generated using two different approximations to the screened Coulomb interaction $W$.  }
\label{Si-val}
\end{centering}
\end{figure}

 \begin{figure}
\begin{centering}
\includegraphics[height=3.1in,trim=0 35 15 70,angle=270]{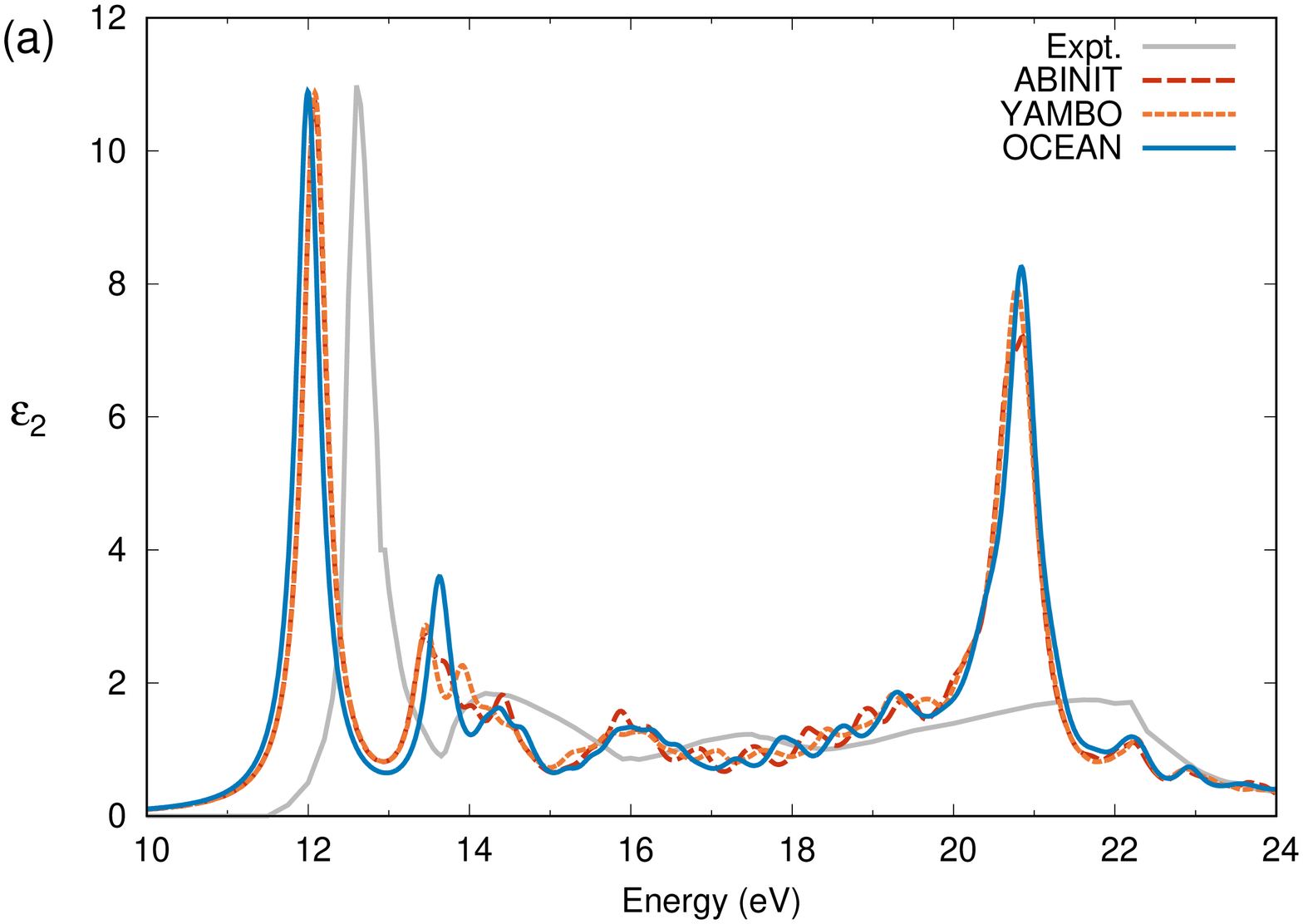}
\includegraphics[height=3.1in,trim=0 35 15 70,angle=270]{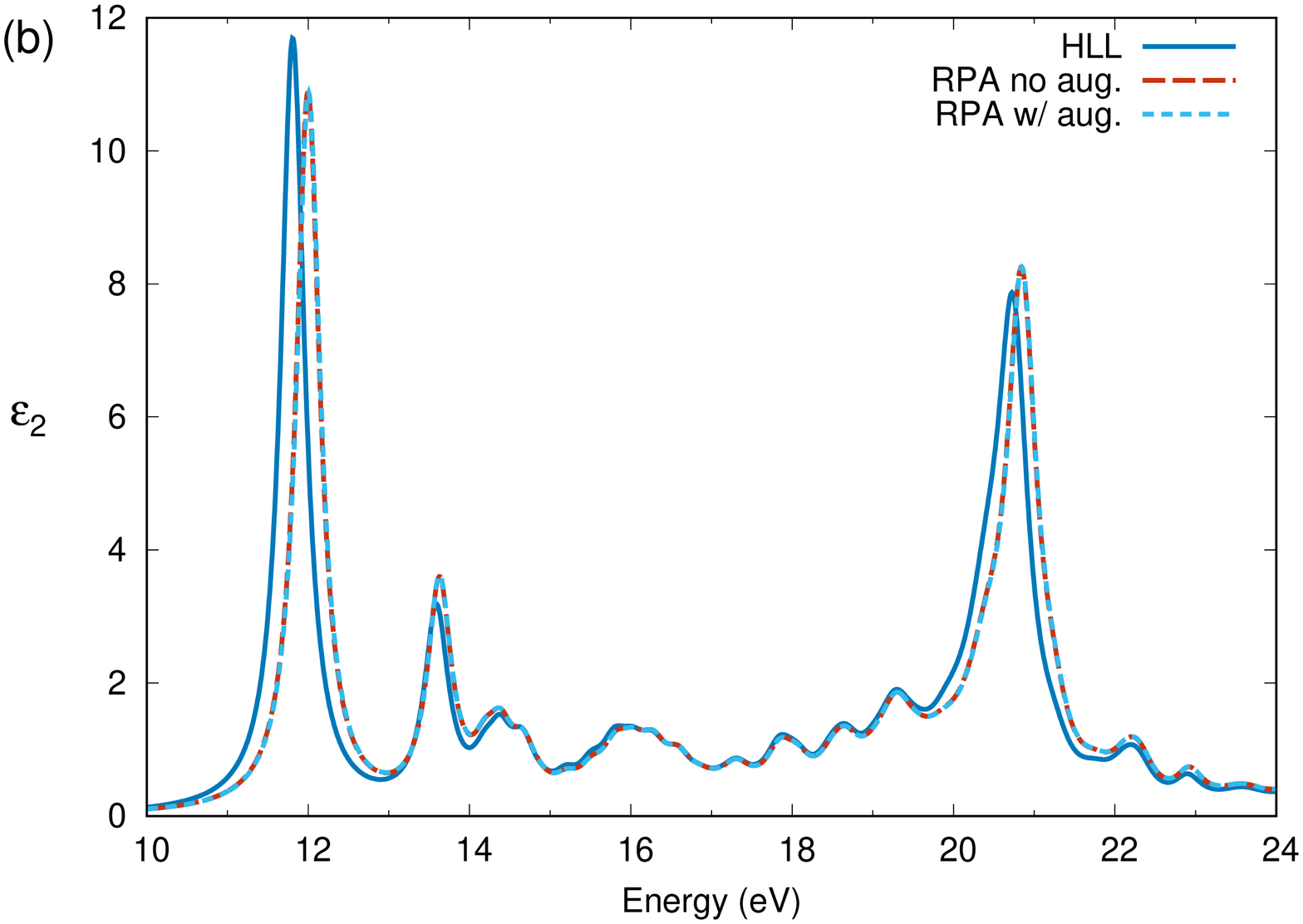}
\caption{ (a) The imaginary part of the dielectric function for LiF as a function of energy calculated with several different BSE codes and compared to experiment taken from Ref.~\cite{Roessler:67}. The {\sc ocean} calculation was carried out using RPA screening and no augmentation to more closely match the methodology of the other two codes. (b) The LiF dielectric function calculated with {\sc ocean} using different approximations to the screened Coulomb interaction $W$. The effect of augmentation is minimal, and the two RPA results are nearly identical. }
\label{LiF-val}
\end{centering}
\end{figure}

To show the effects of using the local RPA, with and without augmentation, instead of the HLL screening on valence calculations we consider calculations of the imaginary part of the dielectric function.

First, we look at bulk silicon --- a standard for valence electronic structure calculations providing a testbed for early DFT, {\it GW}, and valence BSE calculations \cite{PhysRevLett.43.387}. 
As for the lithium halides we use the experimental lattice constant of 0.543~nm \cite{Wyckoff} and the PseudoDojo pseudopotential for silicon. 
Additional input parameters are included in appendix~\ref{conv-param}.
A scissor correction was used to set the DFT band gap to be 1.11~eV \cite{Kittel}. 
Unsurprisingly, we see in Fig.~\ref{Si-val} that in the case of silicon the HLL model performs very well. 
The inclusion of the RPA screening at short range has little effect on the spectrum of bulk silicon, and augmentation has no visible effect. 

Next we return to LiF (also well studied previously within the BSE \cite{PhysRevB.66.165105,ocean0,PhysRevB.88.155113}). 
The simulation details are similar to the x-ray case, but we have included a scissor correction to set the DFT band gap to 14.2~eV \cite{PhysRevB.13.5530}. 
We first compare our calculations BSE results calculated using ABINIT v.~8.10.2 \cite{Gonze2020} and yambo v.~4.5.0 \cite{Sangalli_2019} codes in Fig.~\ref{LiF-val}(a). 
A small, 0.09~eV shift in the position of the first exciton between the {\sc ocean} and yambo indicates slightly weaker screening within {\sc ocean}. The ABINIT results were obtained with a scissor-corrected band gap of only 13.3~eV, chosen to align the exciton position with the yambo calculation. 
The origin of the 1~eV shift in the ABINIT spectrum is not known. 
The three calculations are largely in agreement with each other in shape and intensity as well as previously published calculations, accounting for the choice of scissor correction.

In contrast to bulk silicon, LiF is strongly ionic with the valence orbitals primarily localized on the fluorine site, making it a more difficult system for the model dielectric response. 
In Fig.~\ref{LiF-val}(b), we see discrepancies between the HLL-only and RPA results. 
Both the main exciton near 12~eV and the higher energy peak near 21~eV are red-shifted in the HLL as compared to the RPA calculations. 
Using the HLL model, the main exciton is at 11.8~eV, while the RPA places it at 12.0~eV.
This shift indicates that the HLL model screening is weaker than the RPA calculation for these two peaks. 
This is consistent with a larger $GW$ band-gap correction that often occurs with the HLL approximation is used as compared to when the RPA is used \cite{PhysRevB.43.14142}.   
As for silicon, the differences in the spectra with and without augmentation are very minor and not distinguishable in the plot.

\section{ Implementation Within {\sc ocean} and $N^2 \log N$ scaling}
\label{sec-implementation}

Having shown the utility of the local, real-space screening in BSE calculations of both core-level and valence-level excitations, we now will provide an overview of the implementation and demonstrate the favorable $N^2 \log N$ scaling of our RPA calculations with systems size. 
As previously stated, our goal is to calculate the screened Coulomb interaction, 
starting from the irreducible polarizability within the RPA (Eq.~\ref{eq-rpa}). 
There are a number of costs and bottlenecks associated with this calculation. 
To review, the screened Coulomb interaction $W$ is directly calculable from the reducible polarizability $\chi$. 
The calculation of $\chi$ involves matrix products and matrix inversions of the Coulomb operator and the irreducible polarizability $\chi_0$. 
Within the RPA, $\chi_0$ follows from the one-electron Green's function $g$, which itself can be written from the the electron orbitals. 

In this section, 
we will explicitly outline the method used in version 3 of {\sc ocean} \cite{ocean0,ocean1} in subsections A through F for the each step in calculating the RPA response. 
We note how data and calculations are distributed for parallel computation, and we use bars to indicate when a process has only a subset of the total indices, {\it e.g.}, bands $\bar{b}$ or {\it k}-points $\bar{\mathbf{k}}$. 
In subsection G we examine the scaling behavior with system size and parallel performance.

\subsection{Evaluating wave functions and local basis}

The initial step is determining the electron wave functions and the basis, from which we can generate the Green's functions. 
Density-functional theory (DFT) is used to simulate the electronic Hamiltonian. 
The system is taken to be periodic such that the electron orbitals can be denoted by their band $b$, crystal momentum $\mathbf{k}$, and spin $\sigma$, 
\begin{equation}
H^{\textrm{DFT}} \psi_{b \mathbf{k} \sigma} = \psi_{b \mathbf{k} \sigma} \varepsilon_{b \mathbf{k} \sigma} , 
\end{equation}
where each wave function $\psi_{b \mathbf{k} \sigma}$ has energy $\varepsilon_{b \mathbf{k} \sigma}$. 
The Green's function for energy $E$ can be written in the spectral representation 
\begin{equation}
g_\sigma(\mathbf{r},\mathbf{r}',E) =
\sum_{b=1}^\infty \int_{BZ} \frac{d^3\mathbf{k}}{(2\pi)^3} \, 
\frac{\psi^\dagger_{b \mathbf{k} \sigma}(\mathbf{r}) \psi_{b \mathbf{k} \sigma}(\mathbf{r}')}{ E - \varepsilon_{b \mathbf{k}\sigma}}
\label{eq-green}
\end{equation}
The integral runs over the Brillouin zone.

To construct $g$ we must define the real-space basis. 
As mentioned in Sec.~\ref{sec-localreview}, we calculated the RPA response only for the local potential $v_1$. 
We employ a real-space basis within a sphere $S$ with a radius $r_S$ and centered on a point $\tau$. 
For screening the core hole, $\tau$ is the atomic site, while for valence calculations $\tau$ is one of the grid points in Eq.~\ref{rpa-val}. 
The irreducible polarizability $\chi_0^{(\tau)}$ is then an $N_r \times N_r$ size matrix, as are the Green's functions $g^{(\tau)}(E)$. 
This real-space basis is independent of the size of the system's unit cell, and is discussed further in Appendix~\ref{rad-ang-grids}.

In practice the sum over bands is truncated, and the integral is replaced with a sum over regularly spaced points in {\it k}-space. 
Our approach requires only a few {\it k}-points, often between 1 and 8, and we address this later in Appendix~\ref{sec-kb-conv} on errors and convergence. 
The sum over bands, however, is significant. 
Typically, convergence in the screening is reached when the Green's function is constructed with states up to around 100~eV above the Fermi level.  
This requires on the order of 30 to 50 bands per atom, and the number of bands required scales linearly with system size. 
Unfortunately, aspects of the generation of the one-electron states from DFT scale with the square of the number of bands.

\subsection{Projecting the wave functions}

These DFT eigenstates $\psi$ are generated using an external plane-wave DFT program and are saved to file as Bloch states $u$,
\begin{equation}
\psi_{b \mathbf{k}}(\mathbf{r}) = e^{i \mathbf{k} \cdot \mathbf{r}} u_{b \mathbf{k}}(\mathbf{r}) = e^{i \mathbf{k} \cdot \mathbf{r}} \sum_\mathbf{G} C_{b \mathbf{k}}(\mathbf{G}) e^{i \mathbf{G} \cdot \mathbf{r}} 
\label{bloch}
\end{equation}
which are defined in terms of complex-valued coefficients $C$ of plane waves $\mathbf{G}$. 
The spin index $\sigma$ will be dropped. 
Only the set of coefficients $C$, not the various phases, are written to file. 
We distribute the work for the conversion by band and k-point --- not by plane wave coefficient.

To project the wave functions onto our spherical grid, one option would be to follow the method given by Eq.~\ref{bloch} directly, {\it i.e.}, Fourier interpolation. 
We first create a matrix of the phases, as these will be common across all of the bands at a specific k-point.
\begin{align}
P_{\mathbf{k}}( \mathbf{r}, \mathbf{G} ) &= e^{i (\mathbf{k+G} ) \cdot \mathbf{r}} \\
\psi_{\mathbf{k}}(\mathbf{r},\bar{b}) &=  \sum_{\mathbf{G}} P_{\mathbf{k}}( \mathbf{r}, \mathbf{G} ) C_\mathbf{k}( \mathbf{G}, \bar{b} )
\end{align}
where the bar indicates that we are processing only a subset of the total number of bands. 
The phase matrix requires $\mathcal{O}[N]$ operations from the plane waves, regardless of the number of processors included. 
Projecting the wave functions is  $\mathcal{O}[N^2]$ from the plane waves and bands, but the bands are distributed by processor. 
The summation over k-points is not counted in the estimation of computational cost because it decreases with volume and is usually 8 or 1. 
For a system with more than one site of interest, {\it e.g.}, a disordered, liquid, or amorphous system, 
the number of sites, and therefore the number of local real-space grids, increases with volume as well.
This means that the actual costs increase to $N^2$ and $N^3$.

To avoid $N^3$ scaling, we instead use a fast Fourier transform (FFT), followed by interpolation, 
and completed by applying the complex phase:
\begin{align}
u_{\bar{b} \mathbf{k}} (\mathbf{x}) &= \mathcal{FFT}\left[ u_{\bar{b} \mathbf{k}} (\mathbf{G}) \right] \nonumber \\
u_{\bar{b} \mathbf{k}} (\mathbf{r}) &= \mathrm{Interp.} \left[u_{\bar{b} \mathbf{k}} (\mathbf{x})  \right] \\
\psi_{\bar{b} \mathbf{k}}(\mathbf{r}) &= e^{i \mathbf{k} \cdot \mathbf{r}} u_{\bar{b} \mathbf{k}}(\mathbf{r}) . \nonumber
\end{align}
The real-space grid $\mathbf{x}$ is defined as the Fourier transform dual of $\mathbf{G}$. 
We use 3rd-degree Lagrange polynomial interpolation to generate the wave functions on our desired points $\mathbf{r}$ from the FFT grid $\mathbf{x}$, aided by oversampling the FFT. 
The interpolants are cached to allow reuse within and between sites. 
The costs, including a factor of $N$ sites, are $\mathcal{O}[N^2 \log N]$, $\mathcal{O}[N^2]$, and $\mathcal{O}[N^2]$, respectively. 
All three steps are independent over bands, {\it k}-points, and spins, providing good scaling with the number of processors.

To determine the break-even point between these two methods we must be more specific with the actual costs of each step. 
The $N^3$ term from method 1 is $N_\mathbf{G}  N_b N_\mathbf{r} N_i$, where $i$ are the atomic sites. 
The $N^2 \log N$ term from method 2 is $A_F N_\mathbf{G}  N_b \log N_\mathbf{G}$, where $A_F$ is the FFT prefactor. 
Therefore, method 2 is preferable if $A_F \log N_\mathbf{G} < N_\mathbf{r} N_i$. 
Under the assumption that the logarithm of even a large number is about 10, method 2 is likely preferable, even for single-site calculations \cite{Fermi}. More sophisticated methods for Fourier interpolation onto {\it irregular} grids have been proposed in literature, {\it e.g.}, \cite{Ware} and references therein, but are not explored here.

\subsection{Augmentation}
\label{implementation-augment}

The next step is augmenting the wave functions to recreate the all-electron character, 
\begin{align}
\label{eq-aug1}
 \psi^\textit{ae}_{\bar{b} \bar{\mathbf{k}}}(\mathbf{r}) &=  \psi^\textit{ps}_{\bar{b} \bar{\mathbf{k}}}(\mathbf{r})  \\
 &+ \sum_{\nu,l,m} Y_{lm}(\hat{r}) \left[ \phi^{ae}_{\nu l}(\mathbf{r}) - \phi^{ps}_{\nu l }(\mathbf{r}) \right] \langle \phi^{ps}_{\nu l} Y_{lm} \vert \psi^{\textit{ps}}_{\bar{b} \bar{\mathbf{k}}} \rangle \nonumber \; , 
\end{align}
where $\phi$ are the OPFs and $Y_{lm}$ are spherical harmonics. 
The local basis is substantially coarser than that used in the construction of the OPFs. 
Therefore we enforce unitarity by constructing the overlap matrix $A$,
\begin{equation}
A_{\nu \nu'; l } = \int_0^{r_a} dr r^2  \phi^{\textit{ps}}_{\nu l }(r) \phi^{\textit{ps}}_{\nu' l }(r) \;,
\end{equation}
where any deviation from the identity matrix is due to errors from using a coarser grid. 
The augmentation of Eq.~\ref{eq-aug1} is modified, 
\begin{align}
\label{eq-aug2}
 \psi^\textit{ae}_{\bar{b} \bar{\mathbf{k}}} (\mathbf{r}) =  \psi^\textit{ps}_{\bar{b} \bar{\mathbf{k}}}(\mathbf{r})  
 + \sum_{lm} & \sum_{\nu,\nu' } Y_{lm}(\hat{r}) \left[ \phi^{ae}_{\nu l}(\mathbf{r}) - \phi^{ps}_{\nu l }(\mathbf{r}) \right] \times \nonumber \quad \\ 
 \times &A^{-1}_{\nu \nu';l} \langle \phi^{ps}_{\nu' l} Y_{lm} \vert \psi^{\textit{ps}}_{\bar{b} \bar{\mathbf{k}}} \rangle  ,
\end{align}
preserving unitarity. 

Each process stores a copy of the OPFs and carries out the augmentation for its subset of bands and {\it k}-points. 
The scaling of this section goes as $\mathcal{O}[N^2]$ with a factors of $N_i$ sites and $N_b$ bands.

\subsection{Building $g$ and $\chi_0$}

The RPA polarizability from Eq.~\ref{eq-rpa} can be transformed from the two-time form to a convolution over energy, which can be carried out along the imaginary axis, 
\begin{align}
\chi_0&(\mathbf{r}, \mathbf{r}', \omega=0) =  -i \sum_\sigma \int_{-\infty}^{\infty} \frac{dE}{2\pi}  \, g_\sigma( \mathbf{r}, \mathbf{r}', E )g_\sigma( \mathbf{r}', \mathbf{r}, E ) \nonumber \\
&=   \sum_\sigma \int_{-\infty}^{\infty} \frac{dt}{2\pi}  \, g_\sigma( \mathbf{r}, \mathbf{r}', \mu+it )g_\sigma( \mathbf{r}', \mathbf{r}, \mu+it  ) 
\label{chi1}
\end{align}
where $\mu$ is chosen to be 
in the middle of the gap for insulators or at the Fermi level in metals to avoid poles in $g$.  
[With minimal approximation, a small energy $\Delta$ can be added in quadrature in metals 
to the difference between $\mu$ and the Kohn-Sham eigenvalue $\varepsilon_{b\mathbf{k}}$ according to 
$(\mu - \varepsilon_{b\mathbf{k}}) \rightarrow  \pm \sqrt{( \mu - \varepsilon_{b\mathbf{k}})^2 + \Delta^2}$.]
This same approach can be used for calculating the dynamic polarizability, $\omega \ne 0$. 
However, additional care is needed to avoid the poles in Green's function $g(\omega + \mu +it)$ as $t \rightarrow 0$. 
The integral is replaced by a sum over an energy grid as outlined in appendix~\ref{energyIntegration}.

In principle, partial Green's functions could be constructed using the band and k-point distribution from the previous step. 
However, it is more efficient to redistribute the wave functions into blocks by $\mathbf{r}$ and site.
The processors are split into groups such that each group works on its own site or set of sites. 
Within each group, the processors divide up the $\mathbf{r}$-points. 
This means that the wave functions are now distributed as $\psi_{b \mathbf{k}\sigma}( \bar{\mathbf{r}};\bar{\tau} )$. 
\begin{equation}
g^{(\tau)}_\sigma(\bar{\mathbf{r}},\mathbf{\bar{r}}', \mu + it ) = N_k^{-1} \sum_\mathbf{k} \sum_{b=1}^{N_b} 
\frac{ \psi^\dagger_{b \mathbf{k} \sigma}(\bar{\mathbf{r}}) \psi_{b \mathbf{k} \sigma}(\bar{\mathbf{r}}') } {\mu + it - \varepsilon_{b \mathbf{k} \sigma}}
\end{equation}
where $\mathbf{r}$ implicitly includes only the points for site $\tau$. 
If background communications are enabled, the majority of this data transfer takes place while the conversion process is on-going.  
Within a group of processors cooperatively working on a site $g^{(\tau)}$, the wave functions are shared.
The scaling of this section goes as $\mathcal{O}[N^2]$ with a factors of $N_i$ sites and $N_b$ bands. 

An important consideration in efficiently calculating the Green's function is that it involves an outer-product of the wave functions. 
For each band and k-point, $N_r$ inputs are turned into $N_r^2$ outputs, with $2 N_r^2$ operations. 
In a typical, small calculation the real-space grid has 1600 points per site, which means that at each frequency the Green's function is just under 40~MB in size, making it too large to fit in the local cache of a typical modern processor. 
A na{\" i}ve implementation would be limited by memory bandwidth. 
Instead, the wave functions are broken up, and the Green's function is calculated by tiles in real space.

\subsection{Construction of $\chi$ and $W$}
\label{construct-chi}

In the previous step,  we calculated the irreducible polarizability $\chi^0$. The reducible polarizability is given by 
\begin{align}
\chi = \left[ 1 - \chi^0 v \right]^{-1} \chi^0,
\end{align}
where $v$ is the Coulomb potential operator. We do this by projecting into a spherical basis
\begin{align}
\chi^0(\mathbf{r}, \mathbf{r}') &= \sum_{lm} \sum_{l'm'} \chi^0_{lm;l'm'}(r,r') Y_{lm}(\hat{r}) Y^*_{l'm'}(\hat{r}') \\
v(\mathbf{r}, \mathbf{r}') &= \sum_{lm} \frac{4 \pi}{2l+1} \frac{r_<^l}{r_>^{l+1}} Y_{lm}(\hat{r}) Y^*_{lm}(\hat{r}')
\end{align}
where the Coulomb operator is diagonal in $l,m$. 
We can define
\begin{align}
S_{lm;l'm'}(r,r') &= \delta_{l,l'}\delta_{m,m'}\delta(r-r') \\ 
&- \int \! dx x^2 \, \chi^0_{lm;l'm'}(r,x) \frac{4\pi}{2l'+1} \left[\frac{r_<^{l'}}{r_>^{l'+1}} \right]_{x,r'} \nonumber
\end{align}
by taking advantage of the diagonal nature of $v$ in this basis. 
We therefore have 
\begin{align}
\chi_{lm;l'm'}(r,r') =S^{-1}_{lm;l''m''}(r,r'') \chi^0_{l''m'';l'm'}(r'',r') , 
\end{align}
where the matrices have dimension $N_r(2N_l+1)$.

In this basis the induced change in electron density from the short-range part of the core-hole potential $v^{(1)}$ is 
\begin{align}
\rho^{\textrm{ind}}(\mathbf{r}) &= \sum_{lm} \rho_{lm}^{\textrm{ind}}(r)Y_{lm}(\hat{r}) \\
 \rho_{lm}^{\textrm{ind}}(r) &= \sum_{l'm'} \int \! d^3r' \chi_{lm;l'm'}(r,r') v^{(1)}(r') Y_{l'm'}(\hat{r}') \nonumber \\
 &=  \int dr' r'^2 \chi_{lm;00}(r,r') v^{(1)}(r')
\end{align}
The perturbing  (core-hole) potential is taken to be spherical and therefore only the $l'=0$ part of $\chi$ contributes. Giving a final, induced potential
\begin{align}
v^{\textrm{ind}}(\mathbf{r}) &= \sum_{lm} v_{lm}^{\textrm{ind}}(r) Y_{lm}(\hat{r}) \nonumber \\
 v_{lm}^{\textrm{ind}}(r) &= \int \! dr' r'^2   \rho_{lm}^{\textrm{ind}}(r') \frac{4 \pi}{2l+1}  \frac{r_<^l}{r_>^{l+1}} .
\end{align}

By default, only the $l=0$ part of the response is calculated, 
according to 
\begin{align}
\bar{\chi}(r,r') &= S^{-1}_{00;00}(r,r'') \chi^0_{00;00}(r'',r') \nonumber \\
\bar{\rho}^{\textrm{ind}}(r) &= \int dr' r'^2 \bar{\chi}(r,r') v^{(1)}(r') \, .
\end{align}
The resulting induced potential is approximately the same as the $l=0$ component of the full induced density $\bar{\rho} \approx \rho_{00}$. 
For the core-hole potential, the strong localization means that $l=0$ component of the induced potential is dominant, and this approximation is reasonable. 
For valence calculations, including $l \le 2$ was found to be sufficient.
Because the dimension of $\chi$ is independent of system size, this section scales only with the number of sites $N_i$, linearly with system size or $\mathcal{O}[N]$.

\subsection{$\Gamma$-point}

For large cells, only a single {\it k}-point is required, and the electron orbitals can be calculated at the $\Gamma$-point. 
For systems with time-reversal symmetry the Bloch functions can be treated as real (instead of complex).  
This results a reduction of the required storage by half and substantial time savings in the DFT stage. 
A smaller reduction in runtime is also realized in the screening calculation as shown below in Table~\ref{weakTable}.

\subsection{Timing and Scaling}
\label{sec-time-scale}

\begin{figure}
\begin{centering}
\includegraphics[height=3.1in,trim=0 35 15 70,angle=270]{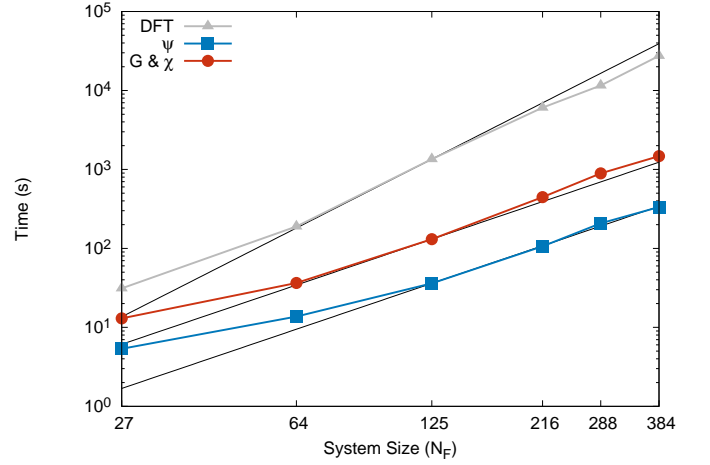}
\caption{ The run time for the DFT and screening as a function of the system size. 
The screening is divided into the wave function projection $\psi$ (blue, squares) and construction of the Green's function and polarizability $G$ \& $\chi$ (red, circles). Guidelines are shown for $\alpha N^2$, normalized to $N_F=125$. 
For the DFT calculation (grey, triangles) the guideline is $N^3$. 
The plotted data are included in the top section of Table~\ref{weakTable}.}
\label{SystemScaling}
\end{centering}
\end{figure}

The calculation of the screening as outlined here is dominated by three steps: calculating the wave functions, 
projecting them onto the radial grid, and constructing the Green's function and $\chi^0$. 
The first, calculating the electron orbitals using DFT, is carried out using the {\sc Quantum} ESPRESSO code \cite{espresso2,espresso1,*espresso0}. 
We report the timing of the DFT step for completeness, however, we are focused on the two steps that are specific to the screening calculation.

To investigate the timing and scaling of screening calculations within {\sc ocean} we use LiF (physical details and convergence parameters are given in appendix~\ref{conv-param}). 
There are two classes of scaling that we are interested in. 
First there is system scaling, by which we mean the increase in run time with an increase in the system size. 
This highlights the inherent simulation size limits of our approach. 
We will also consider strong scaling, the change in execution time due to changing the number of processors.  
We have implemented two levels of parallelism for the screening calculations: internode {\sc mpi} and shared memory {\sc openmp}. 
The testbed for these calculations is a small cluster with 12 nodes. Each node has a dual-socket with 8 processors per CPU (16 per node) \cite{ClusterFootnote}. 
Each timing run was repeated 8 times, and the average value is reported (DFT calculations were run only once).

For the tests, we consider various supercells of LiF, from the unit cell $N_F=1$, to an $6\times8\times8$ supercell $N_F=384$, covering cell volumes from 110~a.u.$^3$ to 42000~a.u.$^3$ (6.2~nm$^3$). For these runs, 32 bands per unit cell were included (12288 for $N_F=384$).
For each supercell the screened Coulomb potential is calculated for all the fluorine sites. 
Each local real-space grid had 2624 points, and the Green's functions were evaluated at 16 imaginary frequencies. 

\begin{table}
\centering
 \begin{tabular}{ c | c |  c | c | c | c }
\,\!  N$_F$ \,\! & Vol.~\!(a.u.$\!^3$) & {\it k}-pts  & DFT (s) &  $\,\psi$ (s) & $\,  g$ \& $\chi$ (s) \\
  \hline  \hline
384 & 42002 & $\Gamma$ &   27421.6  & \, \,  333.6  (8.4) &  1474 \,\, (22)\, \\  
288 &  31502 & $\Gamma$ &  11556.1  & \, \,  208.1  (4.5) & \, \, 891.0 (14.6) \\
216 & 23626 & $\Gamma$ &\,   6063.4 & \, \, 106.4 (3.0)  & \, \,   445.7 \, (5.2) \\
125 & 13673 & $\Gamma$ &\,  1350.2 & \, \,  \, 36.1 (0.1) & \, \,  131.0 \, (1.5) \\
\,  64 & \, 7000 & $\Gamma$ &\, \, 189.6 &  \, \, \,  13.7 (0.1)  & \, \, \, 36.5 \, (1.1) \\
\,  27 & \, 2953 & $\Gamma$ &\, \, \, 31.2 & \, \, \, 5.36 (0.02)  & \, \, \, 13.0 \, (0.4) \\
 \hline  \hline
      125 & 13673 & 1 &\, 2791.2  & \, \, \, 92.0 (1.0) & \, \, 166.1 (3.6) \\
\, 64 & \, 7000 & 1 & \, \,  609.5 & \, \, \, 79.9 (0.2)  & \, \, \, 57.0 (5.1)  \\
\, 27 & \, 2953 & 1 &  \, \, \, 62.5 & \, \, \, 30.3  (0.1) & \, \, \, 15.6  (0.5) \\
\, \, 8  & \, \, 875 & 1 & \, \, \, \, 6.0  & \, \, \, 9.36 (0.41) & \, \, 1.28 (0.04) \\
 \hline  \hline
\, 64 & \, 7000 & 8 &  \, 4585.4  & \, \, 175.5 (5.2) & \, \, 425.4 (72.9) \\
\, 27   & \, 2953& 8& \, \,   408.6 &  \, \, \, 45.8 (1.0) & \, \, 128.0 \, (4.7) \\
\, \, 8  &  \, \, 875& 8 & \, \, \,  17.9 & \, \, \, 17.5 (0.3) & \, \, \, 9.86 (0.03) \\
\, \, 1 & \, \, 110 & 8 &  \, \, \, \, 0.2& \, \, \, 4.39 (0.11) & \, \, \, 0.46 (0.04) \\
\end{tabular}
\caption{ Timing in seconds of selected LiF runs, denoted by the number of fluorine atoms and the {\it k}-point sampling ({\it k}-pts) running on 128 processors (see text).   
The DFT timing includes only the ``electron'' time for the non-self-consistent run to generate the wave functions for the screening. 
The two divisions of the screening calculation, labeled $\psi$ and $g$ \& $\chi$, encompass the totality of the runtime for the screening. 
The timing for $\psi$ includes the time to read in the wave functions, project and augment them, and redistribute them. 
The timing for $g$ \& $\chi$ includes the time to construct the Green's functions, evaluate the polarizability, and screen the core-hole potential. 
The root-mean-square deviations were determined over 8 repeated runs and are included in parentheses.  }
\label{weakTable}
\end{table}

First we look at the scaling with system size or weak scaling.  
In Fig.~\ref{SystemScaling} we show the run time of the projection ($\psi$) and construction of the Green's function and polarizability ($g$ \& $\chi$) steps as a function of super cell size. 
The straight lines on the log-log plot are $\alpha N^2$, where $\alpha$ is set by the timing of $N_F=125$. 
In this set of runs only $\Gamma$-point sampling of the Brillouin zone was used. 
The $\approx \mathcal{O}[N^2]$ growth in calculating the screening is evident by the linear plots, though overhead or inefficiencies dominate the timing of the smallest run. 
Additionally, for large systems the `$g$ \& $\chi$' diverges slightly from the expected $N^2$ behavior which may be indicative of poor cache reuse or other bottlenecks for large system sizes. 
We also include the timing of the DFT portion which follows $\alpha N^3$. 
128 processors across 8 nodes were used for each run, using 128 {\sc mpi} processes.

The timing information for the range of systems from $N_F=1$ to $N_F=384$ is shown in Table~\ref{weakTable}. 
To give a full picture of the scaling, three different settings for the {\it k}-point sampling were used: finite sampling on a $2^3$ {\it k}-point mesh, single {\it k}-point sampling, and $\Gamma$-point sampling. 
Above $N_F=27$, a single {\it k}-point is sufficient to sample the Brillouin zone and gives the same results as the $2^3$ sampling (but more quickly). 
The purpose of timing single {\it k}-point runs in addition to the $\Gamma$-point runs is to distinguish the changes due to the reduction in {\it k}-point sampling from 8 to 1 from the changes in moving between complex and real Bloch functions.  
The single {\it k}-point is taken at $(\sfrac{1}{8}, \sfrac{2}{8}, \sfrac{3}{8})$.

\begin{figure}
\begin{centering}
\includegraphics[height=3.1in,trim=0 35 0 70,angle=270]{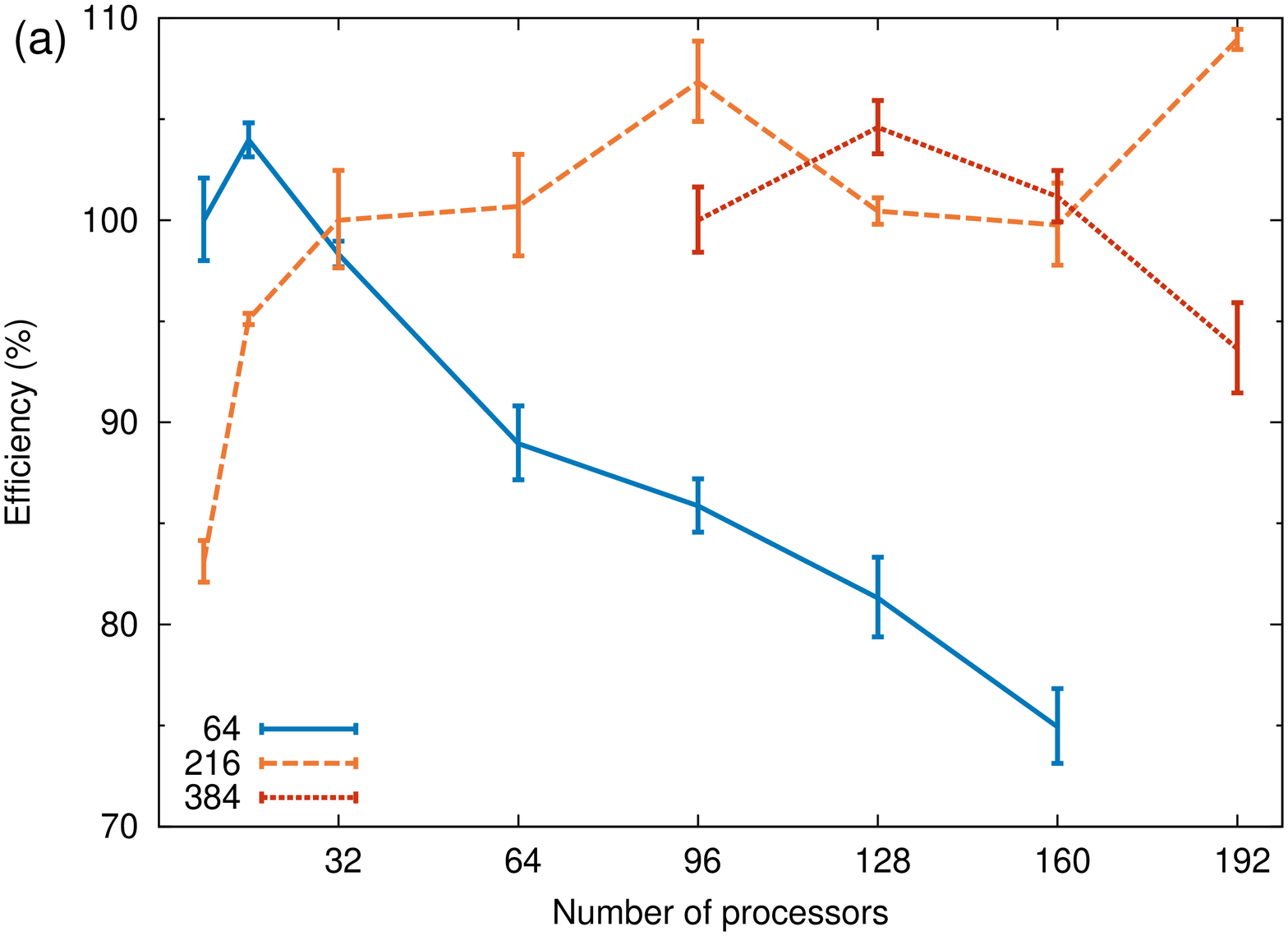}
\includegraphics[height=3.1in,trim=0 35 15 70,angle=270]{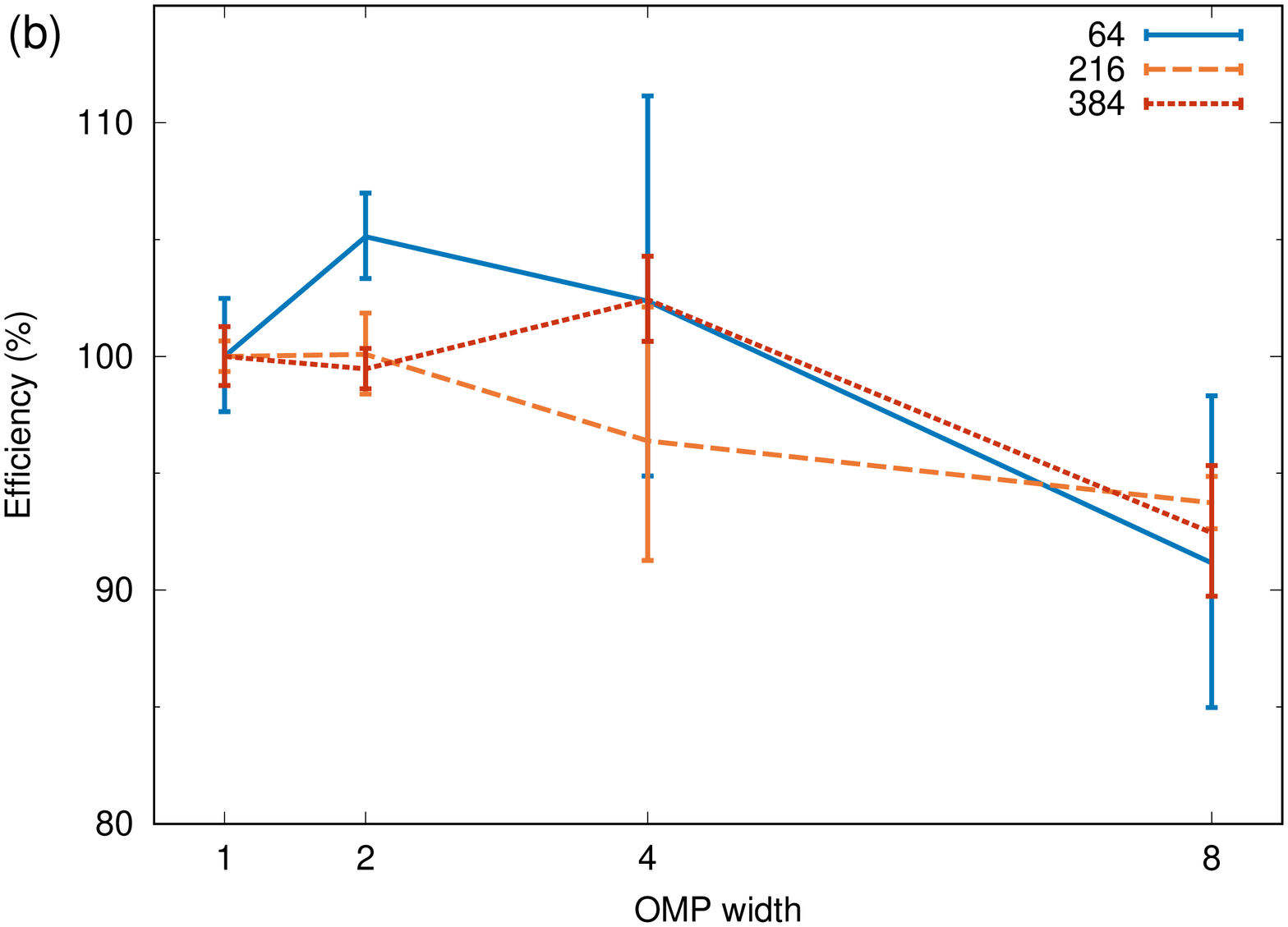}
\caption{ (a) The strong scaling behavior of the complete screening calculation for three different systems sizes, $N_F=64$ (solid, blue) normalized for 8 processors, $N_F=216$ (dashed, orange) normalized for 32 processors, and $N_F=384$ (dotted, red) normalized at 96 processors. 
The error bars reflect the root mean square deviation determined by averaging over 8 runs. 
We see a fall-off in efficiency of the smaller system, but for $N_F=216$ and $N_F=384$ we see good strong scaling up to 192 processors.  
(b) The efficiency of the {\sc omp} parallelization, normalized with one {\sc omp} thread per {\sc mpi} process. The number of processors is kept fixed at 128 and divided between {\sc mpi} and {\sc omp}. 
}
\label{strongScaling}
\end{centering}
\end{figure}

Next, we present the change in run time with changing processor number or strong scaling, Fig.~\ref{strongScaling}. 
Ideally, doubling the number of processors used in a calculation will halve the runtime. 
Longer than expected runtimes may result from serial sections of the code or communication overhead. 
Shorter than expected times may result from better data caching due to each processor working on a smaller data set. 
Here we plot the data as the efficiency $E$ as a function of the number of processors $N$,
\begin{equation}
E( N ) = \frac{N_0}{N} \, \frac{t( N_0 )}{  t( N ) } \times 100\%
\end{equation}
where the efficiency is normalized to the run time with $N_0$ processors. 
Ideal scaling is given by an efficiency of 100~\%.
The efficiency is the measure of merit for planning high-throughput calculations. 
In high-throughput calculations the available hardware resources can be divided between many different calculations, and the runtime of any single calculation should be balanced against the runtime of the complete dataset.

In Fig.~\ref{strongScaling}(a) we show that for a moderately sized system $N_F=64$, there is a drop-off in efficiency above 16 processors. 
In part this is a reflection of the structure of our computer cluster, where each 
addition of 16 processors 
increases the number of nodes in the calculation by 1. 
While the efficiency is quite poor running on 160 processors, the runtime is also very brief. 
The average time is 41.1~sec.\ compared to an idea time of 27.3~sec. 
The larger systems show better scaling. 

\begin{table}
\centering
 \begin{tabular}{ c | c |  c | c | c  }
\,\!  N$_F$ \,\! &  \, {\sc mpi} \, & \,\! {\sc omp} \!\, & $\,\psi$ (s) & $\, g$ \& $\chi$ (s) \\
  \hline
    \hline
  384 &  128  & 1 &   \, \,  333.6 \, (8.4) & \, 1474 \, (22) \\
    & \, 64 & 2 & \, \, 330.9  (11.6)   & \,  1488 \, (15) \\
    & \, 32 & 4 & \, \,   336.4 \, (2.6) & \,   1430 \, (31)  \\
  & \, 16  & 8 & \, \, 504.5 (19.3)   & \, 1452 \, (49) \\
\hline
\hline
  288 &  128  & 1 &   \, \, 208.1  \, (4.5) & \, \, 891.0 (14.6)  \\
    & \, 64 & 2 & \, \, 204.2 \, (4.9)    & \, \, 898.1 (33.0)   \\
      & \, 32 & 4 & \, \, 210.4 \, (2.1)    & \, \,  767.4 \, (5.7)  \\
  & \, 16  & 8 & \, \, 315.0 (12.6)   & \, \,   775.8 \, (6.0)  \\
\hline
\hline
  216 &  128  & 1 &   \, \,  106.4 (3.0)  & \, \,   445.7 \, (5.2)  \\
    & \, 64 & 2 & \, \, 105.8 (2.1)      & \, \,   447.1 \, (9.3) \\
  & \, 32 & 4 & \, \, 108.7 (5.4)   & \, \,  465.9 (28.4)  \\
  & \, 16  & 8 & \, \, 167.3 (2.1)  & \, \,   423.6 \, (7.4)   \\
\hline
\hline
  125 & 128  & 1 &   \, \, \,  36.1  (0.1) & \, \, 131.0 \, (1.5)     \\
    & \, 64 & 2 & \, \, \, 35.4 (1.0)    & \, \,   151.7 \, (2.9) \\
  & \, 32 & 4 & \, \, \,  34.3 (0.2)  & \, \,  148.0 \, (7.2)  \\
  & \, 16  & 8 & \, \, \, 53.6 (0.7)  & \, \,  157.4 (13.3)  \\
\hline
\hline
\, 64 &   128 & 1 & \, \, \, 13.7 (0.1)   & \, \, \,  36.5 (1.1) \\
  & \, 64 & 2 & \, \, \,   11.6   (0.1) & \,  \, \,  37.4 (0.8)  \\
  & \, 32 & 4 & \, \, \,  12.4    (1.7)  & \, \, \, 38.6   (4.1)  \\
  & \, 16  & 8 & \, \, \, 16.3   (0.1) & \, \, \,  40.6 (4.4) \\
\hline
\hline
\, 27 &   128 & 1 & \, \, \,  \, 5.36 (0.02)  & \, \, \, 13.0 (0.4) \\
  & \, 64 & 2 & \, \, \,  \,  2.88 (0.01) & \, \, \, \, 8.4 (0.5)  \\
  & \, 32 & 4 & \, \, \,  \,  2.69 (0.36)   & \, \, \, 10.4 (1.7) \\
  & \, 16  & 8 & \, \, \,  \,  3.35 (0.02)  & \, \, \,   12.2 (0.3)\\
\end{tabular}
\caption{ Timing in seconds of  LiF runs with $\Gamma$-point sampling.
A total of 128 processors are used for each run divided between {\sc mpi} tasks and {\sc openmp} threads. 
The two timing sections are the same as in Table~\ref{weakTable}. }
\label{WeakOMPTable}
\end{table}

Lastly, we examine the {\sc omp} parallelism by repeating the $\Gamma$-centered calculations this time including thread-level parallelism. 
The total number of processors is held fixed at 128, but divided between {\sc mpi} and {\sc omp} with 1, 2, 4, and 8 {\sc omp} threads per {\sc mpi} process. 
The results are shown in Table~\ref{WeakOMPTable} and for three of system sizes in Fig.~\ref{strongScaling}(b). 
We find relatively uniform performance across the first three processor arrangements, but a drop-off in performance using 8 {\sc omp} threads. 
This drop-off indicates an opportunity for further code refinement to better support higher-levels of {\sc openmp} parallelism. 
In Table~\ref{WeakOMPTable} it can be seen that this inefficiency is primarily the result of poor scaling of the `$\psi$' section. 

\section{Discussion and Future Directions}

We have presented a local, real-space method for calculating the RPA polarizability of condensed systems. 
The method scales well with system size $N$, $\mathcal{O}[N^2 \textrm{log} N]$. 
While the method only provides the full RPA response within a restricted real-space range, it is coupled with a model dielectric function to provide the full response. 
This approximation is controlled through a radial cutoff $R_S$, and the contribution of the model goes smoothly to zero as $R_S \rightarrow \infty$. 
This method is implemented within the {\sc ocean} code where the screened Coulomb operator $W$ is used as part of the BSE Hamiltonian for calculating both core-level (near-edge x-ray) and valence-level (UV/vis) spectroscopy. 

In regions near an atom, we have shown that the pseudopotential approximation results in an incorrect RPA polarizability. 
In our screening calculations we correct for this by augmenting the electron orbitals from pseudopotential-based calculations to restore the all-electron character. 
This effect is noticeable in near-edge x-ray spectra, where changes in exciton strength and position due to deficiencies in the screened core-hole potential are similar to changes due to thermal disorder. 
Above the x-ray edge the differences between spectra calculated with and without augmented orbitals fade with increasing photoelectron energy as the photoelectron becomes more and more delocalized. 
In the case of valence-level UV/vis spectroscopy we found that augmentation is not necessary.  

We conclude with a few remarks on improvements and future extensions.  
In particular, the relative ease of the local, real-space method may present an opportunity for developing and testing new model dielectric functions and easily benchmarking them against RPA or TD-DFT quality calculations. 
In the remainder of this section we first detail some enhancements to the current method. 
Next, we discuss computations other than particle-hole spectroscopy that could benefit from our local, real-space polarizability. 
Finally, we show how the real-space method is amenable to higher-order calculations beyond RPA.

\subsection{Refinements}

In the current implementation there is no re-use of the Green's functions between sites. 
The use of site-centered radial grids makes it unlikely that a given $(\mathbf{r,r}')$ pair of one site will exist in the grid of another. 
However, it reasonable to expect that many point pairs will be nearly shared, {\it e.g.}, for sites $\alpha$ and $\beta$ that $\mathbf{r}_\alpha \approx \mathbf{r}_\beta$ and $\mathbf{r}'_\alpha \approx \mathbf{r}'_\beta$. 
In the future, the construction of the grids can be relaxed to maximize the overlaps, decreasing the computational cost of generating the Green's functions.
This would be especially helpful in the case of valence calculations where the site density is high. 

Future improvements to the scalability with system size must focus on generating the electron wave functions.  
For medium to large system sizes, most of the time is in calculating the screening is spent in the DFT (see Sec.~\ref{sec-time-scale}). 
This is exacerbated by the need for unoccupied states in the calculating the Green's function. 
Several methods have been proposed to reduce the number of unoccupied states. 

One option is to directly replace part of the sum over unoccupied states. 
The effective-energy techniques replace the energy denominator in the sum over unoccupied states \cite{PhysRevB.78.085125,PhysRevB.82.041103,PhysRevB.85.085126}. 
The completeness of the Bloch functions then allows the sum over unoccupied bands to be replaced with the identity minus a sum over occupied bands. 
However, these approaches differ in two main ways from ours, not including our local approximation. 
First, the energy convolution is carried out analytically, and the RPA polarizability is constructed via sums over states. 
In our approach the Green's functions are built via a sum over states and the convolution is carried out numerically. 
Second, the effective-energy approaches are formulated in reciprocal space, which has the advantage of a straightforward approximation for the effective energy. 
An easier approach might be to approximate the neglected high-energy bands as plane waves \cite{JAMES1996935,STEINBECK2000105,PhysRevLett.107.186404,PhysRevB.90.075125}.

Alternatively, the induced density response and therefore the screened potential can be more directly calculated using the linear-response Sternheimer equation approach \cite{PhysRevB.81.115105,PhysRevB.88.075117} or eigenvalue decomposition of the polarizability matrix \cite{PhysRevB.78.113303,PhysRevB.85.081101}. 
While these approaches only require the occupied orbitals, they maintain an unfavorable $N^4$ scaling with system size. 
Better scaling might be achievable by adapting these approaches to determine only the local response. 

Lastly, the model dielectric function used to screen $v_2$ (Eq.~\ref{W1}) was designed for bulk systems.
In the case of highly anisotropic systems, like a surface or interface, this may result in slower convergence with respect to the shell radius $R_S$, requiring a larger real-space RPA calculation. 
This could be addressed by modifying the screening model or alleviated through a judicious choice of model parameters, namely choosing the average density $\rho_0$ to more accurately reflect the bulk material (see model details in Appendix~\ref{appendixLL}).

\subsection{Non-BSE Applications}

The screened Coulomb interaction $W$ has many uses in calculations of condensed matter systems other than the use presented here of the direct interaction in the the BSE. 
One such application is in self-energy calculations using the {\it GW} method which requires evaluating the self-energy operator \cite{PhysRev.139.A796},
\begin{equation}
\Sigma(\mathbf{r},\mathbf{r}',E) = i\int \! \frac{ d\omega}{2\pi} \; G(\mathbf{r},\mathbf{r}',E-\omega) W(\mathbf{r},\mathbf{r}',\omega) \quad .
\end{equation}
The local screening approach outlined here can be used to efficiently generate $W$ with RPA quality for small distances $\vert \mathbf{r} - \mathbf{r}'\vert$, just as was shown for valence BSE calculations. 
The terms in the {\it GW} calculations are significantly less localized than core-level excitations, making discrepancies in the short-range part of $W$ less important, and augmentation of the orbitals may not be necessary. 
However, in transition metals the {\it d} orbitals often drive important characteristics of the electronic behavior, forming the top of the valence bands, the bottom of the conduction bands, or both. 
From atomic calculations, it can be seen that the {\it d} orbitals overlap significantly with the semi-core orbitals of the same principle quantum number, {\it e.g.}, the 3{\it d} with the 3{\it s} and 3{\it p}. 
High-accuracy calculations involving localized {\it d} orbitals may require accurately correcting the nodal structure of the {\it s} and {\it p} wave functions in much the same manner as the we have shown for core-level spectroscopy.
Several {\it GW} studies have pointed out discrepancies from using pseudopotentials \cite{PhysRevB.95.155121} or excluding semi-core states \cite{doi:10.1021/acs.jctc.9b00520,PhysRevLett.105.146401}. 
In the present approach only static screening was implemented. 
However, the contour integral in Eq.~\ref{chi1} can be modified to calculate $\omega \ne 0$, and the fundamental scaling of frequency-dependent screening remains the same as that of the static case. 
In addition to standard valence- and conduction-band self-energy calculations, an adaptation of this method could be applicable for determining accurate core-level binding energies \cite{doi:10.1021/acs.jctc.8b00458}.

The local, real-space screening might also be useful for phonon calculations. 
Within the harmonic approximation, the phonons of a  system can be fully described by the dynamical matrix. 
The elements of the dynamical matrix are proportional to the derivative of the force on atom $a$ with respect to changes in the position of $a'$.
This is equivalent to the second derivative of the total energy with respect to the displacement of both. 
The elements of the dynamical matrix can be calculated using density-functional perturbation theory %
 \cite{RevModPhys.73.515,PhysRevB.55.10337,PhysRevB.55.10355}. 
Alternatively, the dielectric response can be used since the polarizability describes how the electron density will change in response to a change in the potential, in this case the motion of the atomic nuclei.
Care would be required as the polarizability gives the density change in response to a local perturbing potential, but standard pseudopotentials include non-local terms \cite{PhysRevB.46.10734}.

\subsection{Beyond RPA Screening}
\label{alda}

The screening calculations in this paper have been carried out only within the RPA approximation. 
From a many-body perturbation theory perspective, the RPA is the lowest-order diagram for the polarizability. 
Higher-order approaches treat interactions between the electron and hole lines in the RPA, or, equivalently, add a vertex correction. 
Unfortunately, additional interaction or vertex terms increase the computational cost and scale worse with system size. 
Our local, real-space approach is an ideal testbed for investigating higher-order approaches because the increase in scaling complexity applies to the dimension of our local dielectric response function which is small and independent of the system size. 

As an example, we have implemented the vertex correction given by the adiabatic local-density approximation (ALDA). 
As shown by Del Sole, Reining, and Godby \cite{PhysRevB.49.8024}, if the first-order approximation to the one-electron self-energy is taken to be the local exchange-correlation potential 
\begin{align}
\Sigma(1,2) = \delta(1,2) v_\textit{xc}(1 )
\end{align}
then the reducible polarizability (and screened interaction $W$) undergo a relatively simple transformation. 
Repeating Eq.~\ref{eq-reducible}, 
\begin{align}
\chi  &= \left( 1 - \chi_0 v  \right)^{-1} \chi_0 \nonumber \\
\tilde{\chi} &= \left( 1 -  \chi_0(v + f_\textit{xc}) \right)^{-1} \chi_0 
\label{chi-alda}
\end{align}
where $f_\textit{xc}$ is the derivative of the exchange-correlation potential with respect to the density, $f_\textit{xc}(1,2) = \delta v_\textit{xc}(1)/\delta n(2)$, and $\tilde{\chi}$ is the ALDA polarizability. 
The use of an ALDA kernel has been investigated within  the {\it GW} approximation \cite{PhysRevB.49.8024,PhysRevB.76.155106,PhysRevLett.94.186402} and for valence BSE calculations of small molecules \cite{PhysRevB.73.205334}. 

Within the ALDA, $f_\textit{xc}$ is a contact interaction, and the expression in Eq.~\ref{chi-alda} is easily evaluated using the {\sc ocean} code as outlined in section~\ref{construct-chi}.   
In the real-space basis $f_{\textit{xc}}$ is diagonal and can be written
\begin{equation}
f_\textit{xc}(\mathbf{r},\mathbf{r}') = \delta(\mathbf{r-r}') \left. \frac{ d v_\textit{xc}(n)}{d n } \right|_{n=n(\mathbf{r})}
\label{kxc_alda}
\end{equation}
where $v_\textit{xc}$ is the LDA exchange-correlation potential and is evaluated at the density $n$ at position $\mathbf{r}$. 
The electron density $n(\mathbf{r})$ is taken from the initial DFT calculation used to generate the electron orbitals for the screening. 
We start with the Perdew-Zunger parameterization for the exchange \cite{PhysRevB.23.5048} and Vosko, Wilk and Nusair parameterization of the correlation energy \cite{doi:10.1139/p80-159} within the local-density approximation fit to the data of Ceperley and Alder \cite{PhysRevLett.45.566}. 
We calculate $f_{xc}$ directly as the second derivative of the exchange-correlation energy with respect to the density using a 5-point finite difference using density differences of 0.01~e$^{-}$ per a.u.$^{3}$. 
Spin-polarized calculations beyond the RPA are not yet supported, but can be included using this same scheme.

 \begin{figure}
\begin{centering}
\includegraphics[height=3.1in,trim=0 35 15 70,angle=270]{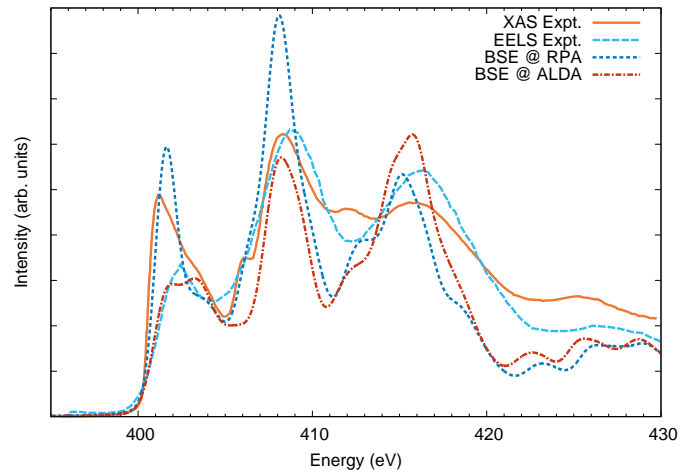}
\caption{ The N K-edge absorption for {\it h}-BN calculated using BSE, but two different approximations to calculate the screening. 
The adiabatic LDA (ALDA) approximation screens the core hole much more efficiently [see Fig.~\ref{shell-distance}(a)] resulting in significantly less excitonic binding. Compared to the RPA results, the ALDA has a weaker exciton and a general shift of spectral weight to higher energies. Two experimental spectra are included showing a large variation between XAS \cite{PREOBRAJENSKI200559} and electron energy loss spectroscopy (EELS) \cite{mcdougall_nicholls_partridge_mcculloch_2014}.
  }
\label{kxc_xas}
\end{centering}
\end{figure}

Once again looking at {\it h}-BN, we can examine how calculating the polarizability with the ALDA instead of RPA changes the XAS. 
In general, the ALDA results in a stronger induced potential [shown for LiF in Fig.~\ref{shell-distance}(a)]. 
This in turn leads to a weaker core-hole potential and correspondingly weaker excitonic effects. 
In Fig.~\ref{kxc_xas} we show the nitrogen K-edge XAS using both the RPA and ALDA for the screening. 
The BSE spectrum calculated using the ALDA is substantially different from the RPA result, but only in the near-edge region, within about 15~eV of the onset. 
The small differences at higher energies would be hidden by broadening if the calculation included the effects of the electron self-energy and vibrational disorder. 
While electron energy loss spectroscopy (EELS) taken within the dipole limit should probe the same excitations as XAS, it is seen in {\it h}-BN that there are large discrepancies between the two \cite{mcdougall_nicholls_partridge_mcculloch_2014}. 
In part, this is due to the different systematic errors such as surface sensitivity and self-absorption effects affecting XAS versus EELS. 
The ALDA screening appears superior to the RPA when comparing to the EELS data, but the RPA appears superior when comparing to the XAS measurement. 
A broad survey of materials and careful quantification of experimental uncertainties is necessary to establish the general applicability of one approximation or the other and should be the subject of future work.

While the only vertex correction that has been implemented in {\sc ocean} is the ALDA, an extension to semi-local exchange-correlation kernels is straightforward, requiring only the additional knowledge of the density gradients. 
Because $f_{\textit{xc}}$ is formed explicitly in real-space, the formulation of Eq.~\ref{chi-alda} is also compatible with non-local exchange-correlation potentials.  
This would require construction of $f_{\textit{xc}}$ as a real-space matrix instead of the diagonal form (Eq.~\ref{kxc_alda}). 
However, the response is still localized and in response to a local perturbing potential with the long-range response handled by a model. 
Therefore, any non-local $f_{\textit{xc}}$ must also be of limited range.

\appendix

\section{Input parameters for x-ray and optical calculations}
\label{conv-param}

We consider the halide K edges of lithium halides, LiF, LiCl, LiBr, and LiI.
All crystallize in the same rock salt \textit{Fm$\bar{3}$m} structure with lattice constants of 0.4017~nm, 0.5130~nm, 0.5501~nm, and 0.6000~nm, respectively \cite{Wyckoff}. 
The plane-wave cut-off energy was set to 100~Ry.\ (increased to 120~Ry.\ for the bromine and iodine pseudopotentials), and the density was converged using a $4^3$ shifted {\it k}-point grid. 
The BSE final states were solved on a $15^3$ grid ($16^3$ for LiI), including 32, 59, 127, or 128 unoccupied bands, respectively, and were downsampled onto a $12^3$ real-space mesh ($10^3$ for LiF). 
The calculations used the local-density approximation for the density functional \cite{PhysRevB.45.13244}, and pseudopotentials are taken from PseudoDojo \cite{pspdojo1,*pspdojo0} and generated with {\sc oncvpsp} \cite{PhysRevB.88.085117,*oncvp}. 
The DFT calculations were carried out using {\sc Quantum ESPRESSO} \cite{espresso2,espresso1,*espresso0}.
We used the ``high-accuracy'' version of the lithium pseudopotential, which includes the Li 1{\it s} as valence. 
For bromine and iodine the standard 3{\it d}4{\it s}4{\it p} (4{\it d}5{\it s}5{\it p}) were used, and additional calculations were carried out with the semi-core iodine pseudopotential.
Note that no valence-level spin-orbit coupling is considered, which would affect the Br 4{\it p} or I 5{\it p} states. 

For the screening calculations of the lithium halides, the orbitals for the screening calculation were generated on a $4^3$ k-point grid, including 72, 150, 197, or 213 bands, for F, Cl, Br, and I respectively, such that energy range from the Fermi level (mid gap) to the highest unoccupied state was approximately 150~eV for each. 
The augmentation radius of each was set by the pseudopotential of each halide, 1.64~a.u., 1.76~a.u., 1.97~a.u., and 2.02~a.u., respectively, and 1.45~a.u.\ for the I semicore. 
For the heavier three the  polarizability was calculated within a sphere of radius 8~a.u.\ on a 160-point uniform radial grid and 64-point angular grid while the neutralizing shell was placed at $R_S = 4$~a.u. 
For the LiF the polarizability was calculated within a sphere of radius 10~a.u.\ with the neutralizing shell placed at $R_S=6$~a.u. 
The real-space grid was divided into three sections. The inner section used a 34-point Gauss-Legendre quadrature for the radial grid and a 64-point angular grid. From the augmentation radius of 1.64~a.u.\ to 2.96~a.u.\ a 27-point, uniformly spaced radial grid and 144-point angular grid was used, and the final grid was an 88-point, uniformly spaced radial grid and 256-point angular grid. 
This grid is excessive for calculations of spectra, but was chosen to accurately show the convergence effects in Fig.~\ref{shell-distance}. 
For the scaling tests of Sec.~\ref{sec-time-scale} a smaller grid was used. A 16-point uniform radial mesh and 36-point angular mesh was used for up to the augmentation radius, a 32-point uniform radial mesh and 64-point angular mesh was used outside it, and the sphere radius was limited to 8~a.u., giving a total of 2624 points.

For hexagonal boron nitride we used the experimentally determined lattice constants of $a = 2.504$~{\AA} and $c = 6.661$~{\AA} \cite{Wyckoff}. 
As for the lithium halides, pseudopotentials were taken from the PseudoDojo collection with a plane-wave cutoff of 100~Ry. 
A $20\times20\times8$ {\it k}-point mesh was used for the BSE with 92 unoccupied bands, while the screening was carried out using a $6\times6\times2$ {\it k}-point mesh and 300 bands. The {\sc exciting} calculation was carried out using a $10\times10\times6$ {\it k}-point mesh and 20 unoccupied bands. 
A sphere of radius 12~a.u.\ was used for the polarizability, divided into 5 sections with cut-offs of 1.39, 3, 4, 5, 9, and 12~a.u.\; 17, 10, 5, 14, and 6 point radial sampling; and angular meshes of 36, 100, 256, 625, 576, and 100. 
These larger grids were to ensure that the small variations in the potential (Fig.~\cite{hBN-pot}) were not due to numerical deficiencies and to accommodate the large shell radius $R_S=7$~a.u.

For the valence calculations, the LiF used a $10^3$ real-space grid for the BSE final states, requiring RPA screening calculations on that grid. 
The polarizability was calculated within a sphere of radius 10~a.u.\ and the neutralizing shell was placed at $R_S = 4$~a.u.
For the LiF valence calculations BSE final states were calculated on a $8^3$ {\it k}-point mesh with 6 conduction bands and 5 valence bands. This lower {\it k}-point mesh was chosen to facilitate the more computationally expensive comparison calculations. 
The RPA screening for the valence used a $3^3$ {\it k}-point mesh and 72 bands. 
Silicon crystalizes in a \textit{Fd$\bar{3}$m} structure with 
experimental lattice constant  0.543~nm \cite{Wyckoff}. 
The PseudoDojo pseudopotential for silicon and a planewave cut-off of 100~Ry were used. 
The BSE final states were calculated on a $16^3$  {\it k}-point grid, including 8 conduction and 4 valence bands, and a real-space grid of $8^3$. 
For the RPA screening 200 bands on a $2^3$ {\it k}-point mesh were used, and the polarizability was calculated with a sphere of radius 8~a.u.\ with the neutralizing shell placed at $R_S = 3.5$~a.u.

\section{Model polarization}
\label{appendixLL}
Here we reproduce the model screening of a spherical shell of charge by the model dielectric function introduced in \cite{Shirley2005} as used in \cite{Ultramicroscopy}:
\begin{align}
\chi_M(\mathbf{r},\mathbf{r}') &= - (2 \rho_0)^{-1}  \times \\
&\times \nabla \cdot \nabla' \big[ \left[ \rho(\mathbf{r}) + \rho(\mathbf{r}') \right] B(\vert \mathbf{r}- \mathbf{r}'\vert) \big], \nonumber
\end{align}
where $\rho$ is the local electron density and $\rho_0$ is the average electron density.
The real-space model $B$ is transform of the Levine-Louie dielectric model $B(q)$. 
\begin{align}
B(\vert \mathbf{r}- \mathbf{r}'\vert) &= \int \frac{d^3q}{(2 \pi)^3} B(q) \exp[ 
i \, \mathbf{q} \cdot (\mathbf{r} - \mathbf{r}')] \\
B(q) &= \frac{1}{4 \pi}\left( \frac{1}{\epsilon_{\textrm{LL}}(q;\rho_0,\epsilon_\infty)} - 1 \right) .
\end{align}
The original formulation of the Levine-Louie model in Ref.~\cite{PhysRevB.25.6310} requires something akin to an average band gap, 
but this can be reformulated using the long-range dielectric constant $\epsilon_\infty$. 
\begin{widetext}
\begin{equation}
 \epsilon_{\textrm{LL}} = 
1 + \frac{2}{\pi q_F }\left[ \frac{1}{Q^2} - \frac{\lambda}{2 Q^3} \left( \tan^{-1}\!\left[\frac{2Q+Q^2}{\lambda}\right] + \tan^{-1}\!\left[\frac{2Q-Q^2}{\lambda}\right]\right) + \left( \frac{\lambda^2}{8 Q^5} + \frac{1}{2Q^3}-\frac{1}{8Q}\right)\ln\!\left( \frac{\lambda^2+(2Q+Q^2)^2}{\lambda^2+(2Q-Q^2)^2}\right) \right] \quad ,
\end{equation}
\end{widetext}
where $Q=q/q_F$, $\lambda^2 = (\epsilon_\infty-1)^{-1} \omega_p^2 \omega_F^{-2}$, $\omega_p$ is the plasmon frequency, and $\omega_F$ and $q_F$ are the Fermi frequency and wave vector of a non-interacting electron gas of density $\rho_0$.

\section{Construction of OPFs}
\label{app-OPFs}

The construction of the projectors is as follows. 
For each angular momentum $l$, self-consistent solutions to Eq.~\ref{eq-atom} are determined for both the all-electron and pseudo-potential systems, {\it e.g.}, for an isolated atom in either the ground-state or a positive ion such that the valence electrons are all bound. 
For this purpose, the desired energy window for the projectors is selected by choice.  
For each $l$, the energy of the most-bound valence state is found. In some cases this would be a semi-core state such as the 3{\it s} and 3{\it p} orbitals in titanium. 
The minimum energy is set below this bound state, $\varepsilon_\textrm{min} = \varepsilon_v - \varepsilon_\textrm{pad}$, where the padding energy is 0.3~Ha.\ by default. 
The energy maximum is set to cover the relevant energy ranges, 50~eV to 100~eV for x-ray absorption transition matrix elements or 100~eV to 200~eV for RPA screening calculations (our default value is 5~Ha.\ $\approx$ 130~eV). 
Strictly speaking, this range depends on the Fermi energy and band gap, but for condensed-matter systems these values only vary by a few eV. 

Having defined the system's effective Hamiltonians
$H^{ae}$ and $H^{ps}$ and an energy window, we can begin to create the projectors. 
First, a set of pseudopotential partial waves are created for 128 energies spanning from $\varepsilon_\textrm{min}$ to  $\varepsilon_\textrm{max}$
\begin{equation}
H^{ps} \phi_i = \varepsilon_i \phi^{ps}_i
\end{equation}
Note, the calculation is only carried out to a finite radius, and, therefore, there is no problem normalizing these states. 
Additionally, these states are not orthogonal, but instead provide an over-determined basis. 

Next, for each pseudopotential partial wave $\phi^{ps}_i$ an all-electron partial wave $\phi^{ae}_i$ is also constructed. 
The $\phi^{ae}$ are not constructed to match exactly the energies of their corresponding pseudopotential partial wave, but instead to match the pseudopotential wave function and scattering properties. 
Specifically, we match the arctangent of the log-derivatives of the partial waves $\beta$, evaluated at the augmentation radius $r_a$
\begin{align}
\beta &= \left. \frac{r}{\phi}\frac{d \phi}{d r } \right|_{r=r_a} \nonumber \\
\bar{\delta} &= \arctan[ \beta ] - \pi \eta
\end{align}
where $\eta$ is the number of nodes in the partial wave, corrected for the lack of core-level resonances in the pseudopotential system. 
We will refer to $\bar{\delta}$ as the phase shift. 
The true phase shift can be related more carefully to the logarithmic derivative and the partial wave (see Sakurai Ch.~7 \cite{Sakurai} among others). 
The pseudopotential properties, matching energy and smoothly matching the wave functions between the pseudo and all-electron systems, are only exact at specific energies. 
At other energies the mapping is only approximate, and we chose to enforce smoothness at the expense of the energy. 
As we are only interested in the spatial behavior of the augmented orbitals, this choice is natural. 

Because we are matching phase shifts the energy of the all-electron partial wave is only approximately the same as that of the pseudopotential partial wave. 
A reference set of all-electron partial waves are constructed within the same energy window. 
Then the energy of each all-electron partial wave is iteratively refined until the phase shifts converge within  $3 \times 10^{-14}$.
Lastly, $\phi^{ae}_i$ is rescaled in a fashion that avoids numerical difficulties in cases of nodes and antinodes approaching $r_a$ for a given energy:   
\begin{align}
\phi^{ae}_i :&= A_i \phi^{ae}_i \nonumber \\
A_i &= \left. \frac{ \left( \phi_i^{ps} \right)^2 + \left( \sfrac{ d\phi_i^{ps}}{dr}\right)^2  }    {  \phi_i^{ps}  \phi_i^{ae} + \left( \sfrac{ d\phi_i^{ps}}{dr}\right) \left( \sfrac{ d\phi_i^{ae}}{dr}\right) } \right|_{r=r_a} . 
\end{align}
Here the partial waves and derivatives are evaluated at the augmentation radius $r_a$. 
In the case where the partial waves and first derivatives are equal we have $A=1$, 
whereas we typically find  $0.95 \le A \le 1.05$.

We now have a set of all-electron and pseudopotential partial waves. 
To generate the optimal projectors we use principle-component analysis (PCA) \cite{doi:10.1080/14786440109462720}. 
We generate eigenvectors and eigenvalues of the overlap matrix $S$, with a matrix element and the $k^{\mathrm{th}}$ eigenvalue and eigenvector denoted by  
\begin{equation}
S_{ij} = \left( \phi^{ps}_i \right)^\dagger \phi^{ps}_j  \; ; \quad S x_k = e_k x_k \,.
\end{equation}
Using normalized partial waves the trace of $S$ is the number of projectors $N \leq 128$. 
The eigenvectors are sorted by eigenvalues, and only a few with the largest eigenvalues are kept such that $\sum_k^n e_k > N(1- \iota)$, 
where $n$ between 3 and 5 is usually sufficient for a small error ($\iota < 10^{-4}$). 
We can now construct the optimal projectors following the prescription,
\begin{equation}
p^{ae}_k = \sum_{i=1}^N x_{ki} \phi^{ae}_k \; ; \quad  p^{ps}_k = \sum_{i=1}^N x_{ki} \phi^{ps}_k  .
\label{eq-final-projectors}
\end{equation}
Here both the pseudo and all-electron projectors are constructed from the same vector $x_k$. 
(In practice, we build the negative of $S$ so that the eigenvalues with the largest absolute magnitude are also the lowest.  
We then only calculate 16 eigenvectors with the smallest eigenvalues using the {\sc syevr} routine provided by the {\sc lapack} library  \cite{DHILLON20041,laug}.) 

\begin{figure}
\begin{centering}
\includegraphics[height=3.1in,trim=0 35 15 70,angle=270]{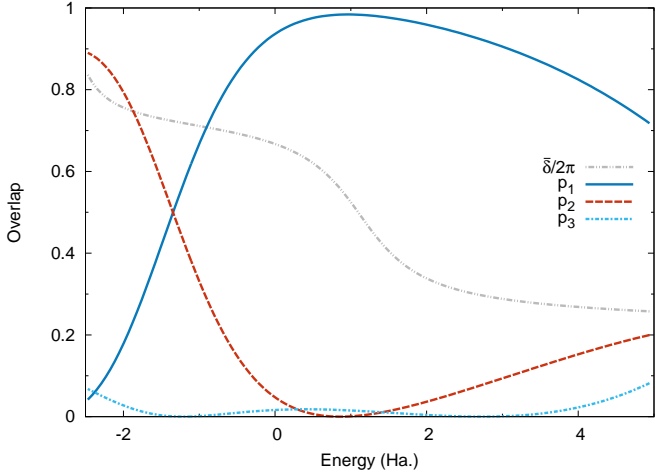}
\caption{ Overlaps between $l=0$ pseudopotential scattering states of titanium and the calculated OPFs as a function of the energy of the scattering state. 
For reference, the rescaled phase shift $\bar{\delta}/2\pi$ is plotted as well. Note the OPFs are not necessarily eigenstates for any energy, {\it i.e.}, the overlap for any single projector may be always less than 1. This is in contrast to pseudopotential or PAW methods that choose projectors that are eigenstates. }
\label{overlaps}
\end{centering}
\end{figure}

We call these projectors optimal because the PCA construction guarantees that the fewest projectors possible are chosen to span the space given by our set of partial waves and target error. 
The strength of this approach is that relatively few projectors per angular momentum are needed to span from the occupied valence bands through 130~eV above the Fermi level. 
The number of projectors generated is typically one more than the number  of scattering resonances of spanned by the OPF energy window.  
In Figure~\ref{overlaps} we show the overlaps between the partial waves and the optimal projectors $\vert \langle \phi^{\textit{ps}}_j \vert p^{\textit{ps}}_i \rangle \vert^2$ as a function of energy for the $l=0$ states of titanium.

The augmentation of the electron wave functions is carried out using the projectors from Eq.~\ref{eq-final-projectors}. 
An all-electron wave function is given as follows (here $\mu$ denotes, say a Bloch state, and the atom's position is taken as the origin): 
\begin{align}
\psi_{\mu}^{ae}(\mathbf{r}) &= \psi_{\mu}^{ps}(\mathbf{r}) \\
&+ \sum_{lm}  \sum_j^{N_l} Y_{lm}(\hat{r}) \left( p^{ae}_{jl}(r) -p^{ps}_{jl}(r) \right) \langle Y_{lm} \, p^{ps}_{jl} \vert \psi_{\mu}^{ps} \rangle_{r_a} \nonumber
\end{align}
where $-l \le m \le +l$, $N_l$ is the number of projectors for a given angular momentum channel, and $Y_{lm}$ are the spherical harmonics. 
The overlap between the wave function and the projectors is taken within the sphere defined by the with radius $r_a$.

\section{Grids, Convergence, and Errors from Approximations}
\label{app-errors}

\subsection{Grids and Integrals}

As outlined previously in Sec.~\ref{sec-implementation} on the implementation within {\sc ocean}, the Green's functions are calculated on a real-space grid determined by separate radial and angular grids, and the internal energy loop integral is calculated for a set of imaginary energies.

\subsubsection{Radial and angular grids}
\label{rad-ang-grids}

The real-space points used for calculating the Green's functions and polarizability are constructed from separate radial and angular grids
\begin{equation}
\mathbf{r}_i = r_j \otimes \hat{\Omega}_k
\end{equation}
The angular grids are taken from the set of extremal points by Womersley and Sloan \cite{Sloan2004}. 
For a given degree $n$ each angular grid has dimension $(n+1)^2$. 
The radial grid has two options, uniform spacing or Gauss-Legendre quadrature. 
Uniform spacing has the advantage that it is directly equivalent to a plane-wave energy cutoff $\vert \mathbf{G}_\textrm{max} \vert =  \pi/ \Delta r$. 
However, testing indicates that the quadrature grids are more efficient, generating converged results with fewer points. 
The radial space can divided into arbitrary parts, each with its own grid spacing or quadrature grid. 

By default, we divide the space in two using $r_a$, the augmentation radius from the OPFs. 
Within this radius we use a 16-point Gauss-Legendre radial grid and the 36-point ($n=5$) angular grid. 
The dense radial grid captures the behavior of the all-electron reconstructed wave functions close to the atomic site. 
For the section outside $r_a$, we use a uniform grid such that $(r_\textrm{max} - r_a)/N < 0.45$~a.u., typically 16 points, and the angular grid is increased to 64 points ($n=7$). 
This gives the Green's functions and polarizability dimensions of $1600\times1600$, independent of the size of the unit cell. 

\subsubsection{Energy integration}
\label{energyIntegration}
By construction, the RPA polarizability requires an integral over the internal loop energy. 
As shown in Eq.~\ref{chi1} this can be transformed from an integral over real energies to one over complex 
energies by closing the contour and realizing that above the Fermi energy all of the poles (single-particle excitation energies) are displaced below the real axis by a small imaginary component. Likewise, below the Fermi level the poles are displaced above the real axis. 
Therefore, the contour is closed by arcs in the upper-right and lower-left quadrants and does not encompass any poles. 
The Green's function is relatively smooth at imaginary values, and we use quadrature to replace the integral with a summation over relatively few energy points.

Following Ref.~\cite{Ultramicroscopy}, we first divide $t \in (-\infty,\infty)$ into four regions, symmetric across $t=0$ by the parameter $\zeta$. 
 such that the number of quadrature points in the region $(0,\zeta)$ will be the same as within $(\zeta,\infty)$. 
This allows Eq.~\ref{chi1} to be rewritten as 
\begin{align}
\chi_0 = & \sum_\sigma \int_{-\infty}^{\infty} \frac{dt}{2\pi}  \, g_\sigma( \mathbf{r}, \mathbf{r}', \mu+it )g_\sigma( \mathbf{r}', \mathbf{r}, \mu+it  ) \nonumber \\
= &\sum_\sigma \frac{\zeta}{\pi} \int_{0}^{1} da \Big[ g_\sigma( \mathbf{r}, \mathbf{r}', \mu+i \zeta a )g_\sigma( \mathbf{r}', \mathbf{r}, \mu+i \zeta a  )  \nonumber \\
&\, \quad +  a^{-2} g_\sigma( \mathbf{r}, \mathbf{r}', \mu+\sfrac{i \zeta }{a^2} )g_\sigma( \mathbf{r}', \mathbf{r}, \mu+\sfrac{i \zeta}{a^2}  ) \Big]
\label{egrid_chi}
\end{align}
The parameter $\zeta$ is chosen to be the geometric mean of the largest and smallest values of $\vert \mu - \varepsilon_{b\mathbf{k}}\vert$, {\it i.e.}, half the band gap and the larger of the distance from $\mu$ to the bottom of the valence bands or the top of the conductions bands. 
To prevent $\zeta$ from going to zero in metallic systems, 0.5~eV is added in quadrature to the minimum (half-gap) value. 

The integral over $a$ in Eq.~\ref{egrid_chi} is replaced by a summation over quadrature points. 
The energy points $a_i$ and weights $w_i$ are taken from Gauss–Legendre quadrature, shifted and scaled by half to match the range. 
Quadrature grids from 4 to 64 points are implemented in the code. 
In Ref.~\cite{Ultramicroscopy}, it was suggested that the two-part integrand be replaced with a single product of Green's functions with energy points at $i \zeta a / (1-a)$ with a prefactor of $(1-a)^{-2}$, giving
\begin{align}
\chi_0 \approx \sum_\sigma \sum_{i}^{N_i} \frac{\zeta w_i}{\pi(1-a_i)^2} & \,g_\sigma( \mathbf{r}, \mathbf{r}', \mu+i \frac{\zeta a_i }{1-a_i} ) \times \nonumber \\
\times & \, g_\sigma( \mathbf{r}', \mathbf{r}, \mu+i \frac{\zeta a_i }{1-a_i} ) \, .
\end{align}
This reproduces the correct large and small $a$ behavior of Eq.~\ref{egrid_chi}, but with only a single set of quadrature points. 
Using this single-grid approximation, a 16-point quadrature grid was found to be sufficient. 
For systems with time-reversal symmetry, the spatial indices on one of the Green's functions can be interchanged. This allows us to calculate only a single Green's function and square it.

\subsection{Bands and {\it k}-points}
\label{sec-kb-conv}

\begin{figure}
\begin{centering}
\includegraphics[height=3.1in,trim=10 22 10 65,angle=270]{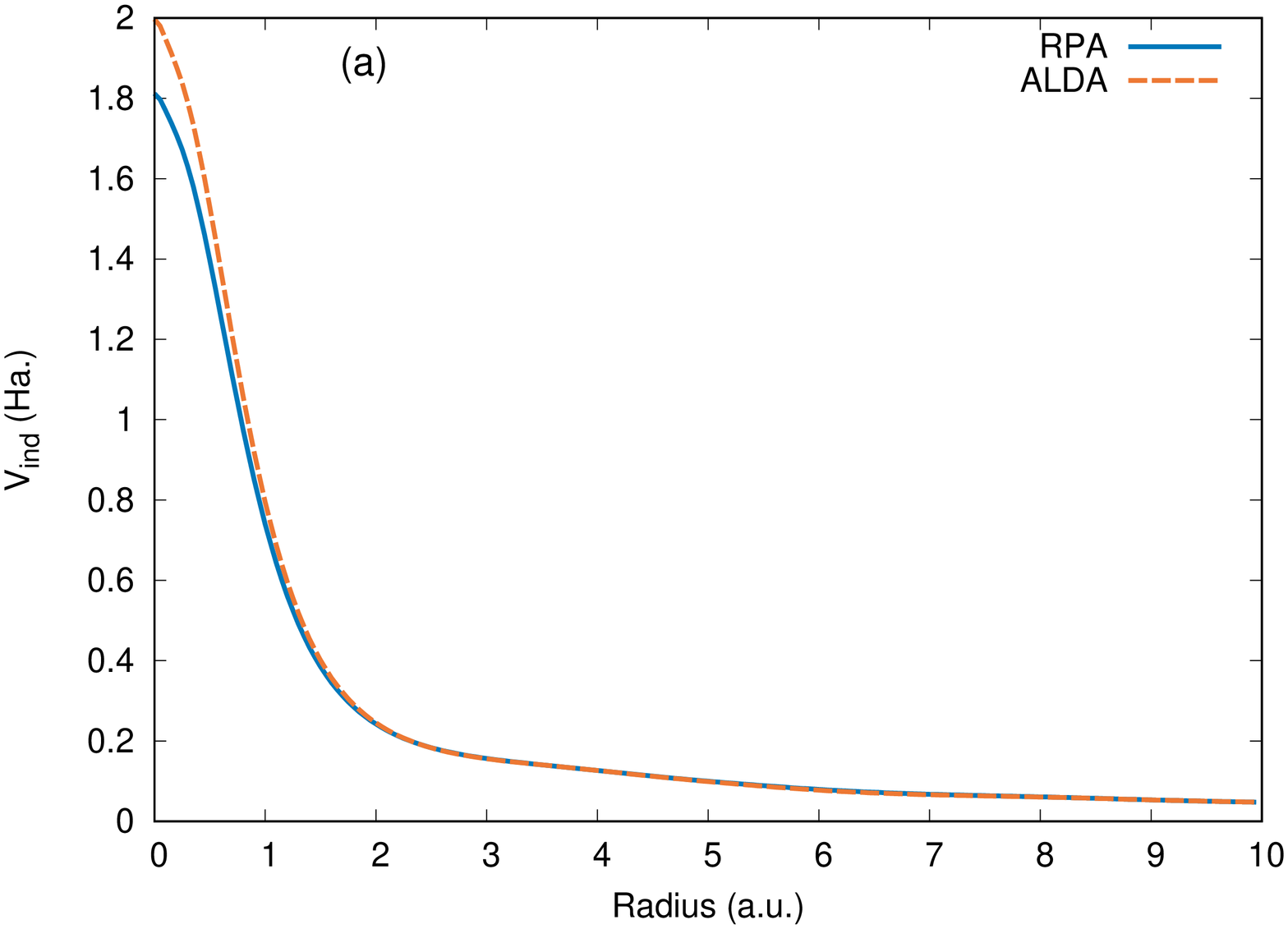}
\includegraphics[height=3.1in,trim=120 24 125 73,angle=270]{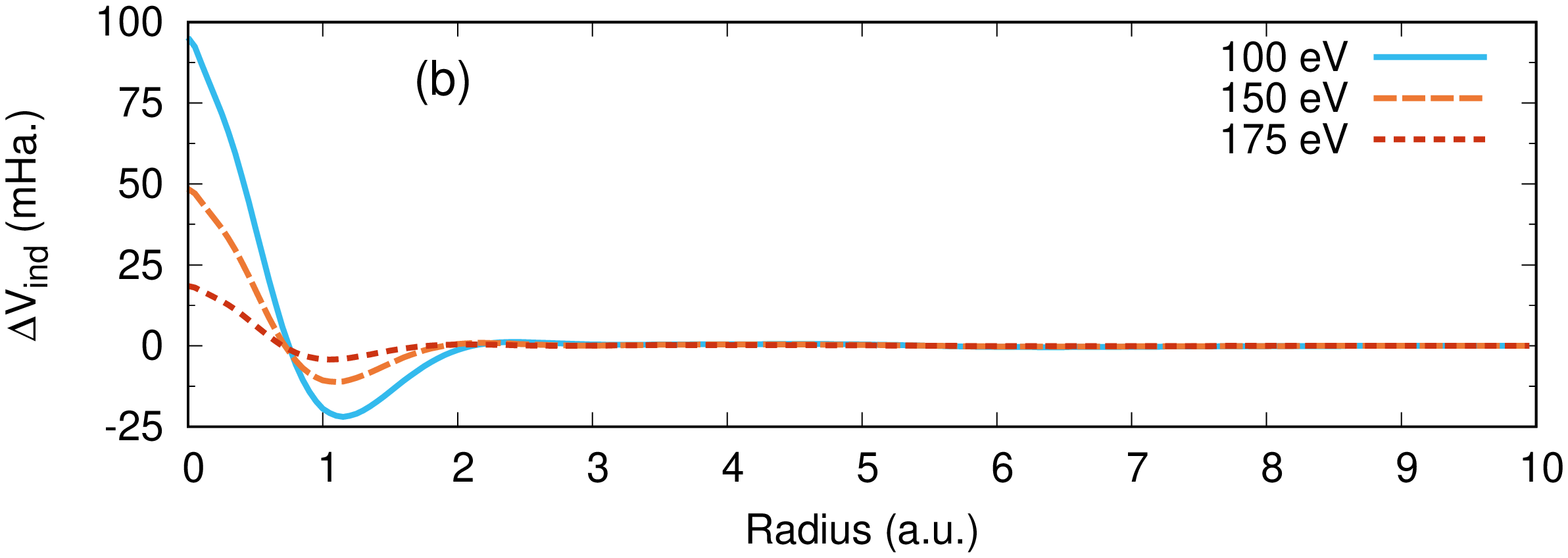}
\includegraphics[height=3.1in,trim=130 24 125 73,angle=270]{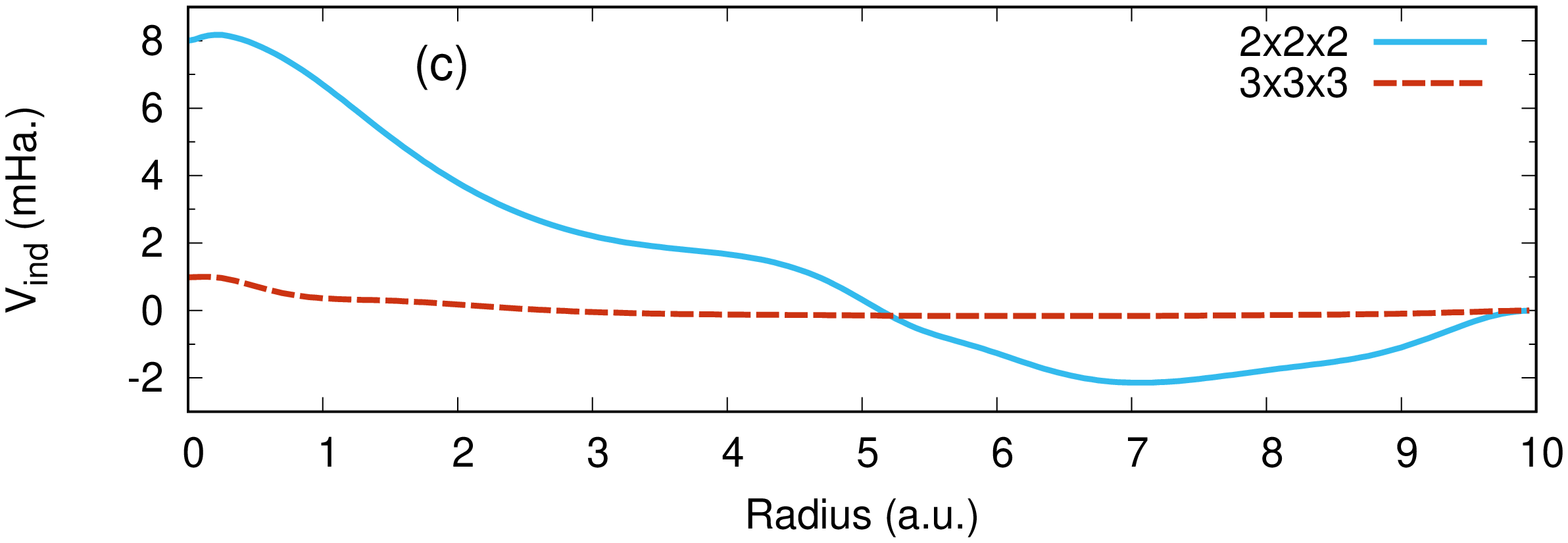}
\includegraphics[height=3.1in,trim=130 24 125 73,angle=270]{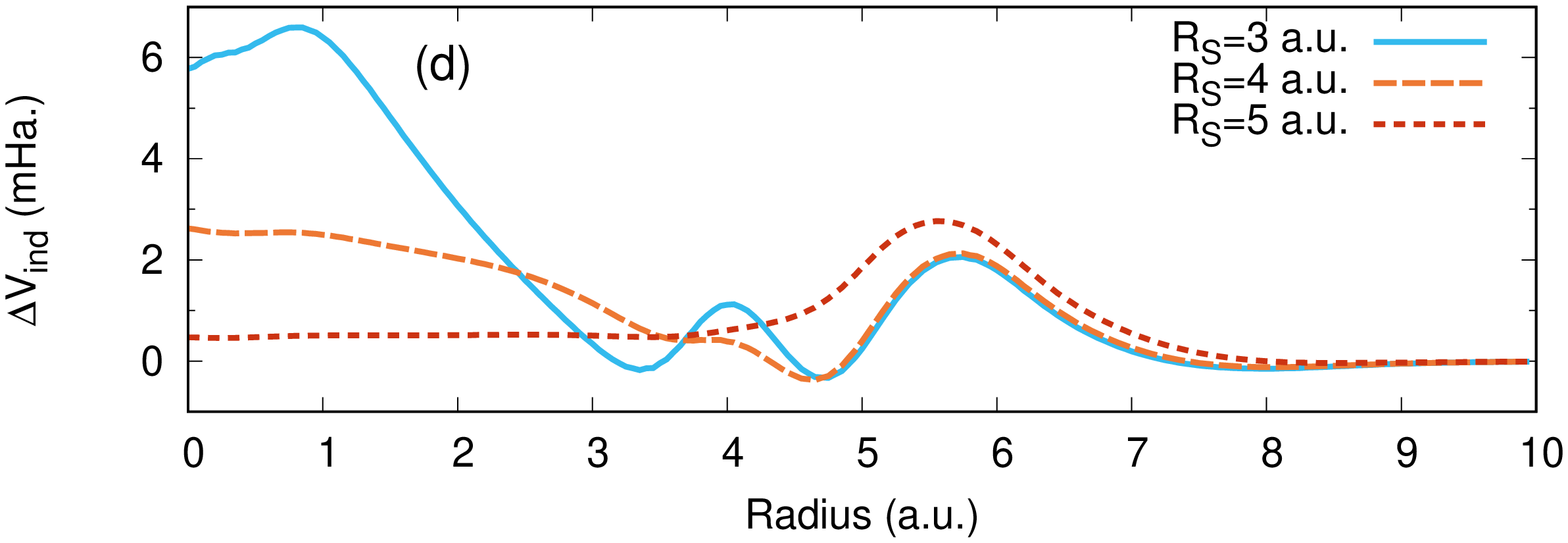}
\caption{ a) The induced potential in response to a fluorine 1{\it s} hole in LiF calculated using a shell radius of 6~a.u.\ and 120 conduction bands, spanning approximately 200~eV, and a $4^3$ {\it k}-point sampling (solid, blue). The orange, dashed line shows the induced potential calculated using $f_\textit{xc}$ within the adiabatic LDA (see text sec.~\ref{alda}). b) The difference plots obtained by subtracting the induced potential calculated with different numbers of bands.
c) The difference plots for changing {\it k}-point grids. 
d) The difference plots for the induced potential changing only the sphere radius $R_S$. Note that the difference plots are in mHa. }
\label{shell-distance}
\end{centering}
\end{figure}

The convergence of the screening calculation also depends on the number of {\it k}-points and bands included in the Green's functions.
The convergence behavior with respect to bands is similar in our approach and other sum over states methods. 
A large number of unoccupied bands may be required, and the error falls as the inverse of the number of bands \cite{PhysRevB.83.081101}. 
To generalize between materials it is preferable to speak of the energy range of the unoccupied bands included in the calculation, {\it e.g.}, the average energy of the highest-band with respect to the conduction band minimum. 
In Fig.~\ref{shell-distance}(b) we show the convergence of the screening potential of the fluorine 1{\it s} hole in LiF with respect to the energy range of unoccupied states. 
This is done by plotting the difference between the calculated induced potential using a conduction band range of 200~eV and that calculated with smaller ranges, {\it e.g.}, $\Delta v_\textit{ind} [100~\textrm{eV}] = v_\textit{ind}[200~\textrm{eV}] - v_\textit{ind}[100~\textrm{eV}]$, etc. 
Typically, the induced potential near the core hole increases with an increase in the number of bands included in the screening calculation.

Like the summation over bands, the {\it k}-point sampling should also be infinite. 
However, while the summation over bands takes the place of an energy integral whose upper bound is positive infinity, the summation over {\it k}-points is, by construction, a properly normalized discretization of the volume integral over the Brillouin zone. 
Errors in finite {\it k}-point sampling arise when the electron wave functions at a given momentum are a poor approximation of other points within the discretization volume. 
In the real-space approach, the convergence with {\it k}-points is rapid. 
Even for systems with small units cells like LiF, only a few {\it k}-points are required. 
In Fig.~\ref{shell-distance}(c) we show the difference plots from reducing the {\it k}-point sampling from $4^3$ down to $3^3$ and $2^3$. 
With only a $2\times2\times2$ {\it k}-point grid, the errors in the induced potential are less than 10~mHa.

\subsection{Real-space Truncation}
\label{sec-truncation}

As introduced in Sec.~\ref{sec-localreview}, the real-space approach relies on partitioning the space around the core hole (or test charge). 
This partitioning is carried out in Eq.~\ref{eq-partition}, where a spherical charge of radius $R_S$ neutralizes the long-range Coulomb tail, allowing the RPA screening to be carried out only locally. 
Our approximation doesn't change the total external potential that is screened. 
However, by using a model dielectric to calculate $W^{(2)}$ we introduce differences with respect to a calculation using the RPA polarizability everywhere. 
Having previously outlined the effect of neglecting the core-hole potential and the effects of various approximations to the augmenting the pseudopotential wave functions in the previous section, we now look to the influence of $R_S$ on the calculated screened core-hole potential and subsequently the absorption spectrum.

To assess the effects of finite $R_S$ on the calculated screened potential we compare the induced potential for the fluorine K edge of LiF. 
In Fig.~\ref{shell-distance}(d) we show the induced potential calculated with a shell radius of 6~a.u.\ and the difference in the potentials between those calculated with shell radii of 5~a.u., 4~a.u., and 3~a.u: $\Delta v_{\textit{ind}}[R_S\!\!=\!\!5] = v_{\textit{ind}}[R_S\!\!=\!\!6]-v_{\textit{ind}}[R_S\!\!=\!\!5]$.  
Near the fluorine site, the difference between the induced potential at $R_S=6$~a.u.\ and  $R_S=5$~a.u.\ is less than 0.013~eV, while the maximum difference between the two, located at 5.55~a.u.\ (approximately the length of a lattice vector), is less than 0.076~eV.

\subsection{Long-range Dielectric Constant}
\label{sec-approx-eps}

\begin{figure}
\begin{centering}
\includegraphics[height=3.1in,trim=0 35 15 70,angle=270]{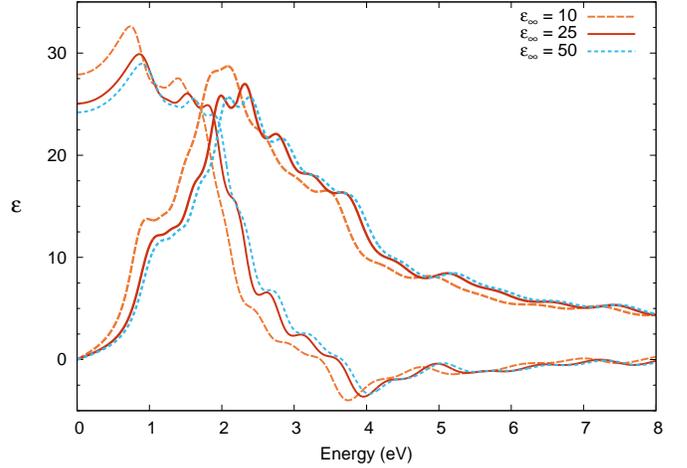}
\caption{ The real (thin) and imaginary (thick) components of the complex dielectric constant of FeS$_2$ plotted for 3 different inputs of the static dielectric constant $\epsilon_\infty$. If the value of $\epsilon_\infty$ is not known from prior calculations or experiment, valence BSE calculations can be carried out to determine it self-consistently: $\epsilon_\infty=25$. }
\label{FeS2_eps_conv}
\end{centering}
\end{figure}

\begin{figure}
\begin{centering}
\includegraphics[height=3.1in,trim=0 35 15 70,angle=270]{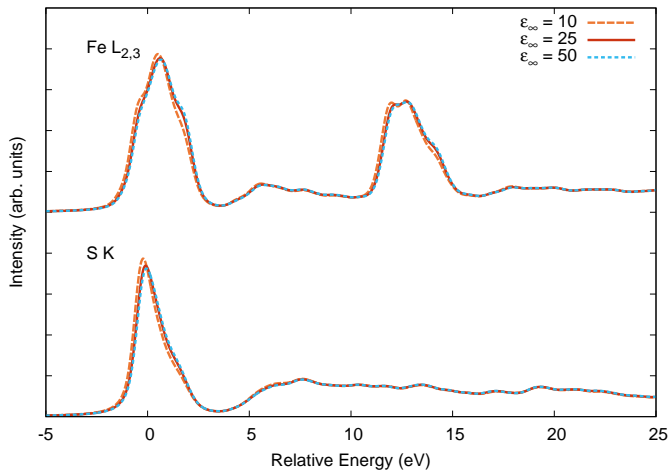}
\caption{ For FeS$_2$ marcasite, the Fe L$_{2,3}$ edge and S K edges are plotted  for 3 different inputs of the static dielectric constant $\epsilon_\infty$. Despite large differences in the input $\epsilon_\infty$, the spectra are largely unchanged. 
The relative positioning takes into account the changes in the effect of the changing screened core-hole potential on the chemical shifts (see text). The error in the position of either edge is less than 0.13~eV for $\epsilon_\infty=10$.}
\label{FeS2_xray}
\end{centering}
\end{figure}

As mentioned previously, the static long-range dielectric constant $\epsilon_\infty$ is a required input for our local, real-space approach. 
The error near the core hole due to an incorrect dielectric constant is approximately
\begin{equation}
\Delta W \approx \left( 1 - \frac{1}{\epsilon_\infty} \right)  R_S^{-1} - \left( 1 - \frac{1}{\widetilde{\epsilon_\infty} } \right) R_S^{-1}
\end{equation}
where $\widetilde{\epsilon_\infty}$ is the input dielectric constant. 
This can be expressed in terms of percentage error in the input dielectric constant 
\begin{equation}
\Delta W \approx  \frac{ 1}{\widetilde{\epsilon_\infty} R_S } \left( \frac{ \epsilon_\infty - \widetilde{\epsilon_\infty} }{\epsilon_\infty} \right) \, .
\end{equation}
As an example, for $\epsilon_\infty=5$, $R_S=5$~a.u., a 10~\% underestimation ($\widetilde{\epsilon_\infty}=4.5$) would lead to an error of 0.12~eV. 
This absolute error directly affects calculations of chemical shifts using the {\sc ocean} code \cite{PhysRevB.90.205207}.

To showcase the errors from an incorrect input value of $\epsilon_\infty$ we look at FeS$_2$ in the cubic {\it Pnnm} phase marcasite. 
As before the cell parameters are taken from experiment \cite{RIEDER2007115},  the pseudopotentials were taken from PseduoDojo, specifically the ``high-accuracy'' iron potential. 
Marcasite crystalizes in the cubic \textit{Pnnm} phase. 
The lattice parameters were set to 0.44446~nm by 0.54246~nm by 0.33864~nm to match experiment \cite{RIEDER2007115}. 
A planewave cut-off of 100~Ry was used and the density was converged on a $4^3$ {\it k}-point mesh. 
The ``high-accuracy" iron and standard sulfur pseudopotentials were taken from PseudoDojo. 
The BSE final states were solved on a $12\times10\times16$ {\it k}-point mesh, including 72 unoccupied bands, and downsampled onto an $8\times10\times6$ real-space mesh. 
For the screening calculations a $2^3$ {\it k}-point mesh and 200 bands were used. 
Absent a previously calculated or experimentally measured value for the static dielectric constant, the input $\epsilon_\infty$ can be determined self-consistently as shown in Fig.~\ref{FeS2_eps_conv}. 
We find that an input value of $\epsilon_\infty=25$ results in a BSE calculation of approximately the same value (25.06) with the photon momentum vector aligned along (111). 
Unsurprisingly, the strength of the static dielectric constant effects the calculated dielectric response, but even comparing $\epsilon_\infty=10$ to $\epsilon_\infty=50$ the spectra are in qualitative agreement.

Next we can compare this effect on the x-ray edges of FeS$_2$ in Fig.~\ref{FeS2_xray}. 
Neither the sulfur K edge nor the iron L edge are strongly dependent on the input value of the dielectric constant. 
The energy scale of both is relative to the conduction band minimum of the $\epsilon_\infty=10$ calculation (of the L$_3$ edge of iron). 
The slight shifts in the onset of the $\epsilon_\infty=25$ and $\epsilon_\infty=50$ spectra are due to changes in the excitonic binding and core-level shift due to differences in the input dielectric. 
As can be seen in Fig.~\ref{FeS2_xray}, for x-ray absorption calculations it may be sufficient to have only a rough estimate of the dielectric constant.

\bibliography{Screen}

\begin{thebibliography}{93}%
\makeatletter
\providecommand \@ifxundefined [1]{%
 \@ifx{#1\undefined}
}%
\providecommand \@ifnum [1]{%
 \ifnum #1\expandafter \@firstoftwo
 \else \expandafter \@secondoftwo
 \fi
}%
\providecommand \@ifx [1]{%
 \ifx #1\expandafter \@firstoftwo
 \else \expandafter \@secondoftwo
 \fi
}%
\providecommand \natexlab [1]{#1}%
\providecommand \enquote  [1]{``#1''}%
\providecommand \bibnamefont  [1]{#1}%
\providecommand \bibfnamefont [1]{#1}%
\providecommand \citenamefont [1]{#1}%
\providecommand \href@noop [0]{\@secondoftwo}%
\providecommand \href [0]{\begingroup \@sanitize@url \@href}%
\providecommand \@href[1]{\@@startlink{#1}\@@href}%
\providecommand \@@href[1]{\endgroup#1\@@endlink}%
\providecommand \@sanitize@url [0]{\catcode `\\12\catcode `\$12\catcode
  `\&12\catcode `\#12\catcode `\^12\catcode `\_12\catcode `\%12\relax}%
\providecommand \@@startlink[1]{}%
\providecommand \@@endlink[0]{}%
\providecommand \url  [0]{\begingroup\@sanitize@url \@url }%
\providecommand \@url [1]{\endgroup\@href {#1}{\urlprefix }}%
\providecommand \urlprefix  [0]{URL }%
\providecommand \Eprint [0]{\href }%
\providecommand \doibase [0]{https://doi.org/}%
\providecommand \selectlanguage [0]{\@gobble}%
\providecommand \bibinfo  [0]{\@secondoftwo}%
\providecommand \bibfield  [0]{\@secondoftwo}%
\providecommand \translation [1]{[#1]}%
\providecommand \BibitemOpen [0]{}%
\providecommand \bibitemStop [0]{}%
\providecommand \bibitemNoStop [0]{.\EOS\space}%
\providecommand \EOS [0]{\spacefactor3000\relax}%
\providecommand \BibitemShut  [1]{\csname bibitem#1\endcsname}%
\let\auto@bib@innerbib\@empty
\bibitem [{\citenamefont {Shirley}(2006)}]{Ultramicroscopy}%
  \BibitemOpen
  \bibfield  {author} {\bibinfo {author} {\bibfnamefont {E.~L.}\ \bibnamefont
  {Shirley}},\ }\bibfield  {title} {\bibinfo {title} {Local screening of a core
  hole: A real-space approach applied to hafnium oxide},\ }\href
  {https://doi.org/https://doi.org/10.1016/j.ultramic.2006.05.008} {\bibfield
  {journal} {\bibinfo  {journal} {Ultramicroscopy}\ }\textbf {\bibinfo {volume}
  {106}},\ \bibinfo {pages} {986 } (\bibinfo {year} {2006})}\BibitemShut
  {NoStop}%
\bibitem [{\citenamefont {Wiser}(1963)}]{PhysRev.129.62}%
  \BibitemOpen
  \bibfield  {author} {\bibinfo {author} {\bibfnamefont {N.}~\bibnamefont
  {Wiser}},\ }\bibfield  {title} {\bibinfo {title} {Dielectric constant with
  local field effects included},\ }\href
  {https://doi.org/10.1103/PhysRev.129.62} {\bibfield  {journal} {\bibinfo
  {journal} {Phys. Rev.}\ }\textbf {\bibinfo {volume} {129}},\ \bibinfo {pages}
  {62} (\bibinfo {year} {1963})}\BibitemShut {NoStop}%
\bibitem [{\citenamefont {Rojas}\ \emph {et~al.}(1995)\citenamefont {Rojas},
  \citenamefont {Godby},\ and\ \citenamefont {Needs}}]{PhysRevLett.74.1827}%
  \BibitemOpen
  \bibfield  {author} {\bibinfo {author} {\bibfnamefont {H.~N.}\ \bibnamefont
  {Rojas}}, \bibinfo {author} {\bibfnamefont {R.~W.}\ \bibnamefont {Godby}},\
  and\ \bibinfo {author} {\bibfnamefont {R.~J.}\ \bibnamefont {Needs}},\
  }\bibfield  {title} {\bibinfo {title} {Space-time method for ab initio
  calculations of self-energies and dielectric response functions of solids},\
  }\href {https://doi.org/10.1103/PhysRevLett.74.1827} {\bibfield  {journal}
  {\bibinfo  {journal} {Phys. Rev. Lett.}\ }\textbf {\bibinfo {volume} {74}},\
  \bibinfo {pages} {1827} (\bibinfo {year} {1995})}\BibitemShut {NoStop}%
\bibitem [{\citenamefont {Blase}\ \emph {et~al.}(1995)\citenamefont {Blase},
  \citenamefont {Rubio}, \citenamefont {Louie},\ and\ \citenamefont
  {Cohen}}]{PhysRevB.52.R2225}%
  \BibitemOpen
  \bibfield  {author} {\bibinfo {author} {\bibfnamefont {X.}~\bibnamefont
  {Blase}}, \bibinfo {author} {\bibfnamefont {A.}~\bibnamefont {Rubio}},
  \bibinfo {author} {\bibfnamefont {S.~G.}\ \bibnamefont {Louie}},\ and\
  \bibinfo {author} {\bibfnamefont {M.~L.}\ \bibnamefont {Cohen}},\ }\bibfield
  {title} {\bibinfo {title} {Mixed-space formalism for the dielectric response
  in periodic systems},\ }\href {https://doi.org/10.1103/PhysRevB.52.R2225}
  {\bibfield  {journal} {\bibinfo  {journal} {Phys. Rev. B}\ }\textbf {\bibinfo
  {volume} {52}},\ \bibinfo {pages} {R2225} (\bibinfo {year}
  {1995})}\BibitemShut {NoStop}%
\bibitem [{\citenamefont {Hung}\ \emph {et~al.}(2016)\citenamefont {Hung},
  \citenamefont {da~Jornada}, \citenamefont {Souto-Casares}, \citenamefont
  {Chelikowsky}, \citenamefont {Louie},\ and\ \citenamefont
  {\"O\u{g}\"ut}}]{PhysRevB.94.085125}%
  \BibitemOpen
  \bibfield  {author} {\bibinfo {author} {\bibfnamefont {L.}~\bibnamefont
  {Hung}}, \bibinfo {author} {\bibfnamefont {F.~H.}\ \bibnamefont
  {da~Jornada}}, \bibinfo {author} {\bibfnamefont {J.}~\bibnamefont
  {Souto-Casares}}, \bibinfo {author} {\bibfnamefont {J.~R.}\ \bibnamefont
  {Chelikowsky}}, \bibinfo {author} {\bibfnamefont {S.~G.}\ \bibnamefont
  {Louie}},\ and\ \bibinfo {author} {\bibfnamefont {S.}~\bibnamefont
  {\"O\u{g}\"ut}},\ }\bibfield  {title} {\bibinfo {title} {Excitation spectra
  of aromatic molecules within a real-space $gw$-bse formalism: Role of
  self-consistency and vertex corrections},\ }\href
  {https://doi.org/10.1103/PhysRevB.94.085125} {\bibfield  {journal} {\bibinfo
  {journal} {Phys. Rev. B}\ }\textbf {\bibinfo {volume} {94}},\ \bibinfo
  {pages} {085125} (\bibinfo {year} {2016})}\BibitemShut {NoStop}%
\bibitem [{\citenamefont {Giustino}\ \emph {et~al.}(2010)\citenamefont
  {Giustino}, \citenamefont {Cohen},\ and\ \citenamefont
  {Louie}}]{PhysRevB.81.115105}%
  \BibitemOpen
  \bibfield  {author} {\bibinfo {author} {\bibfnamefont {F.}~\bibnamefont
  {Giustino}}, \bibinfo {author} {\bibfnamefont {M.~L.}\ \bibnamefont
  {Cohen}},\ and\ \bibinfo {author} {\bibfnamefont {S.~G.}\ \bibnamefont
  {Louie}},\ }\bibfield  {title} {\bibinfo {title} {$gw$ method with the
  self-consistent sternheimer equation},\ }\href
  {https://doi.org/10.1103/PhysRevB.81.115105} {\bibfield  {journal} {\bibinfo
  {journal} {Phys. Rev. B}\ }\textbf {\bibinfo {volume} {81}},\ \bibinfo
  {pages} {115105} (\bibinfo {year} {2010})}\BibitemShut {NoStop}%
\bibitem [{\citenamefont {Nguyen}\ \emph {et~al.}(2012)\citenamefont {Nguyen},
  \citenamefont {Pham}, \citenamefont {Rocca},\ and\ \citenamefont
  {Galli}}]{PhysRevB.85.081101}%
  \BibitemOpen
  \bibfield  {author} {\bibinfo {author} {\bibfnamefont {H.-V.}\ \bibnamefont
  {Nguyen}}, \bibinfo {author} {\bibfnamefont {T.~A.}\ \bibnamefont {Pham}},
  \bibinfo {author} {\bibfnamefont {D.}~\bibnamefont {Rocca}},\ and\ \bibinfo
  {author} {\bibfnamefont {G.}~\bibnamefont {Galli}},\ }\bibfield  {title}
  {\bibinfo {title} {Improving accuracy and efficiency of calculations of
  photoemission spectra within the many-body perturbation theory},\ }\href
  {https://doi.org/10.1103/PhysRevB.85.081101} {\bibfield  {journal} {\bibinfo
  {journal} {Phys. Rev. B}\ }\textbf {\bibinfo {volume} {85}},\ \bibinfo
  {pages} {081101(R)} (\bibinfo {year} {2012})}\BibitemShut {NoStop}%
\bibitem [{\citenamefont {Ben}\ \emph {et~al.}(2019)\citenamefont {Ben},
  \citenamefont {da~Jornada}, \citenamefont {Canning}, \citenamefont
  {Wichmann}, \citenamefont {Raman}, \citenamefont {Sasanka}, \citenamefont
  {Yang}, \citenamefont {Louie},\ and\ \citenamefont
  {Deslippe}}]{DELBEN2019187}%
  \BibitemOpen
  \bibfield  {author} {\bibinfo {author} {\bibfnamefont {M.~D.}\ \bibnamefont
  {Ben}}, \bibinfo {author} {\bibfnamefont {F.~H.}\ \bibnamefont {da~Jornada}},
  \bibinfo {author} {\bibfnamefont {A.}~\bibnamefont {Canning}}, \bibinfo
  {author} {\bibfnamefont {N.}~\bibnamefont {Wichmann}}, \bibinfo {author}
  {\bibfnamefont {K.}~\bibnamefont {Raman}}, \bibinfo {author} {\bibfnamefont
  {R.}~\bibnamefont {Sasanka}}, \bibinfo {author} {\bibfnamefont
  {C.}~\bibnamefont {Yang}}, \bibinfo {author} {\bibfnamefont {S.~G.}\
  \bibnamefont {Louie}},\ and\ \bibinfo {author} {\bibfnamefont
  {J.}~\bibnamefont {Deslippe}},\ }\bibfield  {title} {\bibinfo {title}
  {Large-scale $gw$ calculations on pre-exascale hpc systems},\ }\href
  {https://doi.org/https://doi.org/10.1016/j.cpc.2018.09.003} {\bibfield
  {journal} {\bibinfo  {journal} {Computer Physics Communications}\ }\textbf
  {\bibinfo {volume} {235}},\ \bibinfo {pages} {187 } (\bibinfo {year}
  {2019})}\BibitemShut {NoStop}%
\bibitem [{\citenamefont {Rieger}\ \emph {et~al.}(1999)\citenamefont {Rieger},
  \citenamefont {Steinbeck}, \citenamefont {White}, \citenamefont {Rojas},\
  and\ \citenamefont {Godby}}]{RIEGER1999211}%
  \BibitemOpen
  \bibfield  {author} {\bibinfo {author} {\bibfnamefont {M.~M.}\ \bibnamefont
  {Rieger}}, \bibinfo {author} {\bibfnamefont {L.}~\bibnamefont {Steinbeck}},
  \bibinfo {author} {\bibfnamefont {I.}~\bibnamefont {White}}, \bibinfo
  {author} {\bibfnamefont {H.}~\bibnamefont {Rojas}},\ and\ \bibinfo {author}
  {\bibfnamefont {R.}~\bibnamefont {Godby}},\ }\bibfield  {title} {\bibinfo
  {title} {The gw space-time method for the self-energy of large systems},\
  }\href {https://doi.org/https://doi.org/10.1016/S0010-4655(98)00174-X}
  {\bibfield  {journal} {\bibinfo  {journal} {Computer Physics Communications}\
  }\textbf {\bibinfo {volume} {117}},\ \bibinfo {pages} {211 } (\bibinfo {year}
  {1999})}\BibitemShut {NoStop}%
\bibitem [{\citenamefont {Kaltak}\ \emph {et~al.}(2014)\citenamefont {Kaltak},
  \citenamefont {Klime\v{s}{}},\ and\ \citenamefont
  {Kresse}}]{PhysRevB.90.054115}%
  \BibitemOpen
  \bibfield  {author} {\bibinfo {author} {\bibfnamefont {M.}~\bibnamefont
  {Kaltak}}, \bibinfo {author} {\bibfnamefont {J.}~\bibnamefont
  {Klime\v{s}{}}},\ and\ \bibinfo {author} {\bibfnamefont {G.}~\bibnamefont
  {Kresse}},\ }\bibfield  {title} {\bibinfo {title} {Cubic scaling algorithm
  for the random phase approximation: Self-interstitials and vacancies in si},\
  }\href {https://doi.org/10.1103/PhysRevB.90.054115} {\bibfield  {journal}
  {\bibinfo  {journal} {Phys. Rev. B}\ }\textbf {\bibinfo {volume} {90}},\
  \bibinfo {pages} {054115} (\bibinfo {year} {2014})}\BibitemShut {NoStop}%
\bibitem [{\citenamefont {Kim}\ \emph {et~al.}(2020)\citenamefont {Kim},
  \citenamefont {Martyna},\ and\ \citenamefont
  {Ismail-Beigi}}]{PhysRevB.101.035139}%
  \BibitemOpen
  \bibfield  {author} {\bibinfo {author} {\bibfnamefont {M.}~\bibnamefont
  {Kim}}, \bibinfo {author} {\bibfnamefont {G.~J.}\ \bibnamefont {Martyna}},\
  and\ \bibinfo {author} {\bibfnamefont {S.}~\bibnamefont {Ismail-Beigi}},\
  }\bibfield  {title} {\bibinfo {title} {Complex-time shredded propagator
  method for large-scale $gw$ calculations},\ }\href
  {https://doi.org/10.1103/PhysRevB.101.035139} {\bibfield  {journal} {\bibinfo
   {journal} {Phys. Rev. B}\ }\textbf {\bibinfo {volume} {101}},\ \bibinfo
  {pages} {035139} (\bibinfo {year} {2020})}\BibitemShut {NoStop}%
\bibitem [{\citenamefont {Onida}\ \emph {et~al.}(2002)\citenamefont {Onida},
  \citenamefont {Reining},\ and\ \citenamefont {Rubio}}]{RevModPhys.74.601}%
  \BibitemOpen
  \bibfield  {author} {\bibinfo {author} {\bibfnamefont {G.}~\bibnamefont
  {Onida}}, \bibinfo {author} {\bibfnamefont {L.}~\bibnamefont {Reining}},\
  and\ \bibinfo {author} {\bibfnamefont {A.}~\bibnamefont {Rubio}},\ }\bibfield
   {title} {\bibinfo {title} {Electronic excitations: density-functional versus
  many-body green's-function approaches},\ }\href
  {https://doi.org/10.1103/RevModPhys.74.601} {\bibfield  {journal} {\bibinfo
  {journal} {Rev. Mod. Phys.}\ }\textbf {\bibinfo {volume} {74}},\ \bibinfo
  {pages} {601} (\bibinfo {year} {2002})}\BibitemShut {NoStop}%
\bibitem [{\citenamefont {Levine}\ and\ \citenamefont
  {Louie}(1982)}]{PhysRevB.25.6310}%
  \BibitemOpen
  \bibfield  {author} {\bibinfo {author} {\bibfnamefont {Z.~H.}\ \bibnamefont
  {Levine}}\ and\ \bibinfo {author} {\bibfnamefont {S.~G.}\ \bibnamefont
  {Louie}},\ }\bibfield  {title} {\bibinfo {title} {New model dielectric
  function and exchange-correlation potential for semiconductors and
  insulators},\ }\href {https://doi.org/10.1103/PhysRevB.25.6310} {\bibfield
  {journal} {\bibinfo  {journal} {Phys. Rev. B}\ }\textbf {\bibinfo {volume}
  {25}},\ \bibinfo {pages} {6310} (\bibinfo {year} {1982})}\BibitemShut
  {NoStop}%
\bibitem [{\citenamefont {Shirley}\ \emph
  {et~al.}(2005{\natexlab{a}})\citenamefont {Shirley}, \citenamefont
  {Soininen},\ and\ \citenamefont {Rehr}}]{Shirley2005}%
  \BibitemOpen
  \bibfield  {author} {\bibinfo {author} {\bibfnamefont {E.~L.}\ \bibnamefont
  {Shirley}}, \bibinfo {author} {\bibfnamefont {J.~A.}\ \bibnamefont
  {Soininen}},\ and\ \bibinfo {author} {\bibfnamefont {J.~J.}\ \bibnamefont
  {Rehr}},\ }\bibfield  {title} {\bibinfo {title} {Modeling core-hole screening
  in core-excitation spectroscopies},\ }\href
  {https://doi.org/10.1238/physica.topical.115a00031} {\bibfield  {journal}
  {\bibinfo  {journal} {Physica Scripta}\ }\textbf {\bibinfo {volume} {T115}},\
  \bibinfo {pages} {31} (\bibinfo {year} {2005}{\natexlab{a}})}\BibitemShut
  {NoStop}%
\bibitem [{\citenamefont {Vinson}\ \emph {et~al.}(2011)\citenamefont {Vinson},
  \citenamefont {Rehr}, \citenamefont {Kas},\ and\ \citenamefont
  {Shirley}}]{ocean0}%
  \BibitemOpen
  \bibfield  {author} {\bibinfo {author} {\bibfnamefont {J.}~\bibnamefont
  {Vinson}}, \bibinfo {author} {\bibfnamefont {J.~J.}\ \bibnamefont {Rehr}},
  \bibinfo {author} {\bibfnamefont {J.~J.}\ \bibnamefont {Kas}},\ and\ \bibinfo
  {author} {\bibfnamefont {E.~L.}\ \bibnamefont {Shirley}},\ }\bibfield
  {title} {\bibinfo {title} {Bethe-salpeter equation calculations of core
  excitation spectra},\ }\href {https://doi.org/10.1103/PhysRevB.83.115106}
  {\bibfield  {journal} {\bibinfo  {journal} {Phys. Rev. B}\ }\textbf {\bibinfo
  {volume} {83}},\ \bibinfo {pages} {115106} (\bibinfo {year}
  {2011})}\BibitemShut {NoStop}%
\bibitem [{\citenamefont {Gilmore}\ \emph {et~al.}(2015)\citenamefont
  {Gilmore}, \citenamefont {Vinson}, \citenamefont {Shirley}, \citenamefont
  {Prendergast}, \citenamefont {Pemmaraju}, \citenamefont {Kas}, \citenamefont
  {Vila},\ and\ \citenamefont {Rehr}}]{ocean1}%
  \BibitemOpen
  \bibfield  {author} {\bibinfo {author} {\bibfnamefont {K.}~\bibnamefont
  {Gilmore}}, \bibinfo {author} {\bibfnamefont {J.}~\bibnamefont {Vinson}},
  \bibinfo {author} {\bibfnamefont {E.}~\bibnamefont {Shirley}}, \bibinfo
  {author} {\bibfnamefont {D.}~\bibnamefont {Prendergast}}, \bibinfo {author}
  {\bibfnamefont {C.}~\bibnamefont {Pemmaraju}}, \bibinfo {author}
  {\bibfnamefont {J.}~\bibnamefont {Kas}}, \bibinfo {author} {\bibfnamefont
  {F.}~\bibnamefont {Vila}},\ and\ \bibinfo {author} {\bibfnamefont
  {J.}~\bibnamefont {Rehr}},\ }\bibfield  {title} {\bibinfo {title} {Efficient
  implementation of core-excitation bethe-salpeter equation calculations},\
  }\href {https://doi.org/http://dx.doi.org/10.1016/j.cpc.2015.08.014}
  {\bibfield  {journal} {\bibinfo  {journal} {Comput. Phys. Comm.}\ }\textbf
  {\bibinfo {volume} {197}},\ \bibinfo {pages} {109 } (\bibinfo {year}
  {2015})}\BibitemShut {NoStop}%
\bibitem [{\citenamefont {Martin}(2004)}]{RMartin}%
  \BibitemOpen
  \bibfield  {author} {\bibinfo {author} {\bibfnamefont {R.~M.}\ \bibnamefont
  {Martin}},\ }\href@noop {} {\emph {\bibinfo {title} {Electronic Structure:
  Basic Theory and Practical Methods}}}\ (\bibinfo  {publisher} {Cambridge
  University Press},\ \bibinfo {address} {Cambridge, United Kingdom},\ \bibinfo
  {year} {2004})\BibitemShut {NoStop}%
\bibitem [{\citenamefont {Kleinman}\ and\ \citenamefont
  {Bylander}(1982)}]{PhysRevLett.48.1425}%
  \BibitemOpen
  \bibfield  {author} {\bibinfo {author} {\bibfnamefont {L.}~\bibnamefont
  {Kleinman}}\ and\ \bibinfo {author} {\bibfnamefont {D.~M.}\ \bibnamefont
  {Bylander}},\ }\bibfield  {title} {\bibinfo {title} {Efficacious form for
  model pseudopotentials},\ }\href
  {https://doi.org/10.1103/PhysRevLett.48.1425} {\bibfield  {journal} {\bibinfo
   {journal} {Phys. Rev. Lett.}\ }\textbf {\bibinfo {volume} {48}},\ \bibinfo
  {pages} {1425} (\bibinfo {year} {1982})}\BibitemShut {NoStop}%
\bibitem [{\citenamefont {Shirley}\ and\ \citenamefont
  {Martin}(1993)}]{PhysRevB.47.15404}%
  \BibitemOpen
  \bibfield  {author} {\bibinfo {author} {\bibfnamefont {E.~L.}\ \bibnamefont
  {Shirley}}\ and\ \bibinfo {author} {\bibfnamefont {R.~M.}\ \bibnamefont
  {Martin}},\ }\bibfield  {title} {\bibinfo {title} {Gw quasiparticle
  calculations in atoms},\ }\href {https://doi.org/10.1103/PhysRevB.47.15404}
  {\bibfield  {journal} {\bibinfo  {journal} {Phys. Rev. B}\ }\textbf {\bibinfo
  {volume} {47}},\ \bibinfo {pages} {15404} (\bibinfo {year}
  {1993})}\BibitemShut {NoStop}%
\bibitem [{\citenamefont {Shirley}(2005)}]{SHIRLEY20051187}%
  \BibitemOpen
  \bibfield  {author} {\bibinfo {author} {\bibfnamefont {E.~L.}\ \bibnamefont
  {Shirley}},\ }\bibfield  {title} {\bibinfo {title} {Bethe–salpeter
  treatment of x-ray absorption including core-hole multiplet effects},\ }\href
  {https://doi.org/https://doi.org/10.1016/j.elspec.2005.01.191} {\bibfield
  {journal} {\bibinfo  {journal} {Journal of Electron Spectroscopy and Related
  Phenomena}\ }\textbf {\bibinfo {volume} {144-147}},\ \bibinfo {pages} {1187 }
  (\bibinfo {year} {2005})},\ \bibinfo {note} {proceeding of the Fourteenth
  International Conference on Vacuum Ultraviolet Radiation Physics}\BibitemShut
  {NoStop}%
\bibitem [{\citenamefont {Bl\"ochl}(1994)}]{PhysRevB.50.17953}%
  \BibitemOpen
  \bibfield  {author} {\bibinfo {author} {\bibfnamefont {P.~E.}\ \bibnamefont
  {Bl\"ochl}},\ }\bibfield  {title} {\bibinfo {title} {Projector augmented-wave
  method},\ }\href {https://doi.org/10.1103/PhysRevB.50.17953} {\bibfield
  {journal} {\bibinfo  {journal} {Phys. Rev. B}\ }\textbf {\bibinfo {volume}
  {50}},\ \bibinfo {pages} {17953} (\bibinfo {year} {1994})}\BibitemShut
  {NoStop}%
\bibitem [{\citenamefont {Giannozzi}\ \emph {et~al.}(2017)\citenamefont
  {Giannozzi}, \citenamefont {Andreussi}, \citenamefont {Brumme}, \citenamefont
  {Bunau}, \citenamefont {Nardelli}, \citenamefont {Calandra}, \citenamefont
  {Car}, \citenamefont {Cavazzoni}, \citenamefont {Ceresoli}, \citenamefont
  {Cococcioni}, \citenamefont {Colonna}, \citenamefont {Carnimeo},
  \citenamefont {Corso}, \citenamefont {de~Gironcoli}, \citenamefont {Delugas},
  \citenamefont {DiStasio}, \citenamefont {Ferretti}, \citenamefont {Floris},
  \citenamefont {Fratesi}, \citenamefont {Fugallo}, \citenamefont {Gebauer},
  \citenamefont {Gerstmann}, \citenamefont {Giustino}, \citenamefont {Gorni},
  \citenamefont {Jia}, \citenamefont {Kawamura}, \citenamefont {Ko},
  \citenamefont {Kokalj}, \citenamefont {Kü{\c{c}}ükbenli}, \citenamefont
  {Lazzeri}, \citenamefont {Marsili}, \citenamefont {Marzari}, \citenamefont
  {Mauri}, \citenamefont {Nguyen}, \citenamefont {Nguyen}, \citenamefont {de-la
  Roza}, \citenamefont {Paulatto}, \citenamefont {Ponc{\'{e}}}, \citenamefont
  {Rocca}, \citenamefont {Sabatini}, \citenamefont {Santra}, \citenamefont
  {Schlipf}, \citenamefont {Seitsonen}, \citenamefont {Smogunov}, \citenamefont
  {Timrov}, \citenamefont {Thonhauser}, \citenamefont {Umari}, \citenamefont
  {Vast}, \citenamefont {Wu},\ and\ \citenamefont {Baroni}}]{espresso2}%
  \BibitemOpen
  \bibfield  {author} {\bibinfo {author} {\bibfnamefont {P.}~\bibnamefont
  {Giannozzi}}, \bibinfo {author} {\bibfnamefont {O.}~\bibnamefont
  {Andreussi}}, \bibinfo {author} {\bibfnamefont {T.}~\bibnamefont {Brumme}},
  \bibinfo {author} {\bibfnamefont {O.}~\bibnamefont {Bunau}}, \bibinfo
  {author} {\bibfnamefont {M.~B.}\ \bibnamefont {Nardelli}}, \bibinfo {author}
  {\bibfnamefont {M.}~\bibnamefont {Calandra}}, \bibinfo {author}
  {\bibfnamefont {R.}~\bibnamefont {Car}}, \bibinfo {author} {\bibfnamefont
  {C.}~\bibnamefont {Cavazzoni}}, \bibinfo {author} {\bibfnamefont
  {D.}~\bibnamefont {Ceresoli}}, \bibinfo {author} {\bibfnamefont
  {M.}~\bibnamefont {Cococcioni}}, \bibinfo {author} {\bibfnamefont
  {N.}~\bibnamefont {Colonna}}, \bibinfo {author} {\bibfnamefont
  {I.}~\bibnamefont {Carnimeo}}, \bibinfo {author} {\bibfnamefont {A.~D.}\
  \bibnamefont {Corso}}, \bibinfo {author} {\bibfnamefont {S.}~\bibnamefont
  {de~Gironcoli}}, \bibinfo {author} {\bibfnamefont {P.}~\bibnamefont
  {Delugas}}, \bibinfo {author} {\bibfnamefont {R.~A.}\ \bibnamefont
  {DiStasio}}, \bibinfo {author} {\bibfnamefont {A.}~\bibnamefont {Ferretti}},
  \bibinfo {author} {\bibfnamefont {A.}~\bibnamefont {Floris}}, \bibinfo
  {author} {\bibfnamefont {G.}~\bibnamefont {Fratesi}}, \bibinfo {author}
  {\bibfnamefont {G.}~\bibnamefont {Fugallo}}, \bibinfo {author} {\bibfnamefont
  {R.}~\bibnamefont {Gebauer}}, \bibinfo {author} {\bibfnamefont
  {U.}~\bibnamefont {Gerstmann}}, \bibinfo {author} {\bibfnamefont
  {F.}~\bibnamefont {Giustino}}, \bibinfo {author} {\bibfnamefont
  {T.}~\bibnamefont {Gorni}}, \bibinfo {author} {\bibfnamefont
  {J.}~\bibnamefont {Jia}}, \bibinfo {author} {\bibfnamefont {M.}~\bibnamefont
  {Kawamura}}, \bibinfo {author} {\bibfnamefont {H.-Y.}\ \bibnamefont {Ko}},
  \bibinfo {author} {\bibfnamefont {A.}~\bibnamefont {Kokalj}}, \bibinfo
  {author} {\bibfnamefont {E.}~\bibnamefont {Kü{\c{c}}ükbenli}}, \bibinfo
  {author} {\bibfnamefont {M.}~\bibnamefont {Lazzeri}}, \bibinfo {author}
  {\bibfnamefont {M.}~\bibnamefont {Marsili}}, \bibinfo {author} {\bibfnamefont
  {N.}~\bibnamefont {Marzari}}, \bibinfo {author} {\bibfnamefont
  {F.}~\bibnamefont {Mauri}}, \bibinfo {author} {\bibfnamefont {N.~L.}\
  \bibnamefont {Nguyen}}, \bibinfo {author} {\bibfnamefont {H.-V.}\
  \bibnamefont {Nguyen}}, \bibinfo {author} {\bibfnamefont {A.~O.}\
  \bibnamefont {de-la Roza}}, \bibinfo {author} {\bibfnamefont
  {L.}~\bibnamefont {Paulatto}}, \bibinfo {author} {\bibfnamefont
  {S.}~\bibnamefont {Ponc{\'{e}}}}, \bibinfo {author} {\bibfnamefont
  {D.}~\bibnamefont {Rocca}}, \bibinfo {author} {\bibfnamefont
  {R.}~\bibnamefont {Sabatini}}, \bibinfo {author} {\bibfnamefont
  {B.}~\bibnamefont {Santra}}, \bibinfo {author} {\bibfnamefont
  {M.}~\bibnamefont {Schlipf}}, \bibinfo {author} {\bibfnamefont {A.~P.}\
  \bibnamefont {Seitsonen}}, \bibinfo {author} {\bibfnamefont {A.}~\bibnamefont
  {Smogunov}}, \bibinfo {author} {\bibfnamefont {I.}~\bibnamefont {Timrov}},
  \bibinfo {author} {\bibfnamefont {T.}~\bibnamefont {Thonhauser}}, \bibinfo
  {author} {\bibfnamefont {P.}~\bibnamefont {Umari}}, \bibinfo {author}
  {\bibfnamefont {N.}~\bibnamefont {Vast}}, \bibinfo {author} {\bibfnamefont
  {X.}~\bibnamefont {Wu}},\ and\ \bibinfo {author} {\bibfnamefont
  {S.}~\bibnamefont {Baroni}},\ }\bibfield  {title} {\bibinfo {title} {Advanced
  capabilities for materials modelling with quantum {ESPRESSO}},\ }\href
  {https://doi.org/10.1088/1361-648x/aa8f79} {\bibfield  {journal} {\bibinfo
  {journal} {Journal of Physics: Condensed Matter}\ }\textbf {\bibinfo {volume}
  {29}},\ \bibinfo {pages} {465901} (\bibinfo {year} {2017})}\BibitemShut
  {NoStop}%
\bibitem [{\citenamefont {Giannozzi}\ \emph {et~al.}(2009)\citenamefont
  {Giannozzi}, \citenamefont {Baroni}, \citenamefont {Bonini}, \citenamefont
  {Calandra}, \citenamefont {Car}, \citenamefont {Cavazzoni}, \citenamefont
  {Ceresoli}, \citenamefont {Chiarotti}, \citenamefont {Cococcioni},
  \citenamefont {Dabo} \emph {et~al.}}]{espresso1}%
  \BibitemOpen
  \bibfield  {author} {\bibinfo {author} {\bibfnamefont {P.}~\bibnamefont
  {Giannozzi}}, \bibinfo {author} {\bibfnamefont {S.}~\bibnamefont {Baroni}},
  \bibinfo {author} {\bibfnamefont {N.}~\bibnamefont {Bonini}}, \bibinfo
  {author} {\bibfnamefont {M.}~\bibnamefont {Calandra}}, \bibinfo {author}
  {\bibfnamefont {R.}~\bibnamefont {Car}}, \bibinfo {author} {\bibfnamefont
  {C.}~\bibnamefont {Cavazzoni}}, \bibinfo {author} {\bibfnamefont
  {D.}~\bibnamefont {Ceresoli}}, \bibinfo {author} {\bibfnamefont {G.~L.}\
  \bibnamefont {Chiarotti}}, \bibinfo {author} {\bibfnamefont {M.}~\bibnamefont
  {Cococcioni}}, \bibinfo {author} {\bibfnamefont {I.}~\bibnamefont {Dabo}},
  \emph {et~al.},\ }\bibfield  {title} {\bibinfo {title} {Quantum espresso: a
  modular and open-source software project for quantum simulations of
  materials},\ }\href {http://www.quantum-espresso.org} {\bibfield  {journal}
  {\bibinfo  {journal} {J. Phys. Condens. Matter}\ }\textbf {\bibinfo {volume}
  {21}},\ \bibinfo {pages} {395502} (\bibinfo {year} {2009})}\BibitemShut
  {NoStop}%
\bibitem [{esp()}]{espresso0}%
  \BibitemOpen
  \bibinfo {note} {\texttt{www.quantum-espresso.org}}\BibitemShut {NoStop}%
\bibitem [{\citenamefont {Perdew}\ and\ \citenamefont
  {Wang}(1992)}]{PhysRevB.45.13244}%
  \BibitemOpen
  \bibfield  {author} {\bibinfo {author} {\bibfnamefont {J.~P.}\ \bibnamefont
  {Perdew}}\ and\ \bibinfo {author} {\bibfnamefont {Y.}~\bibnamefont {Wang}},\
  }\bibfield  {title} {\bibinfo {title} {Accurate and simple analytic
  representation of the electron-gas correlation energy},\ }\href
  {https://doi.org/10.1103/PhysRevB.45.13244} {\bibfield  {journal} {\bibinfo
  {journal} {Phys. Rev. B}\ }\textbf {\bibinfo {volume} {45}},\ \bibinfo
  {pages} {13244} (\bibinfo {year} {1992})}\BibitemShut {NoStop}%
\bibitem [{\citenamefont {van Setten}\ \emph {et~al.}(2018)\citenamefont {van
  Setten}, \citenamefont {Giantomassi}, \citenamefont {Bousquet}, \citenamefont
  {Verstraete}, \citenamefont {Hamann}, \citenamefont {Gonze},\ and\
  \citenamefont {Rignanese}}]{pspdojo1}%
  \BibitemOpen
  \bibfield  {author} {\bibinfo {author} {\bibfnamefont {M.}~\bibnamefont {van
  Setten}}, \bibinfo {author} {\bibfnamefont {M.}~\bibnamefont {Giantomassi}},
  \bibinfo {author} {\bibfnamefont {E.}~\bibnamefont {Bousquet}}, \bibinfo
  {author} {\bibfnamefont {M.}~\bibnamefont {Verstraete}}, \bibinfo {author}
  {\bibfnamefont {D.}~\bibnamefont {Hamann}}, \bibinfo {author} {\bibfnamefont
  {X.}~\bibnamefont {Gonze}},\ and\ \bibinfo {author} {\bibfnamefont {G.-M.}\
  \bibnamefont {Rignanese}},\ }\bibfield  {title} {\bibinfo {title} {The
  pseudodojo: Training and grading a 85 element optimized norm-conserving
  pseudopotential table},\ }\href
  {https://doi.org/https://doi.org/10.1016/j.cpc.2018.01.012} {\bibfield
  {journal} {\bibinfo  {journal} {Computer Physics Communications}\ }\textbf
  {\bibinfo {volume} {226}},\ \bibinfo {pages} {39 } (\bibinfo {year}
  {2018})}\BibitemShut {NoStop}%
\bibitem [{psp()}]{pspdojo0}%
  \BibitemOpen
  \bibinfo {note} {\texttt{http://www.pseudo-dojo.org} Scalar-relativstic {\it
  v.}~0.4}\BibitemShut {NoStop}%
\bibitem [{\citenamefont {Hamann}(2013)}]{PhysRevB.88.085117}%
  \BibitemOpen
  \bibfield  {author} {\bibinfo {author} {\bibfnamefont {D.~R.}\ \bibnamefont
  {Hamann}},\ }\bibfield  {title} {\bibinfo {title} {Optimized norm-conserving
  vanderbilt pseudopotentials},\ }\href
  {https://doi.org/10.1103/PhysRevB.88.085117} {\bibfield  {journal} {\bibinfo
  {journal} {Phys. Rev. B}\ }\textbf {\bibinfo {volume} {88}},\ \bibinfo
  {pages} {085117} (\bibinfo {year} {2013})}\BibitemShut {NoStop}%
\bibitem [{onc()}]{oncvp}%
  \BibitemOpen
  \bibinfo {note} {The open-source code {\sc oncvpsp} is avaiable at
  \texttt{http://www.mat-simresearch.com} \, {\it v.}~3.3.1}\BibitemShut
  {NoStop}%
\bibitem [{\citenamefont {Schwartz}\ \emph {et~al.}(2017)\citenamefont
  {Schwartz}, \citenamefont {Ponce}, \citenamefont {Friedrich}, \citenamefont
  {Cramer}, \citenamefont {Vinson},\ and\ \citenamefont
  {Prendergast}}]{SCHWARTZ201730}%
  \BibitemOpen
  \bibfield  {author} {\bibinfo {author} {\bibfnamefont {C.~P.}\ \bibnamefont
  {Schwartz}}, \bibinfo {author} {\bibfnamefont {F.}~\bibnamefont {Ponce}},
  \bibinfo {author} {\bibfnamefont {S.}~\bibnamefont {Friedrich}}, \bibinfo
  {author} {\bibfnamefont {S.~P.}\ \bibnamefont {Cramer}}, \bibinfo {author}
  {\bibfnamefont {J.}~\bibnamefont {Vinson}},\ and\ \bibinfo {author}
  {\bibfnamefont {D.}~\bibnamefont {Prendergast}},\ }\bibfield  {title}
  {\bibinfo {title} {Temperature and radiation effects at the fluorine k-edge
  in lif},\ }\href
  {https://doi.org/https://doi.org/10.1016/j.elspec.2017.05.007} {\bibfield
  {journal} {\bibinfo  {journal} {Journal of Electron Spectroscopy and Related
  Phenomena}\ }\textbf {\bibinfo {volume} {218}},\ \bibinfo {pages} {30 }
  (\bibinfo {year} {2017})}\BibitemShut {NoStop}%
\bibitem [{\citenamefont {Pascal}\ \emph {et~al.}(2014)\citenamefont {Pascal},
  \citenamefont {Boesenberg}, \citenamefont {Kostecki}, \citenamefont
  {Richardson}, \citenamefont {Weng}, \citenamefont {Sokaras}, \citenamefont
  {Nordlund}, \citenamefont {McDermott}, \citenamefont {Moewes}, \citenamefont
  {Cabana},\ and\ \citenamefont {Prendergast}}]{Pascal}%
  \BibitemOpen
  \bibfield  {author} {\bibinfo {author} {\bibfnamefont {T.~A.}\ \bibnamefont
  {Pascal}}, \bibinfo {author} {\bibfnamefont {U.}~\bibnamefont {Boesenberg}},
  \bibinfo {author} {\bibfnamefont {R.}~\bibnamefont {Kostecki}}, \bibinfo
  {author} {\bibfnamefont {T.~J.}\ \bibnamefont {Richardson}}, \bibinfo
  {author} {\bibfnamefont {T.-C.}\ \bibnamefont {Weng}}, \bibinfo {author}
  {\bibfnamefont {D.}~\bibnamefont {Sokaras}}, \bibinfo {author} {\bibfnamefont
  {D.}~\bibnamefont {Nordlund}}, \bibinfo {author} {\bibfnamefont
  {E.}~\bibnamefont {McDermott}}, \bibinfo {author} {\bibfnamefont
  {A.}~\bibnamefont {Moewes}}, \bibinfo {author} {\bibfnamefont
  {J.}~\bibnamefont {Cabana}},\ and\ \bibinfo {author} {\bibfnamefont
  {D.}~\bibnamefont {Prendergast}},\ }\bibfield  {title} {\bibinfo {title}
  {Finite temperature effects on the x-ray absorption spectra of lithium
  compounds: First-principles interpretation of x-ray raman measurements},\
  }\href {https://doi.org/10.1063/1.4856835} {\bibfield  {journal} {\bibinfo
  {journal} {The Journal of Chemical Physics}\ }\textbf {\bibinfo {volume}
  {140}},\ \bibinfo {pages} {034107} (\bibinfo {year} {2014})}\BibitemShut
  {NoStop}%
\bibitem [{\citenamefont {Vinson}\ \emph {et~al.}(2014)\citenamefont {Vinson},
  \citenamefont {Jach}, \citenamefont {Elam},\ and\ \citenamefont
  {Denlinger}}]{PhysRevB.90.205207}%
  \BibitemOpen
  \bibfield  {author} {\bibinfo {author} {\bibfnamefont {J.}~\bibnamefont
  {Vinson}}, \bibinfo {author} {\bibfnamefont {T.}~\bibnamefont {Jach}},
  \bibinfo {author} {\bibfnamefont {W.~T.}\ \bibnamefont {Elam}},\ and\
  \bibinfo {author} {\bibfnamefont {J.~D.}\ \bibnamefont {Denlinger}},\
  }\bibfield  {title} {\bibinfo {title} {Origins of extreme broadening
  mechanisms in near-edge x-ray spectra of nitrogen compounds},\ }\href
  {https://doi.org/10.1103/PhysRevB.90.205207} {\bibfield  {journal} {\bibinfo
  {journal} {Phys. Rev. B}\ }\textbf {\bibinfo {volume} {90}},\ \bibinfo
  {pages} {205207} (\bibinfo {year} {2014})}\BibitemShut {NoStop}%
\bibitem [{\citenamefont {Brouder}\ \emph {et~al.}(2010)\citenamefont
  {Brouder}, \citenamefont {Cabaret}, \citenamefont {Juhin},\ and\
  \citenamefont {Sainctavit}}]{PhysRevB.81.115125}%
  \BibitemOpen
  \bibfield  {author} {\bibinfo {author} {\bibfnamefont {C.}~\bibnamefont
  {Brouder}}, \bibinfo {author} {\bibfnamefont {D.}~\bibnamefont {Cabaret}},
  \bibinfo {author} {\bibfnamefont {A.}~\bibnamefont {Juhin}},\ and\ \bibinfo
  {author} {\bibfnamefont {P.}~\bibnamefont {Sainctavit}},\ }\bibfield  {title}
  {\bibinfo {title} {Effect of atomic vibrations on the x-ray absorption
  spectra at the $k$ edge of al in
  $\ensuremath{\alpha}{\text{-al}}_{2}{\text{o}}_{3}$ and of ti in
  ${\text{tio}}_{2}$ rutile},\ }\href
  {https://doi.org/10.1103/PhysRevB.81.115125} {\bibfield  {journal} {\bibinfo
  {journal} {Phys. Rev. B}\ }\textbf {\bibinfo {volume} {81}},\ \bibinfo
  {pages} {115125} (\bibinfo {year} {2010})}\BibitemShut {NoStop}%
\bibitem [{\citenamefont {Vinson}\ \emph {et~al.}(2017)\citenamefont {Vinson},
  \citenamefont {Jach}, \citenamefont {M\"uller}, \citenamefont
  {Unterumsberger},\ and\ \citenamefont {Beckhoff}}]{PhysRevB.96.205116}%
  \BibitemOpen
  \bibfield  {author} {\bibinfo {author} {\bibfnamefont {J.}~\bibnamefont
  {Vinson}}, \bibinfo {author} {\bibfnamefont {T.}~\bibnamefont {Jach}},
  \bibinfo {author} {\bibfnamefont {M.}~\bibnamefont {M\"uller}}, \bibinfo
  {author} {\bibfnamefont {R.}~\bibnamefont {Unterumsberger}},\ and\ \bibinfo
  {author} {\bibfnamefont {B.}~\bibnamefont {Beckhoff}},\ }\bibfield  {title}
  {\bibinfo {title} {Resonant x-ray emission of hexagonal boron nitride},\
  }\href {https://doi.org/10.1103/PhysRevB.96.205116} {\bibfield  {journal}
  {\bibinfo  {journal} {Phys. Rev. B}\ }\textbf {\bibinfo {volume} {96}},\
  \bibinfo {pages} {205116} (\bibinfo {year} {2017})}\BibitemShut {NoStop}%
\bibitem [{\citenamefont {Olovsson}\ \emph {et~al.}(2019)\citenamefont
  {Olovsson}, \citenamefont {Mizoguchi}, \citenamefont {Magnuson},
  \citenamefont {Kontur}, \citenamefont {Hellman}, \citenamefont {Tanaka},\
  and\ \citenamefont {Draxl}}]{doi:10.1021/acs.jpcc.9b00179}%
  \BibitemOpen
  \bibfield  {author} {\bibinfo {author} {\bibfnamefont {W.}~\bibnamefont
  {Olovsson}}, \bibinfo {author} {\bibfnamefont {T.}~\bibnamefont {Mizoguchi}},
  \bibinfo {author} {\bibfnamefont {M.}~\bibnamefont {Magnuson}}, \bibinfo
  {author} {\bibfnamefont {S.}~\bibnamefont {Kontur}}, \bibinfo {author}
  {\bibfnamefont {O.}~\bibnamefont {Hellman}}, \bibinfo {author} {\bibfnamefont
  {I.}~\bibnamefont {Tanaka}},\ and\ \bibinfo {author} {\bibfnamefont
  {C.}~\bibnamefont {Draxl}},\ }\bibfield  {title} {\bibinfo {title}
  {Vibrational effects in x-ray absorption spectra of two-dimensional layered
  materials},\ }\href {https://doi.org/10.1021/acs.jpcc.9b00179} {\bibfield
  {journal} {\bibinfo  {journal} {The Journal of Physical Chemistry C}\
  }\textbf {\bibinfo {volume} {123}},\ \bibinfo {pages} {9688} (\bibinfo {year}
  {2019})}\BibitemShut {NoStop}%
\bibitem [{\citenamefont {McDougall}\ \emph {et~al.}(2017)\citenamefont
  {McDougall}, \citenamefont {Partridge}, \citenamefont {Nicholls},
  \citenamefont {Russo},\ and\ \citenamefont {McCulloch}}]{PhysRevB.96.144106}%
  \BibitemOpen
  \bibfield  {author} {\bibinfo {author} {\bibfnamefont {N.~L.}\ \bibnamefont
  {McDougall}}, \bibinfo {author} {\bibfnamefont {J.~G.}\ \bibnamefont
  {Partridge}}, \bibinfo {author} {\bibfnamefont {R.~J.}\ \bibnamefont
  {Nicholls}}, \bibinfo {author} {\bibfnamefont {S.~P.}\ \bibnamefont
  {Russo}},\ and\ \bibinfo {author} {\bibfnamefont {D.~G.}\ \bibnamefont
  {McCulloch}},\ }\bibfield  {title} {\bibinfo {title} {Influence of point
  defects on the near edge structure of hexagonal boron nitride},\ }\href
  {https://doi.org/10.1103/PhysRevB.96.144106} {\bibfield  {journal} {\bibinfo
  {journal} {Phys. Rev. B}\ }\textbf {\bibinfo {volume} {96}},\ \bibinfo
  {pages} {144106} (\bibinfo {year} {2017})}\BibitemShut {NoStop}%
\bibitem [{\citenamefont {Geick}\ \emph {et~al.}(1966)\citenamefont {Geick},
  \citenamefont {Perry},\ and\ \citenamefont {Rupprecht}}]{PhysRev.146.543}%
  \BibitemOpen
  \bibfield  {author} {\bibinfo {author} {\bibfnamefont {R.}~\bibnamefont
  {Geick}}, \bibinfo {author} {\bibfnamefont {C.~H.}\ \bibnamefont {Perry}},\
  and\ \bibinfo {author} {\bibfnamefont {G.}~\bibnamefont {Rupprecht}},\
  }\bibfield  {title} {\bibinfo {title} {Normal modes in hexagonal boron
  nitride},\ }\href {https://doi.org/10.1103/PhysRev.146.543} {\bibfield
  {journal} {\bibinfo  {journal} {Phys. Rev.}\ }\textbf {\bibinfo {volume}
  {146}},\ \bibinfo {pages} {543} (\bibinfo {year} {1966})}\BibitemShut
  {NoStop}%
\bibitem [{\citenamefont {Gulans}\ \emph {et~al.}(2014)\citenamefont {Gulans},
  \citenamefont {Kontur}, \citenamefont {Meisenbichler}, \citenamefont {Nabok},
  \citenamefont {Pavone}, \citenamefont {Rigamonti}, \citenamefont
  {Sagmeister}, \citenamefont {Werner},\ and\ \citenamefont
  {Draxl}}]{Gulans_2014}%
  \BibitemOpen
  \bibfield  {author} {\bibinfo {author} {\bibfnamefont {A.}~\bibnamefont
  {Gulans}}, \bibinfo {author} {\bibfnamefont {S.}~\bibnamefont {Kontur}},
  \bibinfo {author} {\bibfnamefont {C.}~\bibnamefont {Meisenbichler}}, \bibinfo
  {author} {\bibfnamefont {D.}~\bibnamefont {Nabok}}, \bibinfo {author}
  {\bibfnamefont {P.}~\bibnamefont {Pavone}}, \bibinfo {author} {\bibfnamefont
  {S.}~\bibnamefont {Rigamonti}}, \bibinfo {author} {\bibfnamefont
  {S.}~\bibnamefont {Sagmeister}}, \bibinfo {author} {\bibfnamefont
  {U.}~\bibnamefont {Werner}},\ and\ \bibinfo {author} {\bibfnamefont
  {C.}~\bibnamefont {Draxl}},\ }\bibfield  {title} {\bibinfo {title} {exciting:
  a full-potential all-electron package implementing density-functional theory
  and many-body perturbation theory},\ }\href
  {https://doi.org/10.1088/0953-8984/26/36/363202} {\bibfield  {journal}
  {\bibinfo  {journal} {Journal of Physics: Condensed Matter}\ }\textbf
  {\bibinfo {volume} {26}},\ \bibinfo {pages} {363202} (\bibinfo {year}
  {2014})}\BibitemShut {NoStop}%
\bibitem [{\citenamefont {Vorwerk}\ \emph {et~al.}(2017)\citenamefont
  {Vorwerk}, \citenamefont {Cocchi},\ and\ \citenamefont
  {Draxl}}]{PhysRevB.95.155121}%
  \BibitemOpen
  \bibfield  {author} {\bibinfo {author} {\bibfnamefont {C.}~\bibnamefont
  {Vorwerk}}, \bibinfo {author} {\bibfnamefont {C.}~\bibnamefont {Cocchi}},\
  and\ \bibinfo {author} {\bibfnamefont {C.}~\bibnamefont {Draxl}},\ }\bibfield
   {title} {\bibinfo {title} {Addressing electron-hole correlation in core
  excitations of solids: An all-electron many-body approach from first
  principles},\ }\href {https://doi.org/10.1103/PhysRevB.95.155121} {\bibfield
  {journal} {\bibinfo  {journal} {Phys. Rev. B}\ }\textbf {\bibinfo {volume}
  {95}},\ \bibinfo {pages} {155121} (\bibinfo {year} {2017})}\BibitemShut
  {NoStop}%
\bibitem [{\citenamefont {Morris}\ \emph {et~al.}(2014)\citenamefont {Morris},
  \citenamefont {Nicholls}, \citenamefont {Pickard},\ and\ \citenamefont
  {Yates}}]{MORRIS20141477}%
  \BibitemOpen
  \bibfield  {author} {\bibinfo {author} {\bibfnamefont {A.~J.}\ \bibnamefont
  {Morris}}, \bibinfo {author} {\bibfnamefont {R.~J.}\ \bibnamefont
  {Nicholls}}, \bibinfo {author} {\bibfnamefont {C.~J.}\ \bibnamefont
  {Pickard}},\ and\ \bibinfo {author} {\bibfnamefont {J.~R.}\ \bibnamefont
  {Yates}},\ }\bibfield  {title} {\bibinfo {title} {Optados: A tool for
  obtaining density of states, core-level and optical spectra from electronic
  structure codes},\ }\href
  {https://doi.org/https://doi.org/10.1016/j.cpc.2014.02.013} {\bibfield
  {journal} {\bibinfo  {journal} {Computer Physics Communications}\ }\textbf
  {\bibinfo {volume} {185}},\ \bibinfo {pages} {1477 } (\bibinfo {year}
  {2014})}\BibitemShut {NoStop}%
\bibitem [{\citenamefont {Clark}\ \emph {et~al.}(2005)\citenamefont {Clark},
  \citenamefont {Segall}, \citenamefont {Pickard}, \citenamefont {Hasnip},
  \citenamefont {Probert}, \citenamefont {Refson},\ and\ \citenamefont
  {Payne}}]{CASTEP}%
  \BibitemOpen
  \bibfield  {author} {\bibinfo {author} {\bibfnamefont {S.~J.}\ \bibnamefont
  {Clark}}, \bibinfo {author} {\bibfnamefont {M.~D.}\ \bibnamefont {Segall}},
  \bibinfo {author} {\bibfnamefont {C.~J.}\ \bibnamefont {Pickard}}, \bibinfo
  {author} {\bibfnamefont {P.~J.}\ \bibnamefont {Hasnip}}, \bibinfo {author}
  {\bibfnamefont {M.~I.~J.}\ \bibnamefont {Probert}}, \bibinfo {author}
  {\bibfnamefont {K.}~\bibnamefont {Refson}},\ and\ \bibinfo {author}
  {\bibfnamefont {M.~C.}\ \bibnamefont {Payne}},\ }\bibfield  {title} {\bibinfo
  {title} {First principles methods using castep},\ }\href
  {https://doi.org/https://doi.org/10.1524/zkri.220.5.567.65075} {\bibfield
  {journal} {\bibinfo  {journal} {Zeitschrift für Kristallographie -
  Crystalline Materials}\ }\textbf {\bibinfo {volume} {220}},\ \bibinfo {pages}
  {567 } (\bibinfo {year} {2005})}\BibitemShut {NoStop}%
\bibitem [{\citenamefont {Preobrajenski}\ \emph {et~al.}(2005)\citenamefont
  {Preobrajenski}, \citenamefont {Vinogradov},\ and\ \citenamefont
  {Mårtensson}}]{PREOBRAJENSKI200559}%
  \BibitemOpen
  \bibfield  {author} {\bibinfo {author} {\bibfnamefont {A.}~\bibnamefont
  {Preobrajenski}}, \bibinfo {author} {\bibfnamefont {A.}~\bibnamefont
  {Vinogradov}},\ and\ \bibinfo {author} {\bibfnamefont {N.}~\bibnamefont
  {Mårtensson}},\ }\bibfield  {title} {\bibinfo {title} {Decay of core
  excitations in bulk h-bn studied with resonant auger spectroscopy},\ }\href
  {https://doi.org/https://doi.org/10.1016/j.elspec.2005.02.005} {\bibfield
  {journal} {\bibinfo  {journal} {Journal of Electron Spectroscopy and Related
  Phenomena}\ }\textbf {\bibinfo {volume} {148}},\ \bibinfo {pages} {59}
  (\bibinfo {year} {2005})}\BibitemShut {NoStop}%
\bibitem [{\citenamefont {McDougall}\ \emph {et~al.}(2014)\citenamefont
  {McDougall}, \citenamefont {Nicholls}, \citenamefont {Partridge},\ and\
  \citenamefont {McCulloch}}]{mcdougall_nicholls_partridge_mcculloch_2014}%
  \BibitemOpen
  \bibfield  {author} {\bibinfo {author} {\bibfnamefont {N.~L.}\ \bibnamefont
  {McDougall}}, \bibinfo {author} {\bibfnamefont {R.~J.}\ \bibnamefont
  {Nicholls}}, \bibinfo {author} {\bibfnamefont {J.~G.}\ \bibnamefont
  {Partridge}},\ and\ \bibinfo {author} {\bibfnamefont {D.~G.}\ \bibnamefont
  {McCulloch}},\ }\bibfield  {title} {\bibinfo {title} {The near edge structure
  of hexagonal boron nitride},\ }\href
  {https://doi.org/10.1017/S1431927614000737} {\bibfield  {journal} {\bibinfo
  {journal} {Microscopy and Microanalysis}\ }\textbf {\bibinfo {volume} {20}},\
  \bibinfo {pages} {1053–1059} (\bibinfo {year} {2014})}\BibitemShut
  {NoStop}%
\bibitem [{\citenamefont {Liang}\ \emph {et~al.}(2017)\citenamefont {Liang},
  \citenamefont {Vinson}, \citenamefont {Pemmaraju}, \citenamefont {Drisdell},
  \citenamefont {Shirley},\ and\ \citenamefont
  {Prendergast}}]{PhysRevLett.118.096402}%
  \BibitemOpen
  \bibfield  {author} {\bibinfo {author} {\bibfnamefont {Y.}~\bibnamefont
  {Liang}}, \bibinfo {author} {\bibfnamefont {J.}~\bibnamefont {Vinson}},
  \bibinfo {author} {\bibfnamefont {S.}~\bibnamefont {Pemmaraju}}, \bibinfo
  {author} {\bibfnamefont {W.~S.}\ \bibnamefont {Drisdell}}, \bibinfo {author}
  {\bibfnamefont {E.~L.}\ \bibnamefont {Shirley}},\ and\ \bibinfo {author}
  {\bibfnamefont {D.}~\bibnamefont {Prendergast}},\ }\bibfield  {title}
  {\bibinfo {title} {Accurate x-ray spectral predictions: An advanced
  self-consistent-field approach inspired by many-body perturbation theory},\
  }\href {https://doi.org/10.1103/PhysRevLett.118.096402} {\bibfield  {journal}
  {\bibinfo  {journal} {Phys. Rev. Lett.}\ }\textbf {\bibinfo {volume} {118}},\
  \bibinfo {pages} {096402} (\bibinfo {year} {2017})}\BibitemShut {NoStop}%
\bibitem [{\citenamefont {Benedict}\ and\ \citenamefont
  {Shirley}(1999)}]{PhysRevB.59.5441}%
  \BibitemOpen
  \bibfield  {author} {\bibinfo {author} {\bibfnamefont {L.~X.}\ \bibnamefont
  {Benedict}}\ and\ \bibinfo {author} {\bibfnamefont {E.~L.}\ \bibnamefont
  {Shirley}},\ }\bibfield  {title} {\bibinfo {title} {Ab initio calculation of
  ${\ensuremath{\epsilon}}_{2}(\ensuremath{\omega})$ including the
  electron-hole interaction: Application to gan and ${\mathrm{caf}}_{2}$},\
  }\href {https://doi.org/10.1103/PhysRevB.59.5441} {\bibfield  {journal}
  {\bibinfo  {journal} {Phys. Rev. B}\ }\textbf {\bibinfo {volume} {59}},\
  \bibinfo {pages} {5441} (\bibinfo {year} {1999})}\BibitemShut {NoStop}%
\bibitem [{\citenamefont {Lawler}\ \emph {et~al.}(2008)\citenamefont {Lawler},
  \citenamefont {Rehr}, \citenamefont {Vila}, \citenamefont {Dalosto},
  \citenamefont {Shirley},\ and\ \citenamefont {Levine}}]{PhysRevB.78.205108}%
  \BibitemOpen
  \bibfield  {author} {\bibinfo {author} {\bibfnamefont {H.~M.}\ \bibnamefont
  {Lawler}}, \bibinfo {author} {\bibfnamefont {J.~J.}\ \bibnamefont {Rehr}},
  \bibinfo {author} {\bibfnamefont {F.}~\bibnamefont {Vila}}, \bibinfo {author}
  {\bibfnamefont {S.~D.}\ \bibnamefont {Dalosto}}, \bibinfo {author}
  {\bibfnamefont {E.~L.}\ \bibnamefont {Shirley}},\ and\ \bibinfo {author}
  {\bibfnamefont {Z.~H.}\ \bibnamefont {Levine}},\ }\bibfield  {title}
  {\bibinfo {title} {Optical to uv spectra and birefringence of
  ${\text{sio}}_{2}$ and ${\text{tio}}_{2}$: First-principles calculations with
  excitonic effects},\ }\href {https://doi.org/10.1103/PhysRevB.78.205108}
  {\bibfield  {journal} {\bibinfo  {journal} {Phys. Rev. B}\ }\textbf {\bibinfo
  {volume} {78}},\ \bibinfo {pages} {205108} (\bibinfo {year}
  {2008})}\BibitemShut {NoStop}%
\bibitem [{\citenamefont {Shirley}\ \emph
  {et~al.}(2005{\natexlab{b}})\citenamefont {Shirley}, \citenamefont
  {Soininen},\ and\ \citenamefont {Rehr}}]{Shirley_2005}%
  \BibitemOpen
  \bibfield  {author} {\bibinfo {author} {\bibfnamefont {E.~L.}\ \bibnamefont
  {Shirley}}, \bibinfo {author} {\bibfnamefont {J.~A.}\ \bibnamefont
  {Soininen}},\ and\ \bibinfo {author} {\bibfnamefont {J.~J.}\ \bibnamefont
  {Rehr}},\ }\bibfield  {title} {\bibinfo {title} {Modeling {CoreHole}
  screening in {CoreExcitation} spectroscopies},\ }\href
  {https://doi.org/10.1238/physica.topical.115a00031} {\bibfield  {journal}
  {\bibinfo  {journal} {Physica Scripta}\ ,\ \bibinfo {pages} {31}} (\bibinfo
  {year} {2005}{\natexlab{b}})}\BibitemShut {NoStop}%
\bibitem [{\citenamefont {Hybertsen}\ and\ \citenamefont
  {Louie}(1988)}]{PhysRevB.37.2733}%
  \BibitemOpen
  \bibfield  {author} {\bibinfo {author} {\bibfnamefont {M.~S.}\ \bibnamefont
  {Hybertsen}}\ and\ \bibinfo {author} {\bibfnamefont {S.~G.}\ \bibnamefont
  {Louie}},\ }\bibfield  {title} {\bibinfo {title} {Model dielectric matrices
  for quasiparticle self-energy calculations},\ }\href
  {https://doi.org/10.1103/PhysRevB.37.2733} {\bibfield  {journal} {\bibinfo
  {journal} {Phys. Rev. B}\ }\textbf {\bibinfo {volume} {37}},\ \bibinfo
  {pages} {2733} (\bibinfo {year} {1988})}\BibitemShut {NoStop}%
\bibitem [{\citenamefont {Lautenschlager}\ \emph {et~al.}(1987)\citenamefont
  {Lautenschlager}, \citenamefont {Garriga}, \citenamefont {Vina},\ and\
  \citenamefont {Cardona}}]{PhysRevB.36.4821}%
  \BibitemOpen
  \bibfield  {author} {\bibinfo {author} {\bibfnamefont {P.}~\bibnamefont
  {Lautenschlager}}, \bibinfo {author} {\bibfnamefont {M.}~\bibnamefont
  {Garriga}}, \bibinfo {author} {\bibfnamefont {L.}~\bibnamefont {Vina}},\ and\
  \bibinfo {author} {\bibfnamefont {M.}~\bibnamefont {Cardona}},\ }\bibfield
  {title} {\bibinfo {title} {Temperature dependence of the dielectric function
  and interband critical points in silicon},\ }\href
  {https://doi.org/10.1103/PhysRevB.36.4821} {\bibfield  {journal} {\bibinfo
  {journal} {Phys. Rev. B}\ }\textbf {\bibinfo {volume} {36}},\ \bibinfo
  {pages} {4821} (\bibinfo {year} {1987})}\BibitemShut {NoStop}%
\bibitem [{\citenamefont {Roessler}\ and\ \citenamefont
  {Walker}(1967)}]{Roessler:67}%
  \BibitemOpen
  \bibfield  {author} {\bibinfo {author} {\bibfnamefont {D.~M.}\ \bibnamefont
  {Roessler}}\ and\ \bibinfo {author} {\bibfnamefont {W.~C.}\ \bibnamefont
  {Walker}},\ }\bibfield  {title} {\bibinfo {title} {Optical constants of
  magnesium oxide and lithium fluoride in the far ultraviolet$\ast$},\ }\href
  {https://doi.org/10.1364/JOSA.57.000835} {\bibfield  {journal} {\bibinfo
  {journal} {J. Opt. Soc. Am.}\ }\textbf {\bibinfo {volume} {57}},\ \bibinfo
  {pages} {835} (\bibinfo {year} {1967})}\BibitemShut {NoStop}%
\bibitem [{\citenamefont {Hanke}\ and\ \citenamefont
  {Sham}(1979)}]{PhysRevLett.43.387}%
  \BibitemOpen
  \bibfield  {author} {\bibinfo {author} {\bibfnamefont {W.}~\bibnamefont
  {Hanke}}\ and\ \bibinfo {author} {\bibfnamefont {L.~J.}\ \bibnamefont
  {Sham}},\ }\bibfield  {title} {\bibinfo {title} {Many-particle effects in the
  optical excitations of a semiconductor},\ }\href
  {https://doi.org/10.1103/PhysRevLett.43.387} {\bibfield  {journal} {\bibinfo
  {journal} {Phys. Rev. Lett.}\ }\textbf {\bibinfo {volume} {43}},\ \bibinfo
  {pages} {387} (\bibinfo {year} {1979})}\BibitemShut {NoStop}%
\bibitem [{\citenamefont {Wyckoff}(1963)}]{Wyckoff}%
  \BibitemOpen
  \bibfield  {author} {\bibinfo {author} {\bibfnamefont {R.~W.~G.}\
  \bibnamefont {Wyckoff}},\ }\href@noop {} {\emph {\bibinfo {title} {Crystal
  Structures}}},\ \bibinfo {edition} {2nd}\ ed.\ (\bibinfo  {publisher}
  {Interscience Publishers},\ \bibinfo {address} {New York},\ \bibinfo {year}
  {1963})\BibitemShut {NoStop}%
\bibitem [{\citenamefont {Kittel}(2005)}]{Kittel}%
  \BibitemOpen
  \bibfield  {author} {\bibinfo {author} {\bibfnamefont {C.}~\bibnamefont
  {Kittel}},\ }\href@noop {} {\emph {\bibinfo {title} {Introduction to Solid
  State Physics}}},\ \bibinfo {edition} {8th}\ ed.\ (\bibinfo  {publisher}
  {John Wiley \& Sons, Inc},\ \bibinfo {year} {2005})\BibitemShut {NoStop}%
\bibitem [{\citenamefont {Puschnig}\ and\ \citenamefont
  {Ambrosch-Draxl}(2002)}]{PhysRevB.66.165105}%
  \BibitemOpen
  \bibfield  {author} {\bibinfo {author} {\bibfnamefont {P.}~\bibnamefont
  {Puschnig}}\ and\ \bibinfo {author} {\bibfnamefont {C.}~\bibnamefont
  {Ambrosch-Draxl}},\ }\bibfield  {title} {\bibinfo {title} {Optical absorption
  spectra of semiconductors and insulators including electron-hole
  correlations: An ab initio study within the lapw method},\ }\href
  {https://doi.org/10.1103/PhysRevB.66.165105} {\bibfield  {journal} {\bibinfo
  {journal} {Phys. Rev. B}\ }\textbf {\bibinfo {volume} {66}},\ \bibinfo
  {pages} {165105} (\bibinfo {year} {2002})}\BibitemShut {NoStop}%
\bibitem [{\citenamefont {Gatti}\ and\ \citenamefont
  {Sottile}(2013)}]{PhysRevB.88.155113}%
  \BibitemOpen
  \bibfield  {author} {\bibinfo {author} {\bibfnamefont {M.}~\bibnamefont
  {Gatti}}\ and\ \bibinfo {author} {\bibfnamefont {F.}~\bibnamefont
  {Sottile}},\ }\bibfield  {title} {\bibinfo {title} {Exciton dispersion from
  first principles},\ }\href {https://doi.org/10.1103/PhysRevB.88.155113}
  {\bibfield  {journal} {\bibinfo  {journal} {Phys. Rev. B}\ }\textbf {\bibinfo
  {volume} {88}},\ \bibinfo {pages} {155113} (\bibinfo {year}
  {2013})}\BibitemShut {NoStop}%
\bibitem [{\citenamefont {Piacentini}\ \emph {et~al.}(1976)\citenamefont
  {Piacentini}, \citenamefont {Lynch},\ and\ \citenamefont
  {Olson}}]{PhysRevB.13.5530}%
  \BibitemOpen
  \bibfield  {author} {\bibinfo {author} {\bibfnamefont {M.}~\bibnamefont
  {Piacentini}}, \bibinfo {author} {\bibfnamefont {D.~W.}\ \bibnamefont
  {Lynch}},\ and\ \bibinfo {author} {\bibfnamefont {C.~G.}\ \bibnamefont
  {Olson}},\ }\bibfield  {title} {\bibinfo {title} {Thermoreflectance of lif
  between 12 and 30 ev},\ }\href {https://doi.org/10.1103/PhysRevB.13.5530}
  {\bibfield  {journal} {\bibinfo  {journal} {Phys. Rev. B}\ }\textbf {\bibinfo
  {volume} {13}},\ \bibinfo {pages} {5530} (\bibinfo {year}
  {1976})}\BibitemShut {NoStop}%
\bibitem [{\citenamefont {Gonze}\ \emph {et~al.}(2020)\citenamefont {Gonze},
  \citenamefont {Amadon}, \citenamefont {Antonius}, \citenamefont {Arnardi},
  \citenamefont {Baguet}, \citenamefont {Beuken}, \citenamefont {Bieder},
  \citenamefont {Bottin}, \citenamefont {Bouchet}, \citenamefont {Bousquet},
  \citenamefont {Brouwer}, \citenamefont {Bruneval}, \citenamefont {Brunin},
  \citenamefont {Cavignac}, \citenamefont {Charraud}, \citenamefont {Chen},
  \citenamefont {Côté}, \citenamefont {Cottenier}, \citenamefont {Denier},
  \citenamefont {Geneste}, \citenamefont {Ghosez}, \citenamefont {Giantomassi},
  \citenamefont {Gillet}, \citenamefont {Gingras}, \citenamefont {Hamann},
  \citenamefont {Hautier}, \citenamefont {He}, \citenamefont {Helbig},
  \citenamefont {Holzwarth}, \citenamefont {Jia}, \citenamefont {Jollet},
  \citenamefont {Lafargue-Dit-Hauret}, \citenamefont {Lejaeghere},
  \citenamefont {Marques}, \citenamefont {Martin}, \citenamefont {Martins},
  \citenamefont {Miranda}, \citenamefont {Naccarato}, \citenamefont {Persson},
  \citenamefont {Petretto}, \citenamefont {Planes}, \citenamefont {Pouillon},
  \citenamefont {Prokhorenko}, \citenamefont {Ricci}, \citenamefont
  {Rignanese}, \citenamefont {Romero}, \citenamefont {Schmitt}, \citenamefont
  {Torrent}, \citenamefont {van Setten}, \citenamefont {Troeye}, \citenamefont
  {Verstraete}, \citenamefont {Zérah},\ and\ \citenamefont
  {Zwanziger}}]{Gonze2020}%
  \BibitemOpen
  \bibfield  {author} {\bibinfo {author} {\bibfnamefont {X.}~\bibnamefont
  {Gonze}}, \bibinfo {author} {\bibfnamefont {B.}~\bibnamefont {Amadon}},
  \bibinfo {author} {\bibfnamefont {G.}~\bibnamefont {Antonius}}, \bibinfo
  {author} {\bibfnamefont {F.}~\bibnamefont {Arnardi}}, \bibinfo {author}
  {\bibfnamefont {L.}~\bibnamefont {Baguet}}, \bibinfo {author} {\bibfnamefont
  {J.-M.}\ \bibnamefont {Beuken}}, \bibinfo {author} {\bibfnamefont
  {J.}~\bibnamefont {Bieder}}, \bibinfo {author} {\bibfnamefont
  {F.}~\bibnamefont {Bottin}}, \bibinfo {author} {\bibfnamefont
  {J.}~\bibnamefont {Bouchet}}, \bibinfo {author} {\bibfnamefont
  {E.}~\bibnamefont {Bousquet}}, \bibinfo {author} {\bibfnamefont
  {N.}~\bibnamefont {Brouwer}}, \bibinfo {author} {\bibfnamefont
  {F.}~\bibnamefont {Bruneval}}, \bibinfo {author} {\bibfnamefont
  {G.}~\bibnamefont {Brunin}}, \bibinfo {author} {\bibfnamefont
  {T.}~\bibnamefont {Cavignac}}, \bibinfo {author} {\bibfnamefont {J.-B.}\
  \bibnamefont {Charraud}}, \bibinfo {author} {\bibfnamefont {W.}~\bibnamefont
  {Chen}}, \bibinfo {author} {\bibfnamefont {M.}~\bibnamefont {Côté}},
  \bibinfo {author} {\bibfnamefont {S.}~\bibnamefont {Cottenier}}, \bibinfo
  {author} {\bibfnamefont {J.}~\bibnamefont {Denier}}, \bibinfo {author}
  {\bibfnamefont {G.}~\bibnamefont {Geneste}}, \bibinfo {author} {\bibfnamefont
  {P.}~\bibnamefont {Ghosez}}, \bibinfo {author} {\bibfnamefont
  {M.}~\bibnamefont {Giantomassi}}, \bibinfo {author} {\bibfnamefont
  {Y.}~\bibnamefont {Gillet}}, \bibinfo {author} {\bibfnamefont
  {O.}~\bibnamefont {Gingras}}, \bibinfo {author} {\bibfnamefont {D.~R.}\
  \bibnamefont {Hamann}}, \bibinfo {author} {\bibfnamefont {G.}~\bibnamefont
  {Hautier}}, \bibinfo {author} {\bibfnamefont {X.}~\bibnamefont {He}},
  \bibinfo {author} {\bibfnamefont {N.}~\bibnamefont {Helbig}}, \bibinfo
  {author} {\bibfnamefont {N.}~\bibnamefont {Holzwarth}}, \bibinfo {author}
  {\bibfnamefont {Y.}~\bibnamefont {Jia}}, \bibinfo {author} {\bibfnamefont
  {F.}~\bibnamefont {Jollet}}, \bibinfo {author} {\bibfnamefont
  {W.}~\bibnamefont {Lafargue-Dit-Hauret}}, \bibinfo {author} {\bibfnamefont
  {K.}~\bibnamefont {Lejaeghere}}, \bibinfo {author} {\bibfnamefont {M.~A.~L.}\
  \bibnamefont {Marques}}, \bibinfo {author} {\bibfnamefont {A.}~\bibnamefont
  {Martin}}, \bibinfo {author} {\bibfnamefont {C.}~\bibnamefont {Martins}},
  \bibinfo {author} {\bibfnamefont {H.~P.~C.}\ \bibnamefont {Miranda}},
  \bibinfo {author} {\bibfnamefont {F.}~\bibnamefont {Naccarato}}, \bibinfo
  {author} {\bibfnamefont {K.}~\bibnamefont {Persson}}, \bibinfo {author}
  {\bibfnamefont {G.}~\bibnamefont {Petretto}}, \bibinfo {author}
  {\bibfnamefont {V.}~\bibnamefont {Planes}}, \bibinfo {author} {\bibfnamefont
  {Y.}~\bibnamefont {Pouillon}}, \bibinfo {author} {\bibfnamefont
  {S.}~\bibnamefont {Prokhorenko}}, \bibinfo {author} {\bibfnamefont
  {F.}~\bibnamefont {Ricci}}, \bibinfo {author} {\bibfnamefont {G.-M.}\
  \bibnamefont {Rignanese}}, \bibinfo {author} {\bibfnamefont {A.~H.}\
  \bibnamefont {Romero}}, \bibinfo {author} {\bibfnamefont {M.~M.}\
  \bibnamefont {Schmitt}}, \bibinfo {author} {\bibfnamefont {M.}~\bibnamefont
  {Torrent}}, \bibinfo {author} {\bibfnamefont {M.~J.}\ \bibnamefont {van
  Setten}}, \bibinfo {author} {\bibfnamefont {B.~V.}\ \bibnamefont {Troeye}},
  \bibinfo {author} {\bibfnamefont {M.~J.}\ \bibnamefont {Verstraete}},
  \bibinfo {author} {\bibfnamefont {G.}~\bibnamefont {Zérah}},\ and\ \bibinfo
  {author} {\bibfnamefont {J.~W.}\ \bibnamefont {Zwanziger}},\ }\bibfield
  {title} {\bibinfo {title} {The abinit project: Impact, environment and recent
  developments},\ }\href {https://doi.org/10.1016/j.cpc.2019.107042} {\bibfield
   {journal} {\bibinfo  {journal} {Comput. Phys. Commun.}\ }\textbf {\bibinfo
  {volume} {248}},\ \bibinfo {pages} {107042} (\bibinfo {year}
  {2020})}\BibitemShut {NoStop}%
\bibitem [{\citenamefont {Sangalli}\ \emph {et~al.}(2019)\citenamefont
  {Sangalli}, \citenamefont {Ferretti}, \citenamefont {Miranda}, \citenamefont
  {Attaccalite}, \citenamefont {Marri}, \citenamefont {Cannuccia},
  \citenamefont {Melo}, \citenamefont {Marsili}, \citenamefont {Paleari},
  \citenamefont {Marrazzo}, \citenamefont {Prandini}, \citenamefont
  {Bonf{\`{a}}}, \citenamefont {Atambo}, \citenamefont {Affinito},
  \citenamefont {Palummo}, \citenamefont {Molina-S{\'{a}}nchez}, \citenamefont
  {Hogan}, \citenamefont {Grüning}, \citenamefont {Varsano},\ and\
  \citenamefont {Marini}}]{Sangalli_2019}%
  \BibitemOpen
  \bibfield  {author} {\bibinfo {author} {\bibfnamefont {D.}~\bibnamefont
  {Sangalli}}, \bibinfo {author} {\bibfnamefont {A.}~\bibnamefont {Ferretti}},
  \bibinfo {author} {\bibfnamefont {H.}~\bibnamefont {Miranda}}, \bibinfo
  {author} {\bibfnamefont {C.}~\bibnamefont {Attaccalite}}, \bibinfo {author}
  {\bibfnamefont {I.}~\bibnamefont {Marri}}, \bibinfo {author} {\bibfnamefont
  {E.}~\bibnamefont {Cannuccia}}, \bibinfo {author} {\bibfnamefont
  {P.}~\bibnamefont {Melo}}, \bibinfo {author} {\bibfnamefont {M.}~\bibnamefont
  {Marsili}}, \bibinfo {author} {\bibfnamefont {F.}~\bibnamefont {Paleari}},
  \bibinfo {author} {\bibfnamefont {A.}~\bibnamefont {Marrazzo}}, \bibinfo
  {author} {\bibfnamefont {G.}~\bibnamefont {Prandini}}, \bibinfo {author}
  {\bibfnamefont {P.}~\bibnamefont {Bonf{\`{a}}}}, \bibinfo {author}
  {\bibfnamefont {M.~O.}\ \bibnamefont {Atambo}}, \bibinfo {author}
  {\bibfnamefont {F.}~\bibnamefont {Affinito}}, \bibinfo {author}
  {\bibfnamefont {M.}~\bibnamefont {Palummo}}, \bibinfo {author} {\bibfnamefont
  {A.}~\bibnamefont {Molina-S{\'{a}}nchez}}, \bibinfo {author} {\bibfnamefont
  {C.}~\bibnamefont {Hogan}}, \bibinfo {author} {\bibfnamefont
  {M.}~\bibnamefont {Grüning}}, \bibinfo {author} {\bibfnamefont
  {D.}~\bibnamefont {Varsano}},\ and\ \bibinfo {author} {\bibfnamefont
  {A.}~\bibnamefont {Marini}},\ }\bibfield  {title} {\bibinfo {title}
  {Many-body perturbation theory calculations using the yambo code},\ }\href
  {https://doi.org/10.1088/1361-648x/ab15d0} {\bibfield  {journal} {\bibinfo
  {journal} {Journal of Physics: Condensed Matter}\ }\textbf {\bibinfo {volume}
  {31}},\ \bibinfo {pages} {325902} (\bibinfo {year} {2019})}\BibitemShut
  {NoStop}%
\bibitem [{\citenamefont {Zhu}\ and\ \citenamefont
  {Louie}(1991)}]{PhysRevB.43.14142}%
  \BibitemOpen
  \bibfield  {author} {\bibinfo {author} {\bibfnamefont {X.}~\bibnamefont
  {Zhu}}\ and\ \bibinfo {author} {\bibfnamefont {S.~G.}\ \bibnamefont
  {Louie}},\ }\bibfield  {title} {\bibinfo {title} {Quasiparticle band
  structure of thirteen semiconductors and insulators},\ }\href
  {https://doi.org/10.1103/PhysRevB.43.14142} {\bibfield  {journal} {\bibinfo
  {journal} {Phys. Rev. B}\ }\textbf {\bibinfo {volume} {43}},\ \bibinfo
  {pages} {14142} (\bibinfo {year} {1991})}\BibitemShut {NoStop}%
\bibitem [{Fer()}]{Fermi}%
  \BibitemOpen
  \href@noop {} {}\bibinfo {note} {A form of this has been attributed to Enrico
  Fermi}\BibitemShut {NoStop}%
\bibitem [{\citenamefont {Ware}(1998)}]{Ware}%
  \BibitemOpen
  \bibfield  {author} {\bibinfo {author} {\bibfnamefont {A.}~\bibnamefont
  {Ware}},\ }\bibfield  {title} {\bibinfo {title} {Fast approximate fourier
  transforms for irregularly spaced data},\ }\href
  {https://doi.org/10.1137/S003614459731533X} {\bibfield  {journal} {\bibinfo
  {journal} {SIAM Review}\ }\textbf {\bibinfo {volume} {40}},\ \bibinfo {pages}
  {838} (\bibinfo {year} {1998})}\BibitemShut {NoStop}%
\bibitem [{Clu()}]{ClusterFootnote}%
  \BibitemOpen
  \bibinfo {note} {Specific hardware is named for identification purposes; it
  does not imply recommendation or endorsement by NIST. The CPUs used were
  Intel Xeon E5-2630 v3.}\BibitemShut {Stop}%
\bibitem [{\citenamefont {Bruneval}\ and\ \citenamefont
  {Gonze}(2008)}]{PhysRevB.78.085125}%
  \BibitemOpen
  \bibfield  {author} {\bibinfo {author} {\bibfnamefont {F.}~\bibnamefont
  {Bruneval}}\ and\ \bibinfo {author} {\bibfnamefont {X.}~\bibnamefont
  {Gonze}},\ }\bibfield  {title} {\bibinfo {title} {Accurate $gw$ self-energies
  in a plane-wave basis using only a few empty states: Towards large systems},\
  }\href {https://doi.org/10.1103/PhysRevB.78.085125} {\bibfield  {journal}
  {\bibinfo  {journal} {Phys. Rev. B}\ }\textbf {\bibinfo {volume} {78}},\
  \bibinfo {pages} {085125} (\bibinfo {year} {2008})}\BibitemShut {NoStop}%
\bibitem [{\citenamefont {Berger}\ \emph {et~al.}(2010)\citenamefont {Berger},
  \citenamefont {Reining},\ and\ \citenamefont {Sottile}}]{PhysRevB.82.041103}%
  \BibitemOpen
  \bibfield  {author} {\bibinfo {author} {\bibfnamefont {J.~A.}\ \bibnamefont
  {Berger}}, \bibinfo {author} {\bibfnamefont {L.}~\bibnamefont {Reining}},\
  and\ \bibinfo {author} {\bibfnamefont {F.}~\bibnamefont {Sottile}},\
  }\bibfield  {title} {\bibinfo {title} {Ab initio calculations of electronic
  excitations: Collapsing spectral sums},\ }\href
  {https://doi.org/10.1103/PhysRevB.82.041103} {\bibfield  {journal} {\bibinfo
  {journal} {Phys. Rev. B}\ }\textbf {\bibinfo {volume} {82}},\ \bibinfo
  {pages} {041103(R)} (\bibinfo {year} {2010})}\BibitemShut {NoStop}%
\bibitem [{\citenamefont {Berger}\ \emph {et~al.}(2012)\citenamefont {Berger},
  \citenamefont {Reining},\ and\ \citenamefont {Sottile}}]{PhysRevB.85.085126}%
  \BibitemOpen
  \bibfield  {author} {\bibinfo {author} {\bibfnamefont {J.~A.}\ \bibnamefont
  {Berger}}, \bibinfo {author} {\bibfnamefont {L.}~\bibnamefont {Reining}},\
  and\ \bibinfo {author} {\bibfnamefont {F.}~\bibnamefont {Sottile}},\
  }\bibfield  {title} {\bibinfo {title} {Efficient $gw$ calculations for
  sno${}_{2}$, zno, and rubrene: The effective-energy technique},\ }\href
  {https://doi.org/10.1103/PhysRevB.85.085126} {\bibfield  {journal} {\bibinfo
  {journal} {Phys. Rev. B}\ }\textbf {\bibinfo {volume} {85}},\ \bibinfo
  {pages} {085126} (\bibinfo {year} {2012})}\BibitemShut {NoStop}%
\bibitem [{\citenamefont {James}\ and\ \citenamefont
  {Woodley}(1996)}]{JAMES1996935}%
  \BibitemOpen
  \bibfield  {author} {\bibinfo {author} {\bibfnamefont {R.}~\bibnamefont
  {James}}\ and\ \bibinfo {author} {\bibfnamefont {S.}~\bibnamefont
  {Woodley}},\ }\bibfield  {title} {\bibinfo {title} {Extraction of green's
  functions from total energy plane wave pseudopotential calculations},\ }\href
  {https://doi.org/https://doi.org/10.1016/0038-1098(95)00815-2} {\bibfield
  {journal} {\bibinfo  {journal} {Solid State Communications}\ }\textbf
  {\bibinfo {volume} {97}},\ \bibinfo {pages} {935 } (\bibinfo {year}
  {1996})}\BibitemShut {NoStop}%
\bibitem [{\citenamefont {Steinbeck}\ \emph {et~al.}(2000)\citenamefont
  {Steinbeck}, \citenamefont {Rubio}, \citenamefont {Reining}, \citenamefont
  {Torrent}, \citenamefont {White},\ and\ \citenamefont
  {Godby}}]{STEINBECK2000105}%
  \BibitemOpen
  \bibfield  {author} {\bibinfo {author} {\bibfnamefont {L.}~\bibnamefont
  {Steinbeck}}, \bibinfo {author} {\bibfnamefont {A.}~\bibnamefont {Rubio}},
  \bibinfo {author} {\bibfnamefont {L.}~\bibnamefont {Reining}}, \bibinfo
  {author} {\bibfnamefont {M.}~\bibnamefont {Torrent}}, \bibinfo {author}
  {\bibfnamefont {I.}~\bibnamefont {White}},\ and\ \bibinfo {author}
  {\bibfnamefont {R.}~\bibnamefont {Godby}},\ }\bibfield  {title} {\bibinfo
  {title} {Enhancements to the gw space-time method},\ }\href
  {https://doi.org/https://doi.org/10.1016/S0010-4655(99)00466-X} {\bibfield
  {journal} {\bibinfo  {journal} {Computer Physics Communications}\ }\textbf
  {\bibinfo {volume} {125}},\ \bibinfo {pages} {105 } (\bibinfo {year}
  {2000})}\BibitemShut {NoStop}%
\bibitem [{\citenamefont {Samsonidze}\ \emph {et~al.}(2011)\citenamefont
  {Samsonidze}, \citenamefont {Jain}, \citenamefont {Deslippe}, \citenamefont
  {Cohen},\ and\ \citenamefont {Louie}}]{PhysRevLett.107.186404}%
  \BibitemOpen
  \bibfield  {author} {\bibinfo {author} {\bibfnamefont {G.}~\bibnamefont
  {Samsonidze}}, \bibinfo {author} {\bibfnamefont {M.}~\bibnamefont {Jain}},
  \bibinfo {author} {\bibfnamefont {J.}~\bibnamefont {Deslippe}}, \bibinfo
  {author} {\bibfnamefont {M.~L.}\ \bibnamefont {Cohen}},\ and\ \bibinfo
  {author} {\bibfnamefont {S.~G.}\ \bibnamefont {Louie}},\ }\bibfield  {title}
  {\bibinfo {title} {Simple approximate physical orbitals for $gw$
  quasiparticle calculations},\ }\href
  {https://doi.org/10.1103/PhysRevLett.107.186404} {\bibfield  {journal}
  {\bibinfo  {journal} {Phys. Rev. Lett.}\ }\textbf {\bibinfo {volume} {107}},\
  \bibinfo {pages} {186404} (\bibinfo {year} {2011})}\BibitemShut {NoStop}%
\bibitem [{\citenamefont {Klime\v{s}}\ \emph {et~al.}(2014)\citenamefont
  {Klime\v{s}}, \citenamefont {Kaltak},\ and\ \citenamefont
  {Kresse}}]{PhysRevB.90.075125}%
  \BibitemOpen
  \bibfield  {author} {\bibinfo {author} {\bibfnamefont {J.}~\bibnamefont
  {Klime\v{s}}}, \bibinfo {author} {\bibfnamefont {M.}~\bibnamefont {Kaltak}},\
  and\ \bibinfo {author} {\bibfnamefont {G.}~\bibnamefont {Kresse}},\
  }\bibfield  {title} {\bibinfo {title} {Predictive $gw$ calculations using
  plane waves and pseudopotentials},\ }\href
  {https://doi.org/10.1103/PhysRevB.90.075125} {\bibfield  {journal} {\bibinfo
  {journal} {Phys. Rev. B}\ }\textbf {\bibinfo {volume} {90}},\ \bibinfo
  {pages} {075125} (\bibinfo {year} {2014})}\BibitemShut {NoStop}%
\bibitem [{\citenamefont {Lambert}\ and\ \citenamefont
  {Giustino}(2013)}]{PhysRevB.88.075117}%
  \BibitemOpen
  \bibfield  {author} {\bibinfo {author} {\bibfnamefont {H.}~\bibnamefont
  {Lambert}}\ and\ \bibinfo {author} {\bibfnamefont {F.}~\bibnamefont
  {Giustino}},\ }\bibfield  {title} {\bibinfo {title} {Ab initio sternheimer-gw
  method for quasiparticle calculations using plane waves},\ }\href
  {https://doi.org/10.1103/PhysRevB.88.075117} {\bibfield  {journal} {\bibinfo
  {journal} {Phys. Rev. B}\ }\textbf {\bibinfo {volume} {88}},\ \bibinfo
  {pages} {075117} (\bibinfo {year} {2013})}\BibitemShut {NoStop}%
\bibitem [{\citenamefont {Wilson}\ \emph {et~al.}(2008)\citenamefont {Wilson},
  \citenamefont {Gygi},\ and\ \citenamefont {Galli}}]{PhysRevB.78.113303}%
  \BibitemOpen
  \bibfield  {author} {\bibinfo {author} {\bibfnamefont {H.~F.}\ \bibnamefont
  {Wilson}}, \bibinfo {author} {\bibfnamefont {F.}~\bibnamefont {Gygi}},\ and\
  \bibinfo {author} {\bibfnamefont {G.}~\bibnamefont {Galli}},\ }\bibfield
  {title} {\bibinfo {title} {Efficient iterative method for calculations of
  dielectric matrices},\ }\href {https://doi.org/10.1103/PhysRevB.78.113303}
  {\bibfield  {journal} {\bibinfo  {journal} {Phys. Rev. B}\ }\textbf {\bibinfo
  {volume} {78}},\ \bibinfo {pages} {113303} (\bibinfo {year}
  {2008})}\BibitemShut {NoStop}%
\bibitem [{\citenamefont {Hedin}(1965)}]{PhysRev.139.A796}%
  \BibitemOpen
  \bibfield  {author} {\bibinfo {author} {\bibfnamefont {L.}~\bibnamefont
  {Hedin}},\ }\bibfield  {title} {\bibinfo {title} {New method for calculating
  the one-particle green's function with application to the electron-gas
  problem},\ }\href {https://doi.org/10.1103/PhysRev.139.A796} {\bibfield
  {journal} {\bibinfo  {journal} {Phys. Rev.}\ }\textbf {\bibinfo {volume}
  {139}},\ \bibinfo {pages} {A796} (\bibinfo {year} {1965})}\BibitemShut
  {NoStop}%
\bibitem [{\citenamefont {Gao}\ and\ \citenamefont
  {Chelikowsky}(2019)}]{doi:10.1021/acs.jctc.9b00520}%
  \BibitemOpen
  \bibfield  {author} {\bibinfo {author} {\bibfnamefont {W.}~\bibnamefont
  {Gao}}\ and\ \bibinfo {author} {\bibfnamefont {J.~R.}\ \bibnamefont
  {Chelikowsky}},\ }\bibfield  {title} {\bibinfo {title} {Real-space based
  benchmark of g0w0 calculations on gw100: Effects of semicore orbitals and
  orbital reordering},\ }\href {https://doi.org/10.1021/acs.jctc.9b00520}
  {\bibfield  {journal} {\bibinfo  {journal} {Journal of Chemical Theory and
  Computation}\ }\textbf {\bibinfo {volume} {15}},\ \bibinfo {pages} {5299}
  (\bibinfo {year} {2019})}\BibitemShut {NoStop}%
\bibitem [{\citenamefont {Shih}\ \emph {et~al.}(2010)\citenamefont {Shih},
  \citenamefont {Xue}, \citenamefont {Zhang}, \citenamefont {Cohen},\ and\
  \citenamefont {Louie}}]{PhysRevLett.105.146401}%
  \BibitemOpen
  \bibfield  {author} {\bibinfo {author} {\bibfnamefont {B.-C.}\ \bibnamefont
  {Shih}}, \bibinfo {author} {\bibfnamefont {Y.}~\bibnamefont {Xue}}, \bibinfo
  {author} {\bibfnamefont {P.}~\bibnamefont {Zhang}}, \bibinfo {author}
  {\bibfnamefont {M.~L.}\ \bibnamefont {Cohen}},\ and\ \bibinfo {author}
  {\bibfnamefont {S.~G.}\ \bibnamefont {Louie}},\ }\bibfield  {title} {\bibinfo
  {title} {Quasiparticle band gap of zno: High accuracy from the conventional
  ${G}^{0}{W}^{0}$ approach},\ }\href
  {https://doi.org/10.1103/PhysRevLett.105.146401} {\bibfield  {journal}
  {\bibinfo  {journal} {Phys. Rev. Lett.}\ }\textbf {\bibinfo {volume} {105}},\
  \bibinfo {pages} {146401} (\bibinfo {year} {2010})}\BibitemShut {NoStop}%
\bibitem [{\citenamefont {Golze}\ \emph {et~al.}(2018)\citenamefont {Golze},
  \citenamefont {Wilhelm}, \citenamefont {van Setten},\ and\ \citenamefont
  {Rinke}}]{doi:10.1021/acs.jctc.8b00458}%
  \BibitemOpen
  \bibfield  {author} {\bibinfo {author} {\bibfnamefont {D.}~\bibnamefont
  {Golze}}, \bibinfo {author} {\bibfnamefont {J.}~\bibnamefont {Wilhelm}},
  \bibinfo {author} {\bibfnamefont {M.~J.}\ \bibnamefont {van Setten}},\ and\
  \bibinfo {author} {\bibfnamefont {P.}~\bibnamefont {Rinke}},\ }\bibfield
  {title} {\bibinfo {title} {Core-level binding energies from gw: An efficient
  full-frequency approach within a localized basis},\ }\href
  {https://doi.org/10.1021/acs.jctc.8b00458} {\bibfield  {journal} {\bibinfo
  {journal} {Journal of Chemical Theory and Computation}\ }\textbf {\bibinfo
  {volume} {14}},\ \bibinfo {pages} {4856} (\bibinfo {year}
  {2018})}\BibitemShut {NoStop}%
\bibitem [{\citenamefont {Baroni}\ \emph {et~al.}(2001)\citenamefont {Baroni},
  \citenamefont {de~Gironcoli}, \citenamefont {Dal~Corso},\ and\ \citenamefont
  {Giannozzi}}]{RevModPhys.73.515}%
  \BibitemOpen
  \bibfield  {author} {\bibinfo {author} {\bibfnamefont {S.}~\bibnamefont
  {Baroni}}, \bibinfo {author} {\bibfnamefont {S.}~\bibnamefont
  {de~Gironcoli}}, \bibinfo {author} {\bibfnamefont {A.}~\bibnamefont
  {Dal~Corso}},\ and\ \bibinfo {author} {\bibfnamefont {P.}~\bibnamefont
  {Giannozzi}},\ }\bibfield  {title} {\bibinfo {title} {Phonons and related
  crystal properties from density-functional perturbation theory},\ }\href
  {https://doi.org/10.1103/RevModPhys.73.515} {\bibfield  {journal} {\bibinfo
  {journal} {Rev. Mod. Phys.}\ }\textbf {\bibinfo {volume} {73}},\ \bibinfo
  {pages} {515} (\bibinfo {year} {2001})}\BibitemShut {NoStop}%
\bibitem [{\citenamefont {Gonze}(1997)}]{PhysRevB.55.10337}%
  \BibitemOpen
  \bibfield  {author} {\bibinfo {author} {\bibfnamefont {X.}~\bibnamefont
  {Gonze}},\ }\bibfield  {title} {\bibinfo {title} {First-principles responses
  of solids to atomic displacements and homogeneous electric fields:
  Implementation of a conjugate-gradient algorithm},\ }\href
  {https://doi.org/10.1103/PhysRevB.55.10337} {\bibfield  {journal} {\bibinfo
  {journal} {Phys. Rev. B}\ }\textbf {\bibinfo {volume} {55}},\ \bibinfo
  {pages} {10337} (\bibinfo {year} {1997})}\BibitemShut {NoStop}%
\bibitem [{\citenamefont {Gonze}\ and\ \citenamefont
  {Lee}(1997)}]{PhysRevB.55.10355}%
  \BibitemOpen
  \bibfield  {author} {\bibinfo {author} {\bibfnamefont {X.}~\bibnamefont
  {Gonze}}\ and\ \bibinfo {author} {\bibfnamefont {C.}~\bibnamefont {Lee}},\
  }\bibfield  {title} {\bibinfo {title} {Dynamical matrices, born effective
  charges, dielectric permittivity tensors, and interatomic force constants
  from density-functional perturbation theory},\ }\href
  {https://doi.org/10.1103/PhysRevB.55.10355} {\bibfield  {journal} {\bibinfo
  {journal} {Phys. Rev. B}\ }\textbf {\bibinfo {volume} {55}},\ \bibinfo
  {pages} {10355} (\bibinfo {year} {1997})}\BibitemShut {NoStop}%
\bibitem [{\citenamefont {Quong}\ and\ \citenamefont
  {Klein}(1992)}]{PhysRevB.46.10734}%
  \BibitemOpen
  \bibfield  {author} {\bibinfo {author} {\bibfnamefont {A.~A.}\ \bibnamefont
  {Quong}}\ and\ \bibinfo {author} {\bibfnamefont {B.~M.}\ \bibnamefont
  {Klein}},\ }\bibfield  {title} {\bibinfo {title} {Self-consistent-screening
  calculation of interatomic force constants and phonon dispersion curves from
  first principles},\ }\href {https://doi.org/10.1103/PhysRevB.46.10734}
  {\bibfield  {journal} {\bibinfo  {journal} {Phys. Rev. B}\ }\textbf {\bibinfo
  {volume} {46}},\ \bibinfo {pages} {10734} (\bibinfo {year}
  {1992})}\BibitemShut {NoStop}%
\bibitem [{\citenamefont {Del~Sole}\ \emph {et~al.}(1994)\citenamefont
  {Del~Sole}, \citenamefont {Reining},\ and\ \citenamefont
  {Godby}}]{PhysRevB.49.8024}%
  \BibitemOpen
  \bibfield  {author} {\bibinfo {author} {\bibfnamefont {R.}~\bibnamefont
  {Del~Sole}}, \bibinfo {author} {\bibfnamefont {L.}~\bibnamefont {Reining}},\
  and\ \bibinfo {author} {\bibfnamefont {R.~W.}\ \bibnamefont {Godby}},\
  }\bibfield  {title} {\bibinfo {title} {$gw$\ensuremath{\Gamma} approximation
  for electron self-energies in semiconductors and insulators},\ }\href
  {https://doi.org/10.1103/PhysRevB.49.8024} {\bibfield  {journal} {\bibinfo
  {journal} {Phys. Rev. B}\ }\textbf {\bibinfo {volume} {49}},\ \bibinfo
  {pages} {8024} (\bibinfo {year} {1994})}\BibitemShut {NoStop}%
\bibitem [{\citenamefont {Morris}\ \emph {et~al.}(2007)\citenamefont {Morris},
  \citenamefont {Stankovski}, \citenamefont {Delaney}, \citenamefont {Rinke},
  \citenamefont {Garc\'{\i}a-Gonz\'alez},\ and\ \citenamefont
  {Godby}}]{PhysRevB.76.155106}%
  \BibitemOpen
  \bibfield  {author} {\bibinfo {author} {\bibfnamefont {A.~J.}\ \bibnamefont
  {Morris}}, \bibinfo {author} {\bibfnamefont {M.}~\bibnamefont {Stankovski}},
  \bibinfo {author} {\bibfnamefont {K.~T.}\ \bibnamefont {Delaney}}, \bibinfo
  {author} {\bibfnamefont {P.}~\bibnamefont {Rinke}}, \bibinfo {author}
  {\bibfnamefont {P.}~\bibnamefont {Garc\'{\i}a-Gonz\'alez}},\ and\ \bibinfo
  {author} {\bibfnamefont {R.~W.}\ \bibnamefont {Godby}},\ }\bibfield  {title}
  {\bibinfo {title} {Vertex corrections in localized and extended systems},\
  }\href {https://doi.org/10.1103/PhysRevB.76.155106} {\bibfield  {journal}
  {\bibinfo  {journal} {Phys. Rev. B}\ }\textbf {\bibinfo {volume} {76}},\
  \bibinfo {pages} {155106} (\bibinfo {year} {2007})}\BibitemShut {NoStop}%
\bibitem [{\citenamefont {Bruneval}\ \emph {et~al.}(2005)\citenamefont
  {Bruneval}, \citenamefont {Sottile}, \citenamefont {Olevano}, \citenamefont
  {Del~Sole},\ and\ \citenamefont {Reining}}]{PhysRevLett.94.186402}%
  \BibitemOpen
  \bibfield  {author} {\bibinfo {author} {\bibfnamefont {F.}~\bibnamefont
  {Bruneval}}, \bibinfo {author} {\bibfnamefont {F.}~\bibnamefont {Sottile}},
  \bibinfo {author} {\bibfnamefont {V.}~\bibnamefont {Olevano}}, \bibinfo
  {author} {\bibfnamefont {R.}~\bibnamefont {Del~Sole}},\ and\ \bibinfo
  {author} {\bibfnamefont {L.}~\bibnamefont {Reining}},\ }\bibfield  {title}
  {\bibinfo {title} {Many-body perturbation theory using the density-functional
  concept: Beyond the $gw$ approximation},\ }\href
  {https://doi.org/10.1103/PhysRevLett.94.186402} {\bibfield  {journal}
  {\bibinfo  {journal} {Phys. Rev. Lett.}\ }\textbf {\bibinfo {volume} {94}},\
  \bibinfo {pages} {186402} (\bibinfo {year} {2005})}\BibitemShut {NoStop}%
\bibitem [{\citenamefont {Tiago}\ and\ \citenamefont
  {Chelikowsky}(2006)}]{PhysRevB.73.205334}%
  \BibitemOpen
  \bibfield  {author} {\bibinfo {author} {\bibfnamefont {M.~L.}\ \bibnamefont
  {Tiago}}\ and\ \bibinfo {author} {\bibfnamefont {J.~R.}\ \bibnamefont
  {Chelikowsky}},\ }\bibfield  {title} {\bibinfo {title} {Optical excitations
  in organic molecules, clusters, and defects studied by first-principles
  green's function methods},\ }\href
  {https://doi.org/10.1103/PhysRevB.73.205334} {\bibfield  {journal} {\bibinfo
  {journal} {Phys. Rev. B}\ }\textbf {\bibinfo {volume} {73}},\ \bibinfo
  {pages} {205334} (\bibinfo {year} {2006})}\BibitemShut {NoStop}%
\bibitem [{\citenamefont {Perdew}\ and\ \citenamefont
  {Zunger}(1981)}]{PhysRevB.23.5048}%
  \BibitemOpen
  \bibfield  {author} {\bibinfo {author} {\bibfnamefont {J.~P.}\ \bibnamefont
  {Perdew}}\ and\ \bibinfo {author} {\bibfnamefont {A.}~\bibnamefont
  {Zunger}},\ }\bibfield  {title} {\bibinfo {title} {Self-interaction
  correction to density-functional approximations for many-electron systems},\
  }\href {https://doi.org/10.1103/PhysRevB.23.5048} {\bibfield  {journal}
  {\bibinfo  {journal} {Phys. Rev. B}\ }\textbf {\bibinfo {volume} {23}},\
  \bibinfo {pages} {5048} (\bibinfo {year} {1981})}\BibitemShut {NoStop}%
\bibitem [{\citenamefont {Vosko}\ \emph {et~al.}(1980)\citenamefont {Vosko},
  \citenamefont {Wilk},\ and\ \citenamefont {Nusair}}]{doi:10.1139/p80-159}%
  \BibitemOpen
  \bibfield  {author} {\bibinfo {author} {\bibfnamefont {S.~H.}\ \bibnamefont
  {Vosko}}, \bibinfo {author} {\bibfnamefont {L.}~\bibnamefont {Wilk}},\ and\
  \bibinfo {author} {\bibfnamefont {M.}~\bibnamefont {Nusair}},\ }\bibfield
  {title} {\bibinfo {title} {Accurate spin-dependent electron liquid
  correlation energies for local spin density calculations: a critical
  analysis},\ }\href {https://doi.org/10.1139/p80-159} {\bibfield  {journal}
  {\bibinfo  {journal} {Canadian Journal of Physics}\ }\textbf {\bibinfo
  {volume} {58}},\ \bibinfo {pages} {1200} (\bibinfo {year}
  {1980})}\BibitemShut {NoStop}%
\bibitem [{\citenamefont {Ceperley}\ and\ \citenamefont
  {Alder}(1980)}]{PhysRevLett.45.566}%
  \BibitemOpen
  \bibfield  {author} {\bibinfo {author} {\bibfnamefont {D.~M.}\ \bibnamefont
  {Ceperley}}\ and\ \bibinfo {author} {\bibfnamefont {B.~J.}\ \bibnamefont
  {Alder}},\ }\bibfield  {title} {\bibinfo {title} {Ground state of the
  electron gas by a stochastic method},\ }\href
  {https://doi.org/10.1103/PhysRevLett.45.566} {\bibfield  {journal} {\bibinfo
  {journal} {Phys. Rev. Lett.}\ }\textbf {\bibinfo {volume} {45}},\ \bibinfo
  {pages} {566} (\bibinfo {year} {1980})}\BibitemShut {NoStop}%
\bibitem [{\citenamefont {Sakurai}(1994)}]{Sakurai}%
  \BibitemOpen
  \bibfield  {author} {\bibinfo {author} {\bibfnamefont {J.~J.}\ \bibnamefont
  {Sakurai}},\ }\href@noop {} {\emph {\bibinfo {title} {Modern Quantum
  Mechanics}}},\ edited by\ \bibinfo {editor} {\bibfnamefont {S.~F.}\
  \bibnamefont {Tuan}}\ (\bibinfo  {publisher} {Addison-Wesley Publishing
  Company, Inc.},\ \bibinfo {address} {Reading, MA},\ \bibinfo {year}
  {1994})\BibitemShut {NoStop}%
\bibitem [{\citenamefont {Pearson}(1901)}]{doi:10.1080/14786440109462720}%
  \BibitemOpen
  \bibfield  {author} {\bibinfo {author} {\bibfnamefont {K.}~\bibnamefont
  {Pearson}},\ }\bibfield  {title} {\bibinfo {title} {On lines and planes of
  closest fit to systems of points in space},\ }\href
  {https://doi.org/10.1080/14786440109462720} {\bibfield  {journal} {\bibinfo
  {journal} {The London, Edinburgh, and Dublin Philosophical Magazine and
  Journal of Science}\ }\textbf {\bibinfo {volume} {2}},\ \bibinfo {pages}
  {559} (\bibinfo {year} {1901})}\BibitemShut {NoStop}%
\bibitem [{\citenamefont {Dhillon}\ and\ \citenamefont
  {Parlett}(2004)}]{DHILLON20041}%
  \BibitemOpen
  \bibfield  {author} {\bibinfo {author} {\bibfnamefont {I.~S.}\ \bibnamefont
  {Dhillon}}\ and\ \bibinfo {author} {\bibfnamefont {B.~N.}\ \bibnamefont
  {Parlett}},\ }\bibfield  {title} {\bibinfo {title} {Multiple representations
  to compute orthogonal eigenvectors of symmetric tridiagonal matrices},\
  }\href {https://doi.org/https://doi.org/10.1016/j.laa.2003.12.028} {\bibfield
   {journal} {\bibinfo  {journal} {Linear Algebra and its Applications}\
  }\textbf {\bibinfo {volume} {387}},\ \bibinfo {pages} {1 } (\bibinfo {year}
  {2004})}\BibitemShut {NoStop}%
\bibitem [{\citenamefont {Anderson}\ \emph {et~al.}(1999)\citenamefont
  {Anderson}, \citenamefont {Bai}, \citenamefont {Bischof}, \citenamefont
  {Blackford}, \citenamefont {Demmel}, \citenamefont {Dongarra}, \citenamefont
  {Du~Croz}, \citenamefont {Greenbaum}, \citenamefont {Hammarling},
  \citenamefont {McKenney},\ and\ \citenamefont {Sorensen}}]{laug}%
  \BibitemOpen
  \bibfield  {author} {\bibinfo {author} {\bibfnamefont {E.}~\bibnamefont
  {Anderson}}, \bibinfo {author} {\bibfnamefont {Z.}~\bibnamefont {Bai}},
  \bibinfo {author} {\bibfnamefont {C.}~\bibnamefont {Bischof}}, \bibinfo
  {author} {\bibfnamefont {S.}~\bibnamefont {Blackford}}, \bibinfo {author}
  {\bibfnamefont {J.}~\bibnamefont {Demmel}}, \bibinfo {author} {\bibfnamefont
  {J.}~\bibnamefont {Dongarra}}, \bibinfo {author} {\bibfnamefont
  {J.}~\bibnamefont {Du~Croz}}, \bibinfo {author} {\bibfnamefont
  {A.}~\bibnamefont {Greenbaum}}, \bibinfo {author} {\bibfnamefont
  {S.}~\bibnamefont {Hammarling}}, \bibinfo {author} {\bibfnamefont
  {A.}~\bibnamefont {McKenney}},\ and\ \bibinfo {author} {\bibfnamefont
  {D.}~\bibnamefont {Sorensen}},\ }\href@noop {} {\emph {\bibinfo {title}
  {{LAPACK} Users' Guide}}},\ \bibinfo {edition} {3rd}\ ed.\ (\bibinfo
  {publisher} {Society for Industrial and Applied Mathematics},\ \bibinfo
  {address} {Philadelphia, PA},\ \bibinfo {year} {1999})\BibitemShut {NoStop}%
\bibitem [{\citenamefont {Sloan}\ and\ \citenamefont
  {Womersley}(2004)}]{Sloan2004}%
  \BibitemOpen
  \bibfield  {author} {\bibinfo {author} {\bibfnamefont {I.~H.}\ \bibnamefont
  {Sloan}}\ and\ \bibinfo {author} {\bibfnamefont {R.~S.}\ \bibnamefont
  {Womersley}},\ }\bibfield  {title} {\bibinfo {title} {Extremal systems of
  points and numerical integration on the sphere},\ }\href
  {https://doi.org/10.1023/B:ACOM.0000016428.25905.da} {\bibfield  {journal}
  {\bibinfo  {journal} {Advances in Computational Mathematics}\ }\textbf
  {\bibinfo {volume} {21}},\ \bibinfo {pages} {107} (\bibinfo {year}
  {2004})}\BibitemShut {NoStop}%
\bibitem [{\citenamefont {Friedrich}\ \emph {et~al.}(2011)\citenamefont
  {Friedrich}, \citenamefont {M\"uller},\ and\ \citenamefont
  {Bl\"ugel}}]{PhysRevB.83.081101}%
  \BibitemOpen
  \bibfield  {author} {\bibinfo {author} {\bibfnamefont {C.}~\bibnamefont
  {Friedrich}}, \bibinfo {author} {\bibfnamefont {M.~C.}\ \bibnamefont
  {M\"uller}},\ and\ \bibinfo {author} {\bibfnamefont {S.}~\bibnamefont
  {Bl\"ugel}},\ }\bibfield  {title} {\bibinfo {title} {Band convergence and
  linearization error correction of all-electron $\mathit{GW}$ calculations:
  The extreme case of zinc oxide},\ }\href
  {https://doi.org/10.1103/PhysRevB.83.081101} {\bibfield  {journal} {\bibinfo
  {journal} {Phys. Rev. B}\ }\textbf {\bibinfo {volume} {83}},\ \bibinfo
  {pages} {081101(R)} (\bibinfo {year} {2011})}\BibitemShut {NoStop}%
\bibitem [{\citenamefont {Rieder}\ \emph {et~al.}(2007)\citenamefont {Rieder},
  \citenamefont {Crelling}, \citenamefont {Šustai}, \citenamefont {Drábek},
  \citenamefont {Weiss},\ and\ \citenamefont {Klementová}}]{RIEDER2007115}%
  \BibitemOpen
  \bibfield  {author} {\bibinfo {author} {\bibfnamefont {M.}~\bibnamefont
  {Rieder}}, \bibinfo {author} {\bibfnamefont {J.~C.}\ \bibnamefont
  {Crelling}}, \bibinfo {author} {\bibfnamefont {O.}~\bibnamefont {Šustai}},
  \bibinfo {author} {\bibfnamefont {M.}~\bibnamefont {Drábek}}, \bibinfo
  {author} {\bibfnamefont {Z.}~\bibnamefont {Weiss}},\ and\ \bibinfo {author}
  {\bibfnamefont {M.}~\bibnamefont {Klementová}},\ }\bibfield  {title}
  {\bibinfo {title} {Arsenic in iron disulfides in a brown coal from the north
  bohemian basin, czech republic},\ }\href
  {https://doi.org/https://doi.org/10.1016/j.coal.2006.07.003} {\bibfield
  {journal} {\bibinfo  {journal} {International Journal of Coal Geology}\
  }\textbf {\bibinfo {volume} {71}},\ \bibinfo {pages} {115 } (\bibinfo {year}
  {2007})}\BibitemShut {NoStop}%
\end{thebibliography}%

\end{document}